\documentclass[11pt]{article}
\usepackage{verbatim}
\usepackage{a4}
\usepackage{a4wide}
\usepackage{amssymb}
\usepackage{amsmath}
\usepackage{color}
\usepackage{array}
\usepackage{lscape}
\usepackage{epic}
\usepackage{graphics,epsfig}
\usepackage{graphicx}
\usepackage{amsfonts}

\def\ie{{\rm i.e.,\/}\ }
\def\etc{{\rm etc.\/}\ }

\def\one{\mbox{\rm 1}\hskip-2.8pt \mbox{\rm l}}

\newcommand{\ZZ}{\mathbb{Z}}

\newcommand{\CC}{\mathbb{C}}

\title{Orders and dimensions for $sl(2)$ or $sl(3)$ module categories and Boundary Conformal Field Theories on a torus}

\author{\bf{R. Coquereaux}\thanks{~Email: \scriptsize{Robert.Coquereaux@cpt.univ-mrs.fr}} $\;${\footnote{{\it Centre de Physique Th\'eorique},{\scriptsize {\it Luminy, Marseille}. UMR 6207,  du CNRS
et des Universit\'es Aix-Marseille I, Aix-Marseille II,
et du Sud Toulon-Var,  affili\'e \`a la FRUMAM (FR 2291).}} }  
  and \bf{G. Schieber}\thanks{~Email: \scriptsize{schieber@cbpf.br}}
 $\;${\footnote{{\it CBPF -- Centro Brasileiro de Pesquisas F\'{\i}sicas},{\scriptsize {\it \, Rio de Janeiro}.}}}
}

\date{}

\begin{document}
\maketitle

\abstract{}

After giving a short description, in terms of action of categories, of some of the structures associated with $sl(2)$ and $sl(3)$  boundary conformal field theories on a torus,  we provide tables of dimensions describing the semisimple and co-semisimple blocks of the corresponding weak bialgebras (quantum groupoids),  tables of quantum dimensions and orders,  and tables describing  induction - restriction.  For reasons of size,  the $sl(3)$ tables of induction are only given for  theories  with self-fusion (existence of a monoidal structure).

\vspace{1.0cm}



\section{Introduction}

Our main and original purpose was to gather tables giving characteristic numbers for $sl(3)$ boundary conformal field theories on a torus:  dimensions of the blocks of the associated weak bialgebras, quantum dimensions, orders ("quantum mass"), induction tables,  \etc a material that is hitherto scattered in a number of publications or  unavailable.
Our tables  contain known results, that we have checked, but they also contain many new explicit formulae.
The same is true for the expressions given in the last section, just before the tables.
  The corresponding data for $sl(2)$ does not use much space, and we could easily summarize it, but the situation is different with $sl(3)$: to keep the size of this paper reasonable, we had sometimes to restrict ourselves to the case of Di Francesco - Zuber graphs with self-fusion (other cases will be described in \cite{DahmaneGil}) and give only partial results for induction tables.
Because we also needed a short introductory section discussing the underlying algebraic structures and giving our notations,  we decided to describe boundary conformal field theories on a torus
in terms of module categories (action of a monoidal category on a category)  mostly adapting the relevant material from  \cite{Ostrik}, while adding  few things like the construction of weak bialgebras in terms of Hom spaces, or the  description of the bimodule structure ${\mathcal A} \times {\mathcal O} \times  {\mathcal A}  \mapsto {\mathcal O}$, where  ${\mathcal A}$ is the fusion algebra and ${\mathcal O}$ is the Ocneanu algebra of quantum symmetries. Independently of the interest of our tables of results, 
we hope that this presentation will provide a bridge between several mathematical or physical communities interested in those topics.

\section{The stage}

In this paper ${\mathcal A}_k$ is the  fusion category of the affine algebra $\widehat{sl}(2)$, or  $\widehat{sl}(3)$, at level $k$, or \mbox{equivalently}, the category of representations with non-zero $q$-dimension for the quantum groups $SL(2)_q$ or $SL(3)_q$ at roots of unity (set $q = exp(i \pi/\kappa)$, with $\kappa = k+2$ for $sl(2)$ and $\kappa = k+3$ for $sl(3)$). This category is additive (existence of $\oplus$), monoidal (existence of $\otimes: {\mathcal A}_k \times {\mathcal A}_k  \mapsto  {\mathcal A}_k$, with associativity constraints, unit object, etc.), tensorial ($\otimes$ is a bifunctor), complex-linear, rigid (existence of duals),  finite (finitely many irreducible objects),  and semisimple, with  irreducible unit object. It is also modular  (braided, with invertible S-matrix) and ribbon (in particular balanced - or tortile). We refer to the literature \cite{MacLane}, \cite{JoyalStreet},  \cite{BakalovKirillov} for a detailed description of these structures. The Grothendieck ring of this monoidal category comes with a special basis (corresponding to simple objects), it is usually called the fusion ring,  or the Verlinde algebra.  The corresponding structure constants, encoded by the so - called fusion matrices $(N_n)^p_q$, are therefore non - negative integers: NIM-reps in CFT terminology. The rigidity property of the category implies that $(N_{\overline n})_{pq} = (N_n)_{qp} $, where $\overline n$ refers to the dual object \ie in our case, to the conjugate representation, so that the fusion ring is automatically a based $\ZZ_+$ ring in the sense of \cite{Ostrik} (maybe it would be better to call it ``rigid'').  In the case of $sl(2)$, this is a ring with one generator (corresponding to the fundamental representation), and fusion matrices are symmetric, because ${\overline n} = n$. In the case of $sl(3)$, it is convenient to use two generators that are conjugate to one another; they correspond to the two fundamental representations. Multiplication by the chosen generator is encoded by a particular fusion matrix $N_1$; it is a finite size matrix of dimension $r \times r$, with $r=k+1$ for $sl(2)$ and $r=k(k+1)/2$ for $sl(3)$. Since its elements are non negative integers, it can be interpreted as the adjacency matrix of a graph, which is the  Cayley graph of multiplication by this generator, that we call the McKay graph of the category.
Edges are non oriented in the case of $sl(2)$ (rather, they carry both orientations) and are oriented in the case of $sl(3)$. Irreducible representations are denoted by $\lambda_p$, with $p\geq 0$, in the first case, and $\lambda_{pq}$, with $p,q\geq 0$ in the next. Notice the shift of indices: the unit $\one$ of the category corresponds to $\lambda_0$ or to  $\lambda_{00}$. One should certainly keep in mind the distinction between the monoidal category, with its objects and morphisms, and its Grothendieck ring, but they will often be denoted by the same symbol. 
Actually, the Mckay graph itself is also denoted ${\mathcal A}_k$. In the $sl(2)$ case, it can be identified with the Coxeter - Dynkin diagram $A_r$, with $r=k+1$. In both $sl(2)$ and $sl(3)$ cases, it is a truncated Weyl chamber at level $k$ (a Weyl alcove). It is often useful to think of ${\mathcal A}_k$ as a category of representations of a would-be quantum object,  that can be also denoted by the same symbol, although this may be quite misleading.

The next ingredient is a category ${\mathcal E}$, not necessarily monoidal, but we suppose that it is additive, semisimple and indecomposable, on which the previous one ${\mathcal A}_k$ (which is monoidal) acts. Action of a monoidal category on a category has been described, under the name ``module categories''  by \cite{EtingofOstrik}, and, in our context, by \cite{Ostrik}. Using a slightly shorter description, we may say that we have such an action when we are given a (monoidal) functor from ${\mathcal A}_k$ to the (monoidal) category of endofunctors of ${\mathcal E}$. The reader can think of this situation as being an analogue of the action of a group on a given space. Actually,  it may be sometimes interesting to think that ${\mathcal E}$ can be acted upon in more than one way, so that we can think of the action of ${\mathcal A}_k$ as a particular ``enrichment'' of ${\mathcal E}$. The word ``module'' being used in so many different ways, we prefer to say that we have an action, or that ${\mathcal E}$ is an actegory (another nice substantive coined by R. Street), and we shall freely use both terminologies.  Irreducible objects of ${\mathcal E}$  are boundary conditions for the corresponding Conformal Field Theory specified by ${\mathcal A}_k$.  It is useful to assume, from now on, that the category ${\mathcal E}$ is indecomposable (it is not equivalent to the direct sum of two non trivial categories with  ${\mathcal A}_k$ action).
Since ${\mathcal E}$ is additive, we have a Grothendieck  group, also denoted by the same symbol. Because of the existence of an action, this (abelian) group has to be a module over the Grothendieck ring of ${\mathcal A}_k$, and it is automatically a $\ZZ_+$ module: the structure constants of the module, usually called annulus coefficients in string theory articles, or in \cite{FuchsRunkelSchweigert-I}, and described by (annular) matrices $F_n = (F_n)_{ab}$, are non negative integers.
Let us consider the class of a particular simple object of ${\mathcal A}_k$, namely the generator $n=1$, for $sl(2)$ or $n=(1,0)$ for $sl(3)$ (one of the two conjugated fundamental irreps).  We interpret  $F_1$ (or $F_{(1,0)}$)  as the adjacency matrix of a graph, called the McKay graph of the category ${\mathcal E}$.
The rigidity property of  ${\mathcal A}_k$ implies that the module ${\mathcal E}$ is rigid (or based \cite{Ostrik}). In other words:  $ (F_{\overline n})_{ab} = (F_ n)_{ba}$.  

In the case of $sl(2)$, $\ZZ_+$ modules for fusion rings at level $k$ have been classified by \cite{DiFrancescoZuber} and \cite{EtingofKhovanov};  McKay graphs are all the Coxeter - Dynkin diagram, plus some diagrams with loops (tadpoles), and  $F_1 = 2 \, \hbox{\rm Cartan matrix\/} - \one$. 
The rigidity condition in the case of $sl(2)$ implies that the matrix $F_1$ is symmetric;  this condition implies that non simply laced diagrams $B_r$, $C_r$, $F_4$ and $G_2$ should be rejected (so they do not fit in the presented framework but it is possible that they could fit in a framework of CFT where the $SL(2,\mathbb{Z})$ invariance constraint is weakened, see also \cite{EstebanRobert} and references therein):  we are therefore left with the $ADE$ diagrams and the tadpoles. A detailed analysis of the situation (\cite{Ocneanu:Unpublished}, \cite{Ostrik}) shows that the tadpole graphs do not give rise to any category endowed with an action of the  monoidal categories of type $sl(2)$. As already mentioned, the category  ${\mathcal E}$ is not required to be monoidal, but there are cases where it is, so that it has a tensor product, compatible with the ${\mathcal A}_k$ action. In another terminology, one says that the corresponding graphs have self - fusion (this is also related to the concept of flatness), or that they define ``quantum subgroups'' of $sl(2)$, whereas the others are only ``quantum modules''. Like in the classical situation, we have a restriction functor ${\mathcal A}_k \mapsto {\mathcal E}$ and an induction functor $ {\mathcal E}  \mapsto {\mathcal A}_k$. The cases where  ${\mathcal E}$ is monoidal correspond to the graphs $A_r$ (with $k=r-1$), 
$D_{even}$, with $k=0 \,  \mbox {\rm mod\/} 4$, $E_6$, with $k=10$, and $E_8$, with $k=28$. This was already known, at the level of rings,  in \cite{Pasquier}, \cite{DiFrancescoZuber} and was proved, at the categorical level, by \cite{KirillovJrOstrik}. The cases $D_{odd}$, with level $k=2 \, \mbox {\rm mod \/} 4$,  and $E_7$, at level $16$, are non monoidal actegories.  ${\mathcal A}_k$ is always modular, but the corresponding actegories are not, even when they happen to be monoidal; however, they always contain a subcategory which is modular (we shall come back to it). At the level of graphs,  the $D$ diagrams (even or odd) are $\ZZ_2$ orbifolds of the $A$ diagrams at the same level. 

In the case of $sl(3)$, the classification of $\ZZ_+$ modules over the corresponding fusion rings at level $k$  is not tractable (or not useful), however there is another route stemming from the classification of $sl(3)$ modular invariants \cite{Gannon-su3}. The graphs encoding all $sl(3)$ module categories  are called the Di Francesco -  Zuber diagrams \cite{DiFrancescoZuber}. Existence of the corresponding categories was shown by A. Ocneanu \cite{Ocneanu:Bariloche}, actually one of the candidates had to be discarded, very much like the tadpole graphs of $sl(2)$. Several $sl(3)$ actegories have monoidal structure (graphs with self - fusion), namely:  ${\mathcal A}_k$ itself, the ${\mathcal D}_k$, whose McKay diagrams are $\ZZ_3$ orbifolds of those of ${\mathcal A}_k$,  when $k$ is divisible by $3$, and three exceptional cases called ${\mathcal E}_5$, ${\mathcal E}_9$ and ${\mathcal E}_{21}$, at levels $5$, $9$ and $21$. 
The other actegories (not monoidal) are: the series ${\mathcal A}_k^c$, for which the number of simple objects is equal to the number of self dual simple objects in  ${\mathcal A}_k$, the ${\mathcal D}_k$ series, when $k=1$ or $2$ mod $3$, the series ${\mathcal D}_k^c$, for all $k$, two modules of exceptionals called ${\mathcal E}_5/3$,
${\mathcal E}_9/3$, and finally the exceptional case ${\mathcal D}_9^t$ (a generalization of $E_7$ that can be obtained from ${\mathcal D}_9$ by using an exceptional twist), along with a ``conjugated case'' called  ${{\mathcal D}_9}^{tc}$.
Some of the graphs of that system have double lines, like ${\mathcal E}_9$, so that it is not appropriate to say that Di Francesco - Zuber diagrams are the ``simply laced'' diagrams of type $sl(3)$: better to call them ``higher ADE''.
In all cases however, with self-fusion or not, the rigidity property implied by ${\mathcal A}_k$ holds (the condition $ (F_{\overline n})_{ab} = (F_ n)_{ba}$ does not forbid double lines). Taking quotients of the above diagrams by discrete groups gives higher analogues of the non $ADE$ Dynkin diagrams which  define modules over the Grothendieck ring of ${\mathcal A}_k$, but  the rigidity condition is not satisfied and the corresponding category should not exist (see however the comment that was made for $sl(2)$). Classification of $sl(4)$ module categories is also claimed to be completed \cite{Ocneanu:Bariloche}.

Let us pause to develop a tentative pedagogical analogy that should make natural the next result.
Consider a finite group, a subgroup, and the corresponding homogenous space (the space of right  cosets, for example). The group can be fibered as a principal bundle over the coset space, the structure group being the chosen subgroup. This subgroup has representations, in particular irreducible ones. For any such, let us say $a$, one can build an associated vector bundle, and consider the space $\Gamma_a$ of its sections. It carries a representation of the big group, although not irreducible (theory of induced representations). For the particular choice of the trivial representation of the small group, call it $0$, the space of sections $\Gamma_0$ is an algebra, namely the space of functions ${\mathcal F}$ on the coset. Moreover every space of sections,  say $\Gamma_a$, is a module over the algebra ${\mathcal F}$. In our case we have a non commutative geometry which is still, in a sense, finite, but the situation is similar.   Simple objects, labelled by $a$,  of the module category  ${\mathcal E}$ can be thought  as points of a graph,  as irreducible representations of a would-be quantum subgroup of $su(2)$ or $su(3)$ at some root of unity, or as spaces of sections $\Gamma_a$ above a quantum space determined by the pair $({\mathcal A}_k , {\mathcal E})$, \ie as modules over some algebra  ${\mathcal F}$ which is an algebra in a monoidal category (${\mathcal A}_k$ in our case), and right modules over ${\mathcal F}$ form an additive category $Mod_{{\mathcal A}_k} ({\mathcal F})$. Reciprocally, we have the following theorem proved in \cite{Ostrik} under actually weaker assumptions than those listed previously for the action of ${\mathcal A}_k$ on  ${\mathcal E}$:  there  exists a semisimple indecomposable algebra ${\mathcal F}$, belonging to the set of objects of ${\mathcal A}_k$ such that the module categories ${\mathcal E}$ and $Mod_{{\mathcal A}_k} ({\mathcal F})$ are equivalent.  It is shown in \cite{KirillovJrOstrik} that  ${\mathcal E}$ is monoidal (self-fusion)
if and only if ${\mathcal F}$ is commutative (terminological warning:
in \cite{KirillovJrOstrik} this property was required, and such
 algebras were called rigid, this requirement and
terminology was abandoned in \cite{Ostrik}).  
Later, we shall give the simple summands of ${\mathcal F}$ (they play the role of quantum Klein invariants) for all $sl(2)$ and $sl(3)$ actegories, in particular for those that are monoidal.
The object ${\mathcal F}$ is  a monoid in ${\mathcal A}$, but  since we shall not use the previous equivalence,  we shall not describe its structure as a monoid (in particular we shall neither give nor study its multiplication). The same algebra ${\mathcal F}$, called Frobenius algebra, or Frobenius monoid,  plays a prominent role in the approach of \cite{FuchsRunkelSchweigert-I}. Notice that module categories associated with $ADE$ Dynkin diagrams for $sl(2)$, or with Di Francesco - Zuber diagrams for $sl(3)$ are never modular (unless ${\mathcal E} ={\mathcal A} $ itself) but they contain a subcategory $J$, also denoted $Mod_{{\mathcal A}_k}^0 ({\mathcal F})$ which is modular. When ${\mathcal E}$ is monoidal,  the subring of its Grothendieck ring associated with this modular subcategory is called the modular subring (or subalgebra) and is also denoted $J$.

The third and final needed ingredient is the centralizer category of ${\mathcal E}$ with respect to the action of ${\mathcal A}_k$. It is sometimes called the ``dual category'' (not a very good name) and is defined as the category of module functors from ${\mathcal E}$ to itself: these endofunctors should be functors $F$  ``commuting'' with the action of ${\mathcal A}_k$, \ie such that $F(\lambda_n \otimes \lambda_a)$ is isomorphic with $\lambda_n \otimes F(\lambda_a)$, for $\lambda_n \in Ob({\mathcal A}_k)$ and $\lambda_a \in Ob({\mathcal E})$,  via a family of  morphisms $c_{\lambda_n, \lambda_m}$ obeying triangular and pentagonal constraints. We simply call  ${\mathcal O} = Fun_{{\mathcal A}_k}({\mathcal E},{\mathcal E})$ this centralizer category\footnote{For $sl(2)$, the structure of $Fun_{{\mathcal A}_k}({\mathcal E}_1,{\mathcal E}_2)$, where 
${\mathcal E}_{1,2}$ can be distinct module categories was obtained by \cite{Ocneanu:Unpublished}.
}, but one should remember that its definition involves both  ${\mathcal A}_k$ and  ${\mathcal E}$. 
Because of the previous compatibility property, if ${\mathcal E}$ is a left actegory over ${{\mathcal A}_k}$, it is probably better to consider it as a right actegory over ${\mathcal O}$ (actually over its opposite, since we are permuting the two factors). The category ${\mathcal O}$ is additive, semi-simple in our case, and monoidal (use composition of functors as tensor product). ${\mathcal E}$ is therefore both a module category over ${\mathcal A}_k$ and over ${\mathcal O}$.  The Grothendieck group of ${\mathcal E}$ is therefore not only a  $\ZZ_+$   module over the fusion ring, but also a $\ZZ_+$   module over the Grothendieck ring of ${\mathcal O}$,  called the Ocneanu ring (or algebra) of quantum symmetries and denoted by the same symbol. Structure constants of the ring of quantum symmetries are encoded by matrices $O_x$, called ``matrices of quantum symmetries'';  structure constants of the module, with respect to the action of quantum symmetries,  are encoded by the so called ``dual annular matrices'' $S_x$.
The next problem is to find a way to describe explicitly this centralizer category.  The solution lies in the construction of a finite dimensional weak bialgebra ${\mathcal B}$, which is going to be such that the monoidal category ${\mathcal A}_k$ can be realized as $Rep({\mathcal B})$, and also such that the monoidal category ${\mathcal O}$ can be realized as $Rep({\widehat{\mathcal B}})$ where $\widehat{\mathcal B}$ is the dual of ${\mathcal B}$. These two algebras are finite dimensional (actually semisimple in our case) and one algebra structure (say $\widehat{\mathcal B}$) can be traded against a coalgebra structure on its dual. ${\mathcal B}$ is a weak bialgebra, not a bialgebra, because $\Delta \one \neq \one \otimes \one$, where $\Delta $ is the coproduct in  ${\mathcal B}$, and $\one$ is its unit. Actually, in our cases, it is not only a weak bialgebra but a weak Hopf algebra (we can define an antipode, with the expected properties \cite{BohmSzlachanyi}, \cite{NikshychVainerman}, \cite{NikshychTuraevVainerman}, \cite{Nill}).  One categorical construction of ${\mathcal B}$ is given in \cite{Ostrik}. We propose another one that should lead to the same bialgebra, and may be simpler. Label irreducible objects $\lambda_{-}$ of categories ${\mathcal A}_k$ by ${m,n,\ldots}$,  of ${\mathcal E}$ by ${a,b,\ldots}$,  and of ${\mathcal O}$ by ${x,y,\ldots}$. Call $H_{ab}^m = Hom(\lambda_n\otimes \lambda_a, \lambda_b)$, the ``horizontal space of type $n$ from $a$ to $b$'' (also called space of essential paths of type $n$ from $a$ to $b$, space of admissible triples, or triangles$\ldots$) Call $V_{ab}^x = Hom(\lambda_a\otimes \lambda_x, \lambda_b)$ the ``vertical space of type $x$ from $a$ to $b$''. We just take these horizontal and vertical spaces as vector spaces and consider the graded sums $H^m = \sum_{ab} H_{ab}^m$ and $V^x = \sum_{ab} V_{ab}^x$. To construct the weak bialgebra, we take the (graded)  endomorphism algebras ${\mathcal B} = \sum_m End(H^m)$ and  $\widehat{{\mathcal B}} = \sum_x End(V^x)$. For obvious reasons, ${\mathcal B}$ and $\widehat{{\mathcal B}}$ are sometimes called ``algebra of double triangles''. Existence of the bialgebra structure (compatibility) rests on the properties of the pairing, or, equivalently, on the properties of the coefficients\footnote{Constructions of ${\mathcal B}$, inspired from \cite{Ocneanu:paths}, and using these properties, were given in \cite{PetkovaZuber:Oc} and \cite{CoqueTrinchero:cells}.} (Ocneanu cells) obtained by pairing two bases of matrix units\footnote{Definition of cells involve normalization choices:  the spaces $H_{ab}^m$ are not always one-dimensional, moreover  one may decide to use bases made of vectors proportional to matrix units rather than  matrix units themselves.} for the two products. Being obtained by pairing double triangles, Ocneanu cells (generalized $6J$ symbols)  are naturally associated with tetrahedra with two types (black ``b'', or white ``w'') of vertices, so that edges $bb$, $bw$ or $ww$ refer to labels $n$, $a$, $x$ of ${\mathcal A}$, ${\mathcal E}$ and ${\mathcal O}$. Using the two product and coproduct structures\footnote{
In the operator algebra community, one would usually define a star operation and a scalar product on ${\mathcal B}$, so that both products could be defined on the same underlying vector space \cite{Ocneanu:paths}.}  one can obtain representation categories $Rep({\mathcal B})$ and $Rep({\widehat{\mathcal B}})$ respectively equivalent to ${\mathcal A}_k$ and ${\mathcal O}$. As already discussed, the character ring of the first (the fusion algebra) is always described by a (generalized) Dynkin diagram of type ${\mathcal A}$.
 The character ring of the next (the algebra of quantum symmetries) is not always commutative, it has  two generators, called chiral,  in the case of $sl(2)$, together with their complex conjugated in the case of $sl(3)$.  The Cayley graph of multiplication by the two chiral generators (two types of lines),  called the Ocneanu graph of ${\mathcal E}$, encodes the structure.
 As already mentioned, ${\mathcal B}$ is weak.  This should not be a surprise  since not any monoidal category arises as representation category of a bialgebra, however (at least when the category is semisimple and has finitely many simple objects, and this is our case), it can be realized  as the category of representations of a weak bialgebra. In the next paragraph, we shall see that the knowledge of a modular invariant is sufficient to reconstruct the character rings of ${\mathcal A}_k$ (that we already know),  of ${\mathcal O}$,  and the semisimple and co semisimple structure of ${\mathcal B}$.
The ${\mathcal A}_k \times {\mathcal O}$ module corresponding to  ${\mathcal E}$ itself can be recovered from the study of the source and target subalgebras of ${\mathcal B}$.
Results obtained in operator algebra by \cite{Ocneanu:Unpublished} and  \cite{Evans-I, Evans-II, Evans&Kawahigashi} have been translated to a categorical language by \cite{Ostrik}: choice of a braiding or of the opposite braiding in the category ${\mathcal A}_k$ can be used to construct two tensor functors  from $({\mathcal A}_k)$ to ${\mathcal O}$, called $\alpha^{L,R}$ (``alpha induction''  in the language of \cite{Evans-II}).  Here our  presentation differs from \cite{Ostrik}, because we find easier to think that  there exists a functor $ {\mathcal A}_k \times {\mathcal O} \times {\mathcal A}_k  \mapsto  {\mathcal O}$, so that the previous $\alpha^{L,R}$ are obtained as particular cases ($\one_A \times \one_O \times  {\mathcal A}_k$ or to  $ {\mathcal A}_k  \times \one_O \times  \one_A $).
 At the level of Grothendieck rings, we have a bimodule property, that reads (we only use labels to denote the corresponding irreducible objects): $m \, x \, n = \sum_y \, (W_{xy})_{mn} \, y$, where $m,n$ refer to irreducible objects of  ${\mathcal A}_k$, $x,y$ to irreducible objects of ${\mathcal O}$, and where $W_{xy}$ constitute a family of so - called toric matrices, with matrix elements $(W_{xy})_{mn}$, again non negative integers.  When both $x$ and $y$ refer to the unit object (that we label $0$), one recovers the modular invariant partition function $Z=W_{00}$ of conformal field theory.
 As explained in \cite{PetkovaZuber:Oc}, when one or two indices $x$ and $y$ are non trivial, toric matrices are interpreted as partition functions on a torus,  in a conformal theory of type $ {\mathcal A}_k$,  with boundary conditions specified by ${\mathcal E}$,  but with defects (one or two) specified by $x$ and $y$. Only $Z$ is modular invariant (it commutes with the generator $S$ and $T$ of $SL(2,\ZZ)$ in the Hurwitz - Verlinde representation). Toric matrices were first introduced and calculated by Ocneanu (unpublished). Various methods to compute or define them can be found in \cite{Coque:Qtetra}, \cite{FuchsRunkelSchweigert-I}, \cite{PetkovaZuber:Oc}. Ref.  \cite{GilCoque:ADE} gives explicit expressions for all $W_{x0}$, for all members of the $sl(2)$ family.
The modular invariant partition functions themselves have been known for many years:  we have the $ADE$ classification of \cite{CIZ} for $sl(2)$, and the classification \cite{Gannon-su3} for $sl(3)$, encoded by Di Francesco - Zuber diagrams. Left and right associativity constraints $(m(n x p) q)=(mn)x(pq)  $ for the $ {\mathcal A} \times {\mathcal A}$ bimodule structure of ${\mathcal O}$ can be written in terms of fusion and toric matrices; a particular case of this equation reads
$
\sum_x (W_{0x})_{\lambda\mu} \, W_{x0} = N_{\lambda}\,Z \,N_{\mu}^{tr} \;,
$
called ``equation of modular splitting'', was presented by A.Ocneanu in Bariloche (2000). Given fusion matrices $N_p$ (known in general) and a modular invariant matrix $Z=W_{00}$, 
solving this equation, \ie finding the $W_{x0}$, allows one to reconstruct the character ring of ${\mathcal O}$. A practical method to solve this equation is given in \cite{EstebanGil}, with several $sl(3)$ examples.
Left and right chiral categories ${\mathcal O}_{L,R}$ are defined, using alpha-induction functors, as additive and monoidal subcategories of ${\mathcal O}$ whose objects are direct summands of $\alpha_{LR} (\lambda)$, for all $\lambda$ in ${\mathcal A}_k$. They are not braided but their intersection, the ambichiral subcategory ${\mathcal J}$ is.  When ${\mathcal E}$ is monoidal, the Grothendieck ring of ${\mathcal J}$, called ambichiral,   is isomorphic with the modular subalgebra $J$ already defined.

\section{Notations and miscellaneous results}

\subsection{Notations (summary)}
From now on we shall work at the level of Grothendieck groups, or rings, but use for them the same notation as for the categories themselves. So, we have a commutative and  associative  algebra ${\mathcal A}$ with\footnote{In the previous section, we used the notation $\lambda_n, \lambda_a, \lambda_x$ for $\lambda_n$, $\sigma_a$, $o_x$.}  a base ${\lambda_n}$, structure constants $(N_n)_{pq}$, an associative algebra ${\mathcal O}$, with a base ${o_x}$, structure constants $(O_x)_{yz}$, a vector space  ${\mathcal E}$ with a base $\sigma_a$ which is a module over ${\mathcal A}$ and ${\mathcal O}$, with structure constants  $(F_n)_{ab}$ and  $(S_x)_{ab}$. When ${\mathcal E}$ has self fusion, its structure constants are $(E_a)_{bc}$. The ring ${\mathcal O}$ is a ${\mathcal A}$ bimodule with structure coefficients $(W_{xy})_{m,n}$. The modular invariant partition function is $Z=W_{00}$.
Like before, the notation ${\mathcal E}$ refers to a generic example, unless it denotes an exceptional case (the context should be clear). 
${\mathcal E}$ being chosen, the numbers of irreducible objects in categories ${\mathcal A}$, ${\mathcal E}$ and ${\mathcal O}$ are respectively denoted $r_A$, $r=r_E$ and $r_O$. They are the number of vertices of the associated graphs. In the case of $\widehat{sl}(2)_k$, $r_A = k+1$. In the case of $\widehat{sl}(3)_k$, $r_A = (k+1)(k+2)/2$.
The script notation using the level $k$ as an index can be generalized to all $sl(N)$ theories but it is incompatible with the traditional notation for $sl(2)$ (the Dynkin diagrams), where  the index refers to the rank. We have the identifications: $E_6={\mathcal E}_{10}$, 
$E_7={\mathcal D}_{16}^t$ (and sometimes $\mathcal{E}_{16}$), $E_8={\mathcal E}_{28}$, 
 $A_r={\mathcal A}_{k=r-1}$, $D_{s+2} = {\mathcal D}_{k=2s}$. There are no ${\mathcal D}$ cases with odd level in the $sl(2)$ family.  Notations for higher ADE diagrams of type $sl(3)$ were given in the previous section. Generalized Coxeter numbers (altitudes) are $\kappa = k+N$  for members of  $sl(N)$ family at level $k$.
More generally, if $\mathfrak{g}$  is a Lie algebra with dual Coxeter number $h$,
we would write $\kappa = k+h$ for
all members of the higher $ADE$ system associated with the affine Lie
algebra $\widehat{\mathfrak{g} }$ at level $k$.

\subsection{Modular blocks}
Call $Z = W_{00}$ the modular invariant. It is a matrix $Z_{mn}$  indexed by (classes of)  irreducible objects $m,n \in Irr({\mathcal A})$. It can also be written as a sesquilinear quadratic form $\sum_{mn} \chi_{\overline{m}} Z_{\overline{m}n} \chi_n$ (the partition function). The following results are attributed to \cite{Ocneanu:Unpublished} and \cite{Evans-I, Evans-II, Evans&Kawahigashi}: 
$$r_E =  Tr(Z) \qquad r_O = Tr(Z Z^t)$$  The Grothendieck ring of ${\mathcal A}$ is commutative but the one of ${\mathcal O}$,  which is not necessarily commutative, is isomorphic to the direct sum of matrix algebras of sizes $Z_{mn}$.
 For example,  take $sl(2)$ at level $8$;  if we are given $Z=\vert  \chi_0 +  \chi_8 \vert^2 + \vert \chi_2 +  \chi_6\vert^2 + 2 \, \vert \chi_4\vert^2 $, we know a priori that $r_A=9$, $r_E=6$, $r_O=12$, and that  the algebra of quantum symmetries is isomorphic with $\bigoplus_{x=1}^{x=8} \CC_x \oplus M(2,\CC)$. Actually this is the $D_6$ module of $A_9$.
The collection $K$ of those irreducible objects $\lambda_n$ of  ${\mathcal A}$ that appear on the diagonal of $Z$,  with multiplicity $Z_{nn}$,  is called the multiset\footnote{Exponents are conventionally defined as indices $n$ shifted by $+1$ for $sl(2)$, or by $+(1,1)$ for $sl(3)$.}  of exponents of $Z$ (or exponents of ${\mathcal E}$).  
Forgetting multiplicities, the matrix $Z$ defines a partition on this set: two exponents $m,n$ are in the same modular block iff $Z_{mn} \neq 0$. For instance, at level 10, in the case of $E_6$, we have 
$Z=\vert  \chi_0 +  \chi_6 \vert^2 + \vert \chi_3 +  \chi_7\vert^2 +\vert \chi_4 + \chi_{10}\vert^2$ and therefore a partition of $K=\{0,6,3,7,4,10\}$ into three subsets $\{0,6\}$, $\{3,7\}$, $\{4,10\}$. At level 17, in the case of $E_7$, we have
$Z=\vert  \chi_0 +  \chi_{16} \vert^2 + \vert \chi_4 +  \chi_{12}\vert^2 +\vert \chi_6 + \chi_{10}\vert^2
+
\vert  \chi_8\vert^2
+
(\overline{\chi}_8)(\chi_2 +  \chi_{14}) +  (\overline{\chi}_2 +  \overline{\chi}_{14}) {\chi}_8
,$
 so we have a partition of $K = \{0,16,4,12,6,10,8\}$ into four subsets $\{0,16\}$, $\{4,12\}$, $\{6,10\}$, $\{8\}$.
 The modular block of the origin (containing $0$) is denoted $K_0$.
We observe that when ${\mathcal E}$  is monoidal (self-fusion), $Z$ is a sum of squares and there is a one to one correspondence between modular blocks $K_a$ and the modules $\Gamma_a$ relative to the irreducible objects $a$ belonging to the modular subalgebra $J$ of the Grothendieck ring of ${\mathcal E}$.
For instance the first modular block $K_0 = \{0,6\}$ of $E_6$ corresponds to the algebra ${\mathcal F} = \Gamma_0 = \lambda_0 \oplus \lambda_6$. This is not so when there is no self-fusion: the first modular block $K_0$ of $E_7$  is  $\{0,16\}$ although ${\mathcal F} = \Gamma_0 = \lambda_0 \oplus \lambda_8 \oplus \lambda_{16}$.

\subsection{Dimensions of horizontal and vertical spaces $H$ and $V$, dimension of the weak Hopf algebra ${\mathcal B}$}

The horizontal space $H = \bigoplus_n H_n$, coming in the construction of the first algebra structure on the weak Hopf algebra ${\mathcal B}$ was defined before, in terms of categorical data. In the $sl(2)$ case  $H$ can be realized as the (Ocneanu) vector space of essential paths  on ADE graphs but also as the vector space underlying the Gelfand-Ponomarev preprojective algebra associated with the corresponding unoriented quiver.  In the first realization $H_n$ is defined as a particular subspace of the space $Paths$ of all paths on the graph ${\mathcal E}$, whereas in the second construction it is defined as a quotient. Identification stems, for instance, from the fact that dimensions of these finite dimensional vector spaces, calculated according to the two definitions, are equal.  In the case of $sl(3)$, the grading label of the  horizontal  space $H_n$ refers to a pair of integers $(n_1,n_2)$, specifying an irreducible representation (it can also be seen as a Young tableau), this suggests a generalization of the notion of preprojective algebras associated with quivers.

In the case of $sl(2)$,  dimensions $d_n = dim H_n$, $d_H= \sum_n d_n$ and $d_{\mathcal B} = \sum_n \, d_n^2$, where $n$ runs in the set of irreducible objects of ${\mathcal A}_k$, and $d_x = dim V_x$, $d_V= \sum_x d_x$ and $d_{\widehat{{\mathcal B}}} = \sum_x \, d_x^2$, where $x$ runs in the set of irreducible objects of ${\mathcal O}$ have been calculated first  by \cite{Ocneanu:Unpublished}, then by   \cite{Coque:Qtetra}, \cite{PetkovaZuber:Oc}, \cite{GilCoque:ADE}. In the case of $sl(3)$, they have been calculated by \cite{Ocneanu:Unpublished}, \cite{GilRobert2}, \cite{Gil:Thesis} and \cite{Dahmane:Thesis}. One check, of course, that $d_{\mathcal B} = d_{\widehat{{\mathcal B}}}$ in all cases, since the underlying vector space is the same. Surprisingly, one also observes\footnote{When it is not so, in particular when the graph ${\mathcal E}$ is a $\ZZ_2$ or $\ZZ_3$ orbifold,  one knows how to ``correct'' this curious linear sum rule, which was first observed in the case of $sl(2)$  by \cite{PetkovaZuber:Oc}.}  that $d_H=d_V$ in most cases.  The collection of  known results giving $d_H$ for all $sl(2)$ cases can be condensed into a closed formula by using a recent result obtained by \cite{MOV} for the dimensions of  preprojective algebras associated with $ADE$ quivers\footnote{Warning: the preprojective algebra is a multiplicative structure on $H$, at least for $sl(2)$ cases, it cannot be identified with either ${\mathcal B}$ or $\widehat{{\mathcal B}}$, see also \cite{ArielRobert}.}.
To compute $d_n$, the pedestrian approach,  that works in all cases, is to calculate the annular matrices $F_n$, describing the module action of ${\mathcal A}_k$ on ${\mathcal E}$,  using recursion formulae giving irreps of $sl(2)$ or $sl(3)$, then to sum over all matrix elements, since $H_{ab}^m = Hom(\lambda_n\otimes \lambda_a, \lambda_b)$.  To compute $d_x$,  one has first to determine ${\mathcal O}$, for instance by solving the modular splitting equation,  then  the dual annular matrices $S_x$ describing the module action of ${\mathcal O}$ on ${\mathcal E}$,  and finally to sum over all matrix elements, since  $V_{ab}^x = Hom(\lambda_a\otimes \lambda_x, \lambda_b)$.

\subsubsection{$sl(2)$}

We shall give the values of $d_n, d_x, d_H, d_V$ and $d_{\mathcal B}$ in tables, but thanks to the above  identification with the vector space underlying the preprojective algebra of quiver theories, we have a closed formula for $d_H$ that work for all $sl(2)$ cases, namely $$d_H = \frac{\kappa (\kappa+1) r}{6}$$ Recall that $k$ is the level, $r=k+1$ is the rank (the number of vertices),  and $\kappa = k+2$ is the Coxeter number. In terms of the dimension $dim(E)$ of the Lie group corresponding to the chosen $ADE$ diagram, and using the Kostant formula $dim(E)= (\kappa+1) r$, we can also write $d_H  = \frac{ \kappa \, dim(E) }{6}$. For instance, $ { E}_6  \rightarrow 156 = 78 \frac{12}{6} \qquad  { E}_7  \rightarrow 399= 133 \frac{18}{6} \qquad  { E}_{8}  \rightarrow 1240 = 248 \frac{30}{6}$.
For $A_r$ graphs, the rank $r = k +1 = \kappa -1$ so that  $d_H(A_r)= (\kappa-1)\kappa (\kappa +1)/6$, this can be obtained directly from the fact that $d_n = (n+1)(k+1-n)$, the trivial representation being labelled $n=0$. Notice that $2 \kappa$ is the period (in $n$) of matrices $F_n$.
It is interesting to summarize how the general formula for $d_H$ is obtained in quiver theories \cite{MOV}: one constructs a generating function for the $d_n$ (matrix Hilbert series) and obtains the relation $X = A^{-1} . \Lambda$ where $A$ is the Cartan matrix of the chosen graph ($A = 2 \one - F_1$), $X$ is the sum of all annular matrices ($X = \sum_n F_n$), and $\Lambda$ is the sum of the two annular matrices corresponding to the endpoints of the fusion graph ($\Lambda  = F_0 + F_k$).
This implies $\sum_m (X)_{mn} = (\rho .  \Lambda)_n$, where $(\rho)_p = \sum_p(A^{-1} )_{mp}$ is the Weyl vector. In particular, $d_H = \sum_{m,n} (X_{mn})$ is twice the sum of matrix elements of the inverse of the Cartan matrix. This, in turn, is given by the Frendenthal - de Vries strange formula, which, using Kostant relation, can be written $\kappa(\kappa+1)r/12$, hence the result.

We also found explicit expressions giving $d_{\mathcal B}$ for the different series (see tables):
in the case of $A_{k+1}$, the given formula comes immediately from the fact that we know that $d_p = (p+1)(k+1-p)$; for the $D$ cases, the results come from the fact that $2 d_p(D) - d_p(A)$ is a superposition of two arithmetic sequences (one has to distinguish  even and odd cases).

\subsubsection{ $sl(3)$}

 There is no Lie group theory associated with the Di Francesco Zuber diagrams, nevertheless it is natural to call ``rank'' $r$, the number of vertices. In particular $r = (k+1)(k+2)/2$ for the ${\cal A}_k$ Weyl alcove. Here again, using generating functions, one obtains the relation $X = A^{-1} . \Lambda$ where $X$ is the sum of all annular matrices ($X = \sum_{mn} F_{mn}$) and $\Lambda$ is the sum of the $3(k+1)$ matrices corresponding to the boundary points of the fusion graph ($\Lambda  = \sum_{m=0}^k  F_{m,0} + \sum_{n=0}^k F_{0,n} + \sum_{m=0}^k F_{m,k-m}$, notice that the three corner vertices have multiplicity $2$), but now,  $A = 3 \one - (F_{10} + F_{01})/2$. This implies $\sum_m (X)_{mn} = (\rho .  \Lambda)_n$, where $(\rho)_p = \sum_p(A^{-1} )_{mp}$ is an higher analogue of the Weyl vector. Those relations hold for all $SU(3)$ cases.  Again, $d_H = \sum_{m,n} (X_{mn})$ and one can show -- but for ${\cal A}_k$ diagrams only -- that  the sum of matrix elements of $A^{-1}$ is equal to $r(\kappa +1)(\kappa +2)/60$.  There is no known closed formula giving $d_H$ in all cases; these dimensions are later given in tables. We however obtained the following closed result for ${\cal A}_k$ diagrams: 
$d_H = (k+1)(k+2)(k+3)(k+4)(k+5)(k^2 +6k + 14)/1680$ or, using  $\kappa=k+3$, $$d_H= (\kappa-2)(\kappa-1)\kappa(\kappa+1)(\kappa+2)(\kappa^2+5)/1680$$ 

Still in the ${\mathcal A}_k$ case, one can show that the recurrence formula $d_{(p \, q )} = d_{(p +1,q -1)}- d_{(p -q ,q -1)}+d_{(p -q ,q )}$ holds when $p\geq q+1 \geq 2$.
We did not find any simple generic formula giving the horizontal dimension $d_n$ of arbitrary $(p,q)$ blocks of ${\mathcal B}$. 
However, there are simple enough (distinct) expressions for blocks of type $(p,0)$, $(p,1)$, $(p,2)$, etc, for instance $ d_{(p,0)} =d_{(0,p)} = (k+2 - p)(k+1 - p)(1 + p)(2 + p)/4 $.  
This was enough to show that $d_{\mathcal B}$,  the sum of the $d_{(p,q)}^2$,  should be a polynomial in $k$ or $\kappa$ with rational
coefficients, and to guess several divisors (the same as for $d_H$); the coefficients of the conjectured 12th degree polynomial
that appears in the $sl(3)$ table were determined from a study of the first cases, but we checked it against direct
calculation of $d_{\mathcal B}$ (\ie determining all the $d_n$) for high values of $\kappa$.

\subsection{Quantum dimensions}
 
\subsubsection{General relations}

As usual, $q$-integers are denoted $[n] = { (q^n - q^{-n})}/{(q-q^{-1})} = \sin(n \pi /k)/\sin(\pi/k)$ with $q = exp[i \pi/\kappa]$. 
One can define quantum dimensions $\mu$ of the vertices of ${\mathcal E}$ from the Perron Frobenius eigenvector of its adjacency matrix.  If the graph ${\mathcal E}$ has self - fusion, they can be calculated  from  the quantum dimensions of the unit (which is $1$) and of the generator, which is 
$$
\beta = [2] = 2 \, cos(\pi/\kappa) \quad  \text{for } sl(2)\;, \qquad \qquad \qquad  \beta = [3] = 1 + 2 \, cos (2 \pi/\kappa) \quad  \text{for } sl(3) \;,
$$
by using the character property of $\mu$. The number $\beta^2$ is  called the Jones index. Let $\mu_a= dim_{\mathcal E}(\sigma_a)$ the quantum dimension of the vertex $\sigma_a$. 
Call also order of ${\mathcal E}$ the quantity
$$
\vert {\mathcal E} \vert = \sum_a \, \mu_a^2 
$$ 
Graphs ${\mathcal A}$ and ${\mathcal O}$ describing fusion and quantum symmetries have self - multiplication,  and quantum dimensions $\mu_n = dim_{\mathcal A}(\lambda_n)$ and $\mu_x=dim_{\mathcal O}(o_x)$  for their vertices can be obtained in a similar way.
In the particular case ${\mathcal E}={\mathcal A}$,  the $\mu_n$ can be obtained from the modular generator $S$ (a unitarizing matrix for the adjacency matrix of the graph):  $\mu_n = S_{0n}/S_{00}$.  For $sl(2)$, $\mu_n=[n+1]$. For $sl(3)$, $\mu_{p,q}=[p+1][q+1][p+q+2]/[2]$. These numbers are  called statistical dimensions in the context
of subfactor theory. Unitarity of $S$ implies 
$$
\vert {\mathcal A} \vert = \sum_n \mu_n^2 = 1/S_{00}^2
$$
and several explicit expressions given later in this section.

Call $\mathfrak{e}$ the intertwiner describing  induction - restriction between ${\mathcal A}$ and ${\mathcal E}$ (also called essential matrix relative to the unit vertex).  It is a rectangular matrix with $r_A$ lines  and $r_E$ columns. Essential matrices $\mathfrak{e}_a$ are obtained from annular matrices $F^n$ as follows: $ (\mathfrak{e}_a)_{nb} = (F^n)_{ab}$. The intertwiner is $\mathfrak{e} = \mathfrak{e}_0$.  
From induction, one obtains  $dim(\Gamma_a) = \sum_n (\mathfrak{e})_{na} \, \mu_n$. It is convenient to write $\lambda_n \uparrow \Gamma_a$ when  the space of sections $\Gamma_a$  contains $\lambda_n$ in its reduction, \ie when the integer $(\mathfrak{e})_{na} \geq 1$. Possible multiplicities being understood, we shall write
$$
dim(\Gamma_a) = \sum_{\lambda_n  \uparrow \Gamma_a} \mu_n 
$$

In particular, for the space of functions ${\mathcal F}$ over the ``discrete non commutative space'': ${\mathcal A}/{\mathcal E} $, we have  $dim(\Gamma_0) = \sum_n (\mathfrak{e})_{n0} \, \mu_n $ and we shall write (see also \cite{coqueKarpacz}):
$$
\vert {\mathcal A}/{\mathcal E} \vert = dim(\Gamma_0) = dim ({\mathcal F})  = \sum_{\lambda_n  \uparrow \Gamma_0} \mu_n
$$
Then, we have the following result : 
$$dim_{\mathcal E}(a) =  \frac{dim \, \Gamma_a}{dim \, \Gamma_0}= \frac{dim \, \Gamma_a}{\vert {\mathcal A}/{\mathcal E} \vert} $$
Moreover,  $\vert {\mathcal A}/{\mathcal E} \vert = \vert {\mathcal A} \vert / \vert {\mathcal E} \vert$.
To obtain the quantum dimensions $\mu_a$ of vertices of ${\mathcal E}$, and therefore the order $\vert {\mathcal E} \vert$,  it is therefore enough to know the $\mu_n$ for  ${\mathcal A}$ diagrams and the induction rules. 
\vskip 0.2cm

{\scriptsize
Example: take ${\mathcal E}=E_6$, and consider the central vertex $\sigma_2$ (the triple point). Its quantum dimension can be calculated using Perron Frobenius, or directly,  using $\sigma_2 = \sigma_1^2 - \sigma_0$:  $$dim_E(\sigma_2)= dim_E(\sigma_1)^2 - dim_E(\sigma_0) = (2 cos(\pi/12))^2 - 1 =   1 + \sqrt{3}$$
but one can also obtain from $A_{11}$: induction from $\sigma_0$ gives  $\Gamma_0 = \lambda_0 \oplus \lambda_6$, so that $\vert A_{11}/E_6 \vert = [1] + [7] = 3+\sqrt{3}$, and induction from $\sigma_2$ gives $\Gamma_2 = \lambda_2\oplus \lambda_4 \oplus \lambda_6 \oplus \lambda_8$, so $$dim_E(\sigma_2) = \frac{[3]+[5]+[7]+[9]}{[1] + [7]}=\frac{6+4\sqrt{3}}{3+\sqrt{3}} =  1 + \sqrt{3},.$$
}

Call $J$ the modular subalgebra of ${\mathcal E}$. For graphs with self - fusion, we have both  $\vert {\mathcal O} \vert  ={\vert {\mathcal E} \vert \times \vert {\mathcal E} \vert}/{\vert J \vert} $ and $\vert {\mathcal O} \vert  = \vert {\mathcal A} \vert $, 
so that we have also $$ \vert   {\mathcal A}_{}  \vert  /  \vert  {\mathcal E}_{}  \vert   =  \vert  {\mathcal E}_{} \vert  /  \vert    { J}\vert$$

 Using quantum dimensions of ${\mathcal E}$, one can compute $\vert J \vert = \sum_{\sigma_c \in J \subset {\mathcal E}} \; \mu_c^2 = \sum_{\sigma_c \in J} \vert \Gamma_c \vert^2 / \vert \Gamma_0 \vert^2$, but the relation for $\vert {\mathcal O} \vert$  can be written $\vert   {J}_{}  \vert =  \vert   {\mathcal A}_{}  \vert / \vert \Gamma_0 \vert^2$, so that comparing the two expressions implies
$$  \vert  A  \vert  = \sum_{c \in J} \vert \Gamma_c \vert^2$$

\subsubsection{Trigonometric identities}

Let $Z$ be a modular invariant matrix of an $sl(N)$ system at level $k$,  with or without self-fusion.   Its matrix elements $(Z)_{m,n}$  are  indexed by a pair of irreps $\lambda_m$ and $\lambda_n$. For $sl(N)$, these are of course multi-indices.
Calling, as before,  $\mu_m$ the quantum dimension of $\lambda_m$, we know that the vector $\mu$ of quantum dimensions is proportional to the first row of the $S$ matrix since $\mu_m = S_{0m}/S_{00}$. Therefore, one obtains the following (known) identity:
 
  $$  \sum_{m,n}   \, \mu_m \,  (Z)_{m,n} \, \mu_n = \sum_m  \, \mu_m^2$$
  
 Indeed, the left hand side is proportional to $(S.Z. \tilde S)_{00}$, which is equal to $(Z.S. \tilde S)_{00}$  by modular invariance, and
the result follows from unitarity of $S$ and the fact that $Z_{00}=1$.

  The right hand side of this identity $\vert {\mathcal A}_k \vert $ can be calculated from the modular matrix $S$. 
$$
\vert   {\mathcal A}_{} \vert  = \frac{\kappa}{2}  
\frac{1}{\sin ^2 \left(\frac{\pi }{\kappa }\right)} \quad \mbox{for $sl(2)$, with $\kappa = k+2$} $$
$$
\vert   {\mathcal A}_{} \vert  = \frac{3}{256} \kappa ^2 
\frac{1}{\sin ^6 \left(\frac{\pi }{\kappa }\right)} \frac{1}{\cos ^2\left(\frac{\pi }{\kappa }\right)}  \quad \mbox{for $sl(3)$, with $\kappa = k+3$}$$

Using known expressions for the quantum dimensions $\mu_n$ and  the above values for the order $\vert   {\mathcal A}_{} \vert$,  the previous identity involving the partition function $Z$ leads, after simplification by a non trivial common denominator, to a set of  trigonometric identities.

\noindent
 $ADE$ trigonometric identities, \ie identities of $sl(2)$ type, read\footnote{Warning: In these expressions, we shifted by $+(1)$, or $+(1,1)$, the indices labelling irreducible representations.}
$$  \sum_{m,n=1}^{r}   (Z)_{m,n} \, \sin(m\, \pi/\kappa) \,    \sin(n \, \pi/\kappa)  =  \kappa/2$$

\noindent
Higher  $ADE$ trigonometric identities of $sl(3)$ type read\footnotemark[\value{footnote}]:
$$  
\sum_{{}_{m=(m_1,m_2)}^{n=(n_1,n_2)}} (Z)_{m,n} \,
sin(m_1\frac{ \pi}{\kappa}) \, sin(m_2 \frac{\pi}{\kappa}) \, sin((m_1+m_2) \frac{\pi}{\kappa}) \, sin(n_1 \frac{\pi}{\kappa}) \, sin(n_2 \frac{\pi}{\kappa}) \, sin((n_1+n_2) \frac{\pi}{\kappa}) = 3 \kappa^2/64
$$
\vskip 0.3cm
  \subsubsection{Products of quantum dimensions}
Products of quantum dimensions are also of interest. In the $sl(2)$ case at level $k$, the general expression is easy to obtain. In the $sl(3)$ cases, or for other Lie algebras, they can be obtained as a by - product of a discriminant formula recently discovered by  \cite{Gepner}.
\noindent
Define: $$D = {(\sum_{\lambda_n \in  {\mathcal A}_{} } \mu_n^2)^r}/{\prod_{\lambda_n \in  {\mathcal A}_{} } \mu_n^2}$$ where $r$ is the total number of irreps of the fusion algebra,  $r = k+1$ for $sl(2)$, $r=(k+1)(k+2)/2!$ for $sl(3)$. In the case of $sl(2)$, using previous results for $\sum_{\lambda_n \in  {\mathcal A}_{} } \mu_n^2$ and a classical trigonometric identity giving   $${\prod_{\lambda_n \in  {\mathcal A}_{} }\;  \mu_n^2} = 
[{2^{-{(\kappa - 1)}} {\kappa}  \, {{(Sin(\pi/\kappa))}^{-{(\kappa - 1)}}}}]^2$$ where results are written in terms of the Coxeter number $\kappa$, one finds $$ D = 2^{\kappa -1} \kappa ^{\kappa -3} \quad \mbox{ for } sl(2)$$ 
More generally, it was shown recently in \cite{Gepner}, that $D$ is the square of the determinant of the matrix $S_{mn}/S_{m0}$ (this is the matrix of ``quantum conjugacy classes'') which, in the case of $sl(2)$ could also be defined as the discriminant of the characteristic polynomial of the adjacency matrix of the graph $A_r$.  Example:   Take $A_{11}$ of the $sl(2)$ system, the characteristic polynomial of the adjacency matrix is $-s^{11}+10 s^9-36 s^7+56 s^5-35 s^3+6 s$ and its discriminant is $10567230160896$,  indeed equal to $2^{11} \, 12^9$.  General expressions for $D$, which is always an integer, are obtained in the quoted reference.
For $sl(2)$, one recovers the previous expression, whereas \cite{Gepner} gives  $$D = 3^{ ( \kappa-2)(\kappa-1)/2} \,  \kappa^{(\kappa-4)(\kappa-2)}\quad \mbox{ for } sl(3)$$ where $\kappa = k+3$. Since $\vert {\mathcal A}_k \vert$ is already known, one can use this Gepner formula for $D$ to provide an efficient way for obtaining the square product of quantum dimensions. For $sl(3)$ we find:
$${\prod_{\lambda_n \in  {\mathcal A}^{} } \; \mu_n^2} = 
16^{-(\kappa -2) (\kappa -1)} \kappa ^{3 (\kappa -2)} \left(\cos \left(\frac{\pi }{\kappa }\right) \sin ^3\left(\frac{\pi
   }{\kappa }\right)\right)^{-(\kappa -2) (\kappa -1)}$$

\subsection{Methodological remarks}

When the McKay graph of the category ${\mathcal E}$, \ie the matrix $F_1$,  is known, all annular
matrices are obtained from the  $sl(2)$ or $sl(3)$ recurrence formulae (see for instance \cite{RobertDahmaneGilHassan}). This  determines the horizontal dimensions $d_n$ and therefore 
the block structure of the associated quantum groupoid ${\mathcal B}$ for the multiplication $\circ$.
This also gives, by induction, the quantum dimensions $\mu_a$ of irreducible objects and the order $\vert {\mathcal E} \vert$. 

In most cases  the graph of ${\mathcal E}$ is not known {\it a priori\/}:  one has to start from a given modular invariant, \ie the partition function $Z$, and solve the modular splitting equation (see for instance \cite{EstebanGil}). This determines the Ocneanu quantum symmetries, \ie the ring corresponding to the centralizer category  ${\mathcal O}$, and therefore the vertical dimensions $d_x$ describing  the block structure of  $\widehat{\mathcal B}$, for the multiplication $\widehat{\circ}$. Usually the McKay graph of ${\mathcal E}$ is itself obtained as a by product of this calculation, from the requirement that the vector space spanned by its vertices should be a module on both the fusion ring and the ring of quantum symmetries.

There is a huge number of structure constants (cells) of different kinds for ${\mathcal B}$,  respectively associated with tetrahedra with black and white vertices. The above method bypasses their calculation,  and this is fortunate because they are only known explicitly  when ${\mathcal E} = {\mathcal A}_k$, in the $sl(2)$ case. Their determination in the other cases would anyway involve   non canonical choices (bases in horizontal or vertical spaces). 

Given a graph, candidate to be the McKay graph of an $sl(3)$ actegory ${\mathcal E}$, there is a method, due to A. Ocneanu \cite{Ocneanu:Bariloche}, that allows one to see whether or not it is so.
The condition is that the graph should carry an $sl(3)$ self-connection, \ie a family of complex numbers (one for every oriented triangle) obeying two sets of conditions known as the ``small pocket'' and the ``large pocket'' equations. If these conditions fail,  either  the corresponding category does not exist, or if it exists, it is not an $sl(3)$ actegory.  Self - connections (that can also be used to  define $sl(3)$ - or higher -  analogues of  preprojective algebras)  have been determined for all $sl(3)$ cases by \cite{Ocneanu:Unpublished2000}.  These calculations, that we have later reproduced in some cases,  are not given in the tables.
 
 As mentioned in the first section, another possibility  to obtain ${\mathcal E}$ is to define an algebra ${\mathcal F}$ in the category ${\mathcal A}_k$. Our methods tell us what ${\mathcal F}$ is, as an object of ${\mathcal A}_k$, and this information appears in our $sl(3)$ tables  (and of course for $sl(2)$ as well), but we neither give nor study the multiplication map ${\mathcal F} \otimes {\mathcal F} \mapsto{\mathcal F}$. In many $sl(3)$ cases, and like for $sl(2)$,  it could be obtained from the study of conformal embeddings of affine algebras. In particular the three exceptional $sl(3)$ examples with self-fusion (monoidal structure), namely ${\mathcal E}_5$, ${\mathcal E}_9$ and
 ${\mathcal E}_{21}$ correspond to the embeddings of $A_2$, at levels $k=5,9,21$ into $A_5$, 
 $E_6$ and $E_7$ respectively, at level 1,
and the ${\mathcal D}_3$ example to an embedding
 into $spin(8)$.
The above four embeddings have been studied in the context of
subfactor theory by \cite{FengXu}. See also \cite{Wassermann}.
We remind the reader that for a given Lie algebra $\mathfrak{g}$ and a given level $k$, the conformal charge is $c = \frac{dim(\mathfrak{g}) \, k}{k+h}$ where $h$ is  the dual Coxeter number of $\mathfrak{g}$, not to be confused with the generalized Coxeter number or altitude $\kappa = k+h$ of a graph (or an actegory) associated with $\mathfrak{g}$  at level $k$. In particular, $c=3k/(k+2)$ for $sl(2)$ and $c=8k/(k+3)$ for $sl(3)$.

\subsection{Tables}

\subsubsection{The $sl(2)$ family}
There are three tables. Notations should be clear. Most results are hardly new. In particular, the dimensions $d_n$ and $d_v$ given in table $1$ were obtained already in \cite{Ocneanu:Unpublished}, \cite{Coque:Qtetra}, \cite{PetkovaZuber:Oc} and \cite{GilCoque:ADE}. 
The compact expression for $d_H$  can be seen in \cite{MOV}, but it is obtained there as the solution of a problem given in the context of quivers theory. Several quantum dimensions and orders given in table $2$ can also be found in \cite{GilCoque:ADE} and \cite{Gil:Thesis}. Nevertheless, both tables contain new entries and explicit generic formulae, for instance the closed expressions giving $d_B$.
 We also list the well known partition functions $Z$, but to save space, we use  short notations like $\vert \ldots + [n] + \ldots \vert^2$ where $[n]=(n-1)$, stands for the $\widehat{sl}(2)$ characters $\chi_n$. The same remarks hold for 
$\widehat{sl}(3)$ partition functions (also well known) recalled in the next tables.

\subsubsection{The $sl(3)$ family}

Many results concerning individual $sl(3)$ graphs can be found in the original article \cite{DiFrancescoZuber}
and in the book \cite{YellowBook}, chap.17.
Other aspects of members of this system, in particular their quantum  symmetries, are described in \cite{Ocneanu:Bariloche}, \cite{GilDahmaneHassan}, \cite{RobertDahmaneGilHassan}, \cite{EstebanGil} and \cite{Dahmane:Thesis}.
Here we are mostly interested in giving tables describing the block structure of the quantum groupoid ${\mathcal B}$, \ie dimensions $d_n$, $d_x$, quantum dimensions $\mu_n$ and orders. Many such results were certainly obtained by \cite{Ocneanu:Unpublished},  and several examples are studied in \cite{GilRobert2} and \cite{Gil:Thesis}. Results presented here are complete only for the cases with self - fusion, for which we also give tables describing induction rules; the other cases should be made available in \cite{DahmaneGil} and \cite{Dahmane:Thesis}.  Some of these results already appeared in \cite{GilRobert2}.
In both $sl(2)$ and $sl(3)$ tables, cases with self-fusion (monoidal structure) are underlined.

\section*{Acknowledgments}
One of us (R.C.) is grateful to A. Andruskiewitsch, A. Davydov , H. Montani and R. Street  for many discussions and patient explanations about categories. R.C. thanks CONICET, CBPF, IMPA  and Macquarie University for partial support.  G.S. was supported by a fellowship of FAPERJ and thanks IMPA for its kind hospitality.




\begin{landscape}
\begin{table}[H]
{\tiny$$\begin{array}{|c|c|c|c|c|c|}
\hline
{} & {} & {} & {}  & {} & {} \\
Graph  & d_n  & d_x & d_H =  \kappa(\kappa+1)r/6 & d_V - d_H  & d_{\mathcal B} = d_{\hat{\mathcal{B}}} \\
{} & {}  & {} & {} & {} & {} \\
\hline
\underline{\mathcal{A}_1} = \underline{A_2} &  (2,2) & (2,2) & 4 & 0 & 8=2^3 \\
\underline{\mathcal{A}_2} = \underline{A_3} &  (3,4,3) & (3,4,3) & 10 & 0 & 34=2^1 17^1 \\
\underline{\mathcal{A}_3} = \underline{A_4} &  (4,6,6,4) & (4,6,6,4) & 20 & 0 & 104 =2^3 13^1\\
\underline{\mathcal{A}_4} = \underline{A_5} &  (5,8,9,8,5) & (5,8,9,8,5) & 35 & 0 & 259=7^1 37^1 \\
\underline{\mathcal{A}_5} = \underline{A_6} &  (6,10,12,12,10,6) & (6,10,12,12,10,6) & 56 & 0 & 560=2^4 5^1 7^1 \\
\underline{\mathcal{A}_{10}} = \underline{A_{11}}  & (11, 20, 27, 32, 35, 36, 35, 32, 27, 20, 11) &  \ldots & 286 = 2^1 11^1 13^1 & 0 & 8294=2^1 11^1 13^1 29^1 \\
\underline{\mathcal{A}_{16}} = \underline{A_{17}} &  (17, 32, 45, 56, 65, 72, 77, 80, 81, 80, 77, 72, 65, 56, 45, 32, 17) &  \ldots& 3^1 17^1 19^1 & 0 & 3^1 5^1 13^1 17^1 19^1 \\
\underline{\mathcal{A}_{28}} = \underline{A_{29}} &   (29, 56, 81, 104, 125, 144, 161, 176, 189, 200, 209, 216, 221, 224; 225;  \mbox{sym.}) &  \ldots    & 5^1 29^1 31^1& 0 & 17^1 29^1 31^1 53^1 \\
\underline{\mathcal{A}_k} = \underline{A_{r=k+1}}  & d_n=(n+1)(k+1-n), \;\; n=0,...,k  & d_x=d_n & (\kappa-1)\kappa(\kappa+1)/6 & 0 & \kappa (\kappa^4 -1)/30  \\
\hline
\underline{\mathcal{D}_4} = \underline{D_4}  & (4,6;8;6,4) & (4,6; 4,4)\,(4,6;4,4) & 28=4^1 7^1 & (4+4) & 168=2^3 3^1 7^1 \\
\underline{\mathcal{D}_8} = \underline{D_6}  & (6,10,14,16;18;16,14,10,6) & (6,10,14,16;9,9)\,(6,10,14,16;9,9) & 110 = 2^1 5^1 11^1 & (9+9) & 1500=2^2 3^1 5^3 \\
\underline{\mathcal{D}_{16}} = \underline{D_{10}}  &  (10, 18, 26, 32, 38, 42, 46, 48; 50;  \mbox{sym.}) & (10,18,26,32,38,42,46,48;25,25)_2 & 570=2^1 3^1 5^1 19^1&  (25 + 25)  & 2^2 5501^1 \\
\underline{\mathcal{D}_{28}} = \underline{D_{16}}  & \begin{tabular}{c} (16, 30, 44, 56, 68, 78, 88, 96, 104, 110, 116, 120, 124, 126 ; 128 ;  \mbox{sym.}) \end{tabular} &  \ldots \,    \ldots & 2480=2^4 5^1 31^1 &  (64+64)  & 2^5 7757^1  \\
\underline{\mathcal{D}_k} = \underline{D_{r_{e}=\frac{k}{2}+2}}  & (r,\kappa, \ldots  ; \frac{1}{2}(1 + \frac{\kappa}{2})^2 ; \ldots , \kappa,r)   & (r,\kappa, \ldots ;  \frac{1}{4}(1 + \frac{\kappa}{2})^2, \frac{1}{4}(1 + \frac{\kappa}{2})^2)_2 & \kappa (\kappa+1)(\kappa+2)/12 & \frac{1}{2}(1 + \frac{\kappa}{2})^2  & \frac{(2+\kappa)(120+\kappa(28+\kappa(26+\kappa(17+4\kappa))))}{480}  \\
\hline
\mathcal{D}_6 = D_5  & (5,8,11,12,11,8,5) & (5,8,11,12,11,8,5) & 60=2^2 3^1 5^1 & 0 & 564=2^2 3^1 47^1 \\
\mathcal{D}_{10} = D_7  & (7,12,17,20,23,24,23,20,17,12,7) & (7,12,17,20,23,24,23,20,17,12,7) & 182 = 2^1 7^1 13^1& 0 & 3398=2^1 1699^1 \\
\mathcal{D}_k = D_{r_{o}=\frac{k}{2}+2}  & (r,\kappa, \ldots ,\kappa,r)   &  (r,k+2, \ldots ,k+2,r)  & \kappa (\kappa+1)(\kappa+2)/12 & 0 &  \frac{\kappa(176+\kappa(80+\kappa(60+\kappa(25+4\kappa))))}{480}  \\
\hline
\underline{\mathcal{E}_{10}} = \underline{E_6}  &  (6,10,14,18,20,20,20,18,14,10,6)  & \unitlength 0.40mm 
\parbox{70pt}{\begin{picture}(50,72) 
\multiput(25,5)(0,10){3}{\circle*{2}} 
\multiput(25,45)(0,10){3}{\circle{2}} 
\multiput(5,25)(0,10){3}{\circle*{2}} 
\multiput(45,25)(0,10){3}{\circle*{2}} 
\thicklines 
\put(5,45){\line(1,1){20}} 
\put(5,35){\line(1,1){20}} 
\put(5,25){\line(1,1){20}} 
\put(5,25){\line(0,1){20}}  
\put(45,45){\line(-1,-1){20}} 
\put(45,35){\line(-1,-1){20}} 
\put(45,25){\line(-1,-1){20}} 
\put(25.3,5){\line(0,1){20}} 
\dashline[20]{2}(45,45)(25,65) 
\dashline[20]{2}(45,35)(25,55) 
\dashline[20]{2}(45,25)(25,45) 
\dashline[90]{2}(45,25)(45,45) 
\dashline[20]{2}(5,45)(25,25) 
\dashline[20]{2}(5,35)(25,15) 
\dashline[20]{2}(5,25)(25,5) 
\dashline[20]{2}(26.5,5)(26.5,25) 
\put(25,68){\makebox(0,0){{\tiny 6}}} 
\put(25,58){\makebox(0,0){{\tiny 8}}} 
\put(25,48){\makebox(0,0){{\tiny 6}}} 
\put(1,45){\makebox(0,0){\tiny 10}} 
\put(1,35){\makebox(0,0){\tiny 14}} 
\put(1,25){\makebox(0,0){\tiny 10}} 
\put(49,45){\makebox(0,0){\tiny 10}} 
\put(49,35){\makebox(0,0){\tiny 14}} 
\put(49,25){\makebox(0,0){\tiny 10}} 
\put(21,25){\makebox(0,0){\tiny 20}} 
\put(21,15){\makebox(0,0){\tiny 28}} 
\put(21,5){\makebox(0,0){\tiny 20}} 
\end{picture}} & 156 = 2^2 3^1 13^1 & 0 & 2512 = 2^4 157^1\\
\hline
\mathcal{E}_{16} = E_7  & (7,12,17,22,27,30,33,34,35,34,33,30,27,22,17,12,7)  & \unitlength0.35mm
\parbox{60pt}{\begin{picture}(50,93) 
\put(25,5){\circle*{2}} 
\multiput(25,15)(0,10){2}{\circle{2}} 
\put(25,35){\circle*{2}} 
\put(25,45){\circle{2}} 
\put(25,75){\circle*{2}} 
\put(25,85){\circle{2}} 
\multiput(5,10)(0,20){2}{\circle*{2}} 
\multiput(5,60)(0,20){2}{\circle*{2}} 
\multiput(45,10)(0,20){2}{\circle*{2}} 
\multiput(45,60)(0,20){2}{\circle*{2}} 
\put(20,60){\circle{2}} 
\put(30,60){\circle{2}} 
\thicklines 
\put(5,80){\line(4,1){20}} 
\put(5,80){\line(5,-4){25}} 
\put(5,60){\line(1,0){15}} 
\put(5,60){\line(4,-3){20}} 
\dashline[100]{8}(5,60)(25,25) 
\put(5,30){\line(5,6){25}} 
\put(5,30){\line(4,-3){20}} 
\put(5,10){\line(4,3){20}} 
\put(5,10){\line(4,1){20}} 
\put(45,80){\line(-4,-1){20}} 
\put(45,60){\line(-4,3){20}} 
\put(45,60){\line(-4,-5){20}} 
\put(45,30){\line(-4,1){20}} 
\put(45,10){\line(-4,5){20}} 
\put(45,10){\line(-4,-1){20}}  
\thicklines 
\dashline[50]{2}(45,80)(25,85) 
\dashline[50]{2}(45,80)(20,60) 
\dashline[50]{2}(45,60)(30,60) 
\dashline[50]{2}(45,60)(25,45) 
\dashline[50]{2}(45,60)(25,25) 
\dashline[50]{2}(45,30)(20,60) 
\dashline[50]{2}(45,30)(25,15) 
\dashline[50]{2}(45,10)(25,25) 
\dashline[50]{2}(45,10)(25,15) 
\dashline[50]{2}(5,80)(25,75) 
\dashline[50]{2}(5,60)(25,75) 
\dashline[50]{2}(5,60)(25,35) 
\dashline[50]{2}(5,30)(25,35) 
\dashline[50]{2}(5,10)(25,35) 
\dashline[50]{2}(5,10)(25,5) 
\put(25,89){\makebox(0,0){\tiny 7}} 
\put(25,50){\makebox(0,0){\tiny 17}} 
\put(31,64){\makebox(0,0){\tiny 17}} 
\put(19,64){\makebox(0,0){\tiny 18}} 
\put(25,29){\makebox(0,0){\tiny 33}} 
\put(25,19){\makebox(0,0){\tiny 27}} 
\put(0,10){\makebox(0,0){\tiny 30}} 
\put(0,30){\makebox(0,0){\tiny 22}} 
\put(0,60){\makebox(0,0){\tiny 34}} 
\put(0,80){\makebox(0,0){\tiny 12}} 
\put(50,10){\makebox(0,0){\tiny 22}} 
\put(50,30){\makebox(0,0){\tiny 30}} 
\put(50,60){\makebox(0,0){\tiny 34}} 
\put(50,80){\makebox(0,0){\tiny 12}} 
\put(25,40){\makebox(0,0){\tiny 44}} 
\put(25,8){\makebox(0,0){\tiny 16}} 
\put(25,79){\makebox(0,0){\tiny 24}} 
\end{picture}}   & 399 = 3^1 7^1 19^1 & 0 & 10905 = 3^1 5^1 727^1 \\
\hline
\underline{\mathcal{E}_{28}} = \underline{E_8}  &    (8,14,20,26,32,38,44,48,52,56,60,62,64,64,64; \mbox{sym.})  & \unitlength 0.50mm
\parbox{120pt}{\begin{picture}(70,83) 
\multiput(5,35)(0,10){2}{\circle*{2}} 
\multiput(15,25)(0,10){4}{\circle*{2}} 
\multiput(25,15)(0,10){6}{\circle*{2}} 
\multiput(35,5)(0,10){6}{\circle*{2}} 
\multiput(35,65)(0,10){2}{\circle{2}} 
\multiput(45,15)(0,10){6}{\circle*{2}} 
\multiput(55,25)(0,10){4}{\circle*{2}} 
\multiput(65,35)(0,10){2}{\circle*{2}} 
\thicklines 
\put(35,75){\line(-1,-1){20}} 
\put(5,35){\line(1,1){30}} 
\put(5,35){\line(1,2){10}} 
\put(5,45){\line(1,0){10}} 
\put(45,65){\line(-1,-1){20}} 
\put(15,25){\line(1,1){30}} 
\put(15,25){\line(1,2){10}} 
\put(15,35){\line(1,0){10}} 
\put(55,55){\line(-1,-1){20}} 
\put(25,15){\line(1,1){30}} 
\put(25,15){\line(1,2){10}} 
\put(25,25){\line(1,0){10}} 
\put(65,45){\line(-1,-1){20}} 
\put(35,5){\line(1,1){30}} 
\put(35,5){\line(1,2){10}} 
\put(35,15){\line(1,0){10}} 
\thicklines 
\dashline[50]{2}(35,75)(55,55) 
\dashline[50]{2}(65,35)(35,65) 
\dashline[50]{2}(65,35)(55,55) 
\dashline[50]{2}(65,45)(55,45) 
\dashline[50]{2}(25,65)(45,45) 
\dashline[50]{2}(55,25)(25,55) 
\dashline[50]{2}(55,25)(45,45) 
\dashline[50]{2}(55,35)(45,35) 
\dashline[50]{2}(15,55)(35,35) 
\dashline[50]{2}(45,15)(15,45) 
\dashline[50]{2}(45,15)(35,35) 
\dashline[50]{2}(45,25)(35,25) 
\dashline[50]{2}(5,45)(25,25) 
\dashline[50]{2}(35,5)(5,35) 
\dashline[50]{2}(35,5)(25,25) 
\dashline[50]{2}(35,15)(25,15) 
\scriptsize 
\put(35,79){\makebox(0,0){{\tiny 8}}} 
\put(35,69){\makebox(0,0){\tiny 12}} 
\put(35,59){\makebox(0,0){\tiny 28}} 
\put(35,49){\makebox(0,0){\tiny 32}} 
\put(35,39){\makebox(0,0){\tiny 60}} 
\put(35,29){\makebox(0,0){\tiny 96}} 
\put(35,19){\makebox(0,0){\tiny 52}} 
\put(35.5,10){\makebox(0,0){\tiny 64}} 
\put(25,69){\makebox(0,0){\tiny 14}} 
\put(25,59){\makebox(0,0){\tiny 16}} 
\put(25,49){\makebox(0,0){\tiny 40}} 
\put(25,39){\makebox(0,0){\tiny 64}} 
\put(25,29){\makebox(0,0){\tiny 64}} 
\put(24.5,19){\makebox(0,0){\tiny 78}} 
\put(46,69){\makebox(0,0){\tiny 14}} 
\put(45,59){\makebox(0,0){\tiny 22}} 
\put(45,49){\makebox(0,0){\tiny 40}} 
\put(45,39){\makebox(0,0){\tiny 48}} 
\put(45,29){\makebox(0,0){\tiny 48}} 
\put(46,19){\makebox(0,0){\tiny 78}} 
\put(15,59){\makebox(0,0){\tiny 20}} 
\put(15,49){\makebox(0,0){\tiny 32}} 
\put(15,39){\makebox(0,0){\tiny 44}} 
\put(15,29){\makebox(0,0){\tiny 52}} 
\put(56,59){\makebox(0,0){\tiny 20}} 
\put(55,49){\makebox(0,0){\tiny 32}} 
\put(55,39){\makebox(0,0){\tiny 32}} 
\put(56,29){\makebox(0,0){\tiny 40}} 
\put(5,49){\makebox(0,0){\tiny 22}} 
\put(5,39){\makebox(0,0){\tiny 26}}  
\put(66,49){\makebox(0,0){\tiny 16}} 
\put(66.5,39){\makebox(0,0){\tiny 26}} 
\end{picture}}  & 1240 = 2^3 5^1 31& 0 & 63136 = 2^5 1973^1 \\
\hline
\end{array}$$}
\caption{Horizontal dimensions, vertical dimensions and bialgebra dimensions for $sl(2)$ cases.}
\end{table}
\normalsize
\end{landscape}



\begin{landscape}
\begin{table}[H]
{\tiny$$\begin{array}{|c|c|c|c|c|c|}
\hline
{} & {} & {} & {} & {} & {} \\
Graph & \textrm{induction} &  q-dim  &   \vert {\cal E} \vert &  \vert {\cal A}/{\cal E} \vert  &\vert {\cal J} \vert    \\
{} & {} & {} & {} & {} & {} \\
\hline
\underline{\mathcal{A}_2} = \underline{A_3} & . & 1,\sqrt{2},1 &  4 & 1 &    \vert  A_3 \vert  \\
\underline{\mathcal{A}_3} = \underline{A_4} & . & 1,\frac{1}{2}(1+\sqrt{5}),\frac{1}{2}(1+\sqrt{5}),1 &   5+\sqrt{5}  & 1&    \vert  A_4 \vert \\
\underline{\mathcal{A}_4} = \underline{A_5} & . & 1,\sqrt{3},2,\sqrt{3},1  &  12 & 1 &    \vert  A_5 \vert \\
\underline{\mathcal{A}_5} = \underline{A_6} & .  & 1,2\cos(\frac{\pi}{7}),1+2\cos(\frac{\pi}{7}),1+2\cos(\frac{\pi}{7}),1 &  18.59 & 1 &    \vert  A_6 \vert   \\
\underline{\mathcal{A}_k} = \underline{A_{r=k+1}} & . & [1],[2],\ldots,[2],[1] &  (\kappa/2) \csc ^2\left({\pi }/{\kappa}\right)  & 1 &    \vert  A_{k+1} \vert   \\
\underline{\mathcal{A}_{10}} = \underline{A_{11}} & . & 1,\sqrt{2+\sqrt{3}},1+\sqrt{3},\sqrt{3(2+\sqrt{3})},2+\sqrt{3};2\sqrt{2+\sqrt{3}};\mbox{sym.}  & 24 \left(2+\sqrt{3}\right) & 1 &   \vert  A_{11} \vert  \\
\underline{\mathcal{A}_{16}} = \underline{A_{17}} &  . &  \ldots &  9 \csc ^2\left({\pi }/{18}\right) = 298.47& 1 &   \vert  A_{17} \vert \\
\underline{\mathcal{A}_{28}} = \underline{A_{29}} &  . & \ldots  &   30 \left(12+5 \sqrt{5}+\sqrt{3 \left(85+38 \sqrt{5}\right)}\right) & 1  &   \vert  A_{29} \vert  \\
\hline
\underline{\mathcal{D}_4} = \underline{D_4} & \parbox{40pt}{\begin{picture}(40,25)(0,0)
\put(0,5){\begin{picture}(0,0)
\multiput(5,10)(15,0){2}{\circle*{1.5}}
\put(35,17,5){\circle*{1.5}}
\put(35,2,5){\circle*{1.5}}
\put(5,12){$\ast$}
\thinlines
\put(5,10){\line(1,0){15}}
\put(20,10){\line(2,1){15}}
\put(20,10){\line(2,-1){15}}
\put(5,5){\makebox(0,0){0}}
\put(5,0){\makebox(0,0){4}}
\put(20,5){\makebox(0,0){1}}
\put(20,0){\makebox(0,0){3}}
\put(35,13){\makebox(0,0){2}}
\put(35,-2){\makebox(0,0){2}}
\end{picture}}
\end{picture}} & \parbox{40pt}{\begin{picture}(40,25)(0,0)
\put(0,5){\begin{picture}(0,0)
\multiput(5,10)(15,0){2}{\circle*{1.5}}
\put(35,17,5){\circle*{1.5}}
\put(35,2,5){\circle*{1.5}}
\put(5,12){$\ast$}
\thinlines
\put(5,10){\line(1,0){15}}
\put(20,10){\line(2,1){15}}
\put(20,10){\line(2,-1){15}}
\put(5,5){\makebox(0,0){1}}
\put(20,5){\makebox(0,0){$\sqrt{3}$}}
\put(35,13){\makebox(0,0){1}}
\put(35,-2){\makebox(0,0){1}}
\end{picture}}
\end{picture}} & \frac{|A_5|}{2}=6 & 2  & \frac{|A_5|}{4}  \\
\underline{\mathcal{D}_8} = \underline{D_6} & \parbox{70pt}{\begin{picture}(70,25)(0,0)
\put(0,5){\begin{picture}(0,0)
\multiput(5,10)(15,0){4}{\circle*{1.5}}
\put(65,17,5){\circle*{1.5}}
\put(65,2,5){\circle*{1.5}}
\put(5,12){$\ast$}
\thinlines
\put(5,10){\line(1,0){45}}
\put(50,10){\line(2,1){15}}
\put(50,10){\line(2,-1){15}}
\put(5,5){\makebox(0,0){0}}
\put(5,0){\makebox(0,0){8}}
\put(20,5){\makebox(0,0){1}}
\put(20,0){\makebox(0,0){7}}
\put(35,5){\makebox(0,0){2}}
\put(35,0){\makebox(0,0){6}}
\put(50,5){\makebox(0,0){3}}
\put(50,0){\makebox(0,0){5}}
\put(65,13){\makebox(0,0){4}}
\put(65,-2){\makebox(0,0){4}}
\end{picture}}
\end{picture}}  & \sqrt{\frac{1}{2}(5+\sqrt{5})} , \frac{1}{2}(3+\sqrt{5}) , \sqrt{5+2\sqrt{5}} ; 
\begin{array}{c} \frac{1}{2}(1+\sqrt{5}) \\ \frac{1}{2}(1+\sqrt{5}) \\ {} \\ \end{array} & \frac{|A_9|}{2}=5(3+\sqrt{5})  & 2 & \frac{|A_9|}{4}   \\
\underline{\mathcal{D}_{16}} = \underline{D_{10}} &   \ldots &   [1] , [2], [3] ,[4], [5],[6], [7],[8];
[9]/2 , [9]/2  & \frac{|A_{17}|}{2}=149.23 & 2  & \frac{|A_{17}|}{4} \\
\underline{\mathcal{D}_{28}} = \underline{D_{16}} &  \ldots &  \ldots \,    \ldots & \frac{|A_{29}|}{2} & 2 & \frac{|A_{29}|}{4}  \\
{} & {} & {} & {} & {} & {} \\
\hline
{} & {} & {} & {} & {} & {} \\
\mathcal{D}_6 = D_5 & \parbox{55pt}{\begin{picture}(55,25)(0,0)
\put(0,5){\begin{picture}(0,0)
\multiput(5,10)(15,0){3}{\circle*{1.5}}
\put(50,17,5){\circle*{1.5}}
\put(50,2,5){\circle*{1.5}}
\put(5,12){$\ast$}
\thinlines
\put(5,10){\line(1,0){30}}
\put(35,10){\line(2,1){15}}
\put(35,10){\line(2,-1){15}}
\put(5,5){\makebox(0,0){0}}
\put(5,0){\makebox(0,0){6}}
\put(20,5){\makebox(0,0){1}}
\put(20,0){\makebox(0,0){5}}
\put(35,5){\makebox(0,0){2}}
\put(35,0){\makebox(0,0){4}}
\put(50,12.5){\makebox(0,0){3}}
\put(50,-2.5){\makebox(0,0){3}}
\end{picture}}
\end{picture}}  & 1,\sqrt{2+\sqrt{2}},1+\sqrt{2}; \begin{array}{c} \sqrt{1+\frac{1}{\sqrt{2}}} \\ \sqrt{1+\frac{1}{\sqrt{2}}} \end{array}  & \frac{| A_{7} |}{2} = 2 csc^2(\pi/8)
=13.65  & 2 & |A_7| \\
\mathcal{D}_{10} = D_7 & \parbox{90pt}{\begin{picture}(90,30)(0,-5)
\multiput(5,10)(15,0){5}{\circle*{1.5}}
\put(80,17,5){\circle*{1.5}}
\put(80,2,5){\circle*{1.5}}
\put(5,12){$\ast$}
\put(5,10){\line(1,0){60}}
\put(65,10){\line(2,1){15}}
\put(65,10){\line(2,-1){15}}
\put(5,5){\makebox(0,0){0}}
\put(5,0){\makebox(0,0){10}}
\put(20,5){\makebox(0,0){1}}
\put(20,0){\makebox(0,0){9}}
\put(35,5){\makebox(0,0){2}}
\put(35,0){\makebox(0,0){8}}
\put(50,5){\makebox(0,0){3}}
\put(50,0){\makebox(0,0){7}}
\put(65,5){\makebox(0,0){4}}
\put(65,0){\makebox(0,0){6}}
\put(80,12.5){\makebox(0,0){5}}
\put(80,-2.5){\makebox(0,0){5}}
\end{picture}} & 1,\sqrt{2+\sqrt{3}},1+\sqrt{3},\sqrt{3(2+\sqrt{3})},2+\sqrt{3};\begin{array}{c} \sqrt{2+\sqrt{3}} \\ \sqrt{2+\sqrt{3}} \end{array} & \frac{| A_{11} |}  {2} = 3 csc^2(\pi/12)  & 2 &  |A_{11}|  \\
{} & {} & {} & {} & {} & {} \\
\hline
\underline{\mathcal{E}_{10}} = \underline{E_6} &  \parbox{70pt}{\begin{picture}(70,50)
\put(-20,15){\begin{picture}(0,0)
\multiput(25,10)(15,0){5}{\circle*{2}}
\put(55,25){\circle*{2}}
\put(25,12){$\ast$}
\put(25,10){\line(1,0){60}}
\put(55,10){\line(0,1){15}}
\put(25,3){\makebox(0,0){$0$}}
\put(25,-2){\makebox(0,0){$6$}}
\put(40,3){\makebox(0,0){$1$}}
\put(40,-2){\makebox(0,0){$5$}}
\put(40,-7){\makebox(0,0){$7$}}
\put(55,3){\makebox(0,0){$2$}}
\put(55,-2){\makebox(0,0){$4$}}
\put(55,-7){\makebox(0,0){$6$}}
\put(55,-12){\makebox(0,0){$8$}}
\put(70,3){\makebox(0,0){$3$}}
\put(70,-2){\makebox(0,0){$5$}}
\put(70,-7){\makebox(0,0){$9$}}
\put(85,3){\makebox(0,0){$4$}}
\put(85,-2){\makebox(0,0){$10$}}
\put(63,27){\makebox(0,0){$3,7$}}
\end{picture}}
\end{picture}} &  \parbox{140pt}{\begin{picture}(140,50)
\put(-20,15){\begin{picture}(0,0)
\multiput(25,10)(30,0){5}{\circle*{2}}
\put(85,25){\circle*{2}}
\put(25,10){\line(1,0){120}}
\put(85,10){\line(0,1){15}}
\put(25,3){\makebox(0,0){$1$}}
\put(55,3){\makebox(0,0){$\sqrt{2+\sqrt{3}}\quad$}}
\put(85,3){\makebox(0,0){$1+\sqrt{3}$}}
\put(115,3){\makebox(0,0){$\quad\sqrt{2+\sqrt{3}}$}}
\put(145,3){\makebox(0,0){$1$}}
\put(93,27){\makebox(0,0){$\sqrt{2}$}}
\end{picture}}
\end{picture}} & 4 \left(3+\sqrt{3}\right)  & 3+\sqrt{3} & 4  \\
{} & {} & {} & {} & {} & {} \\
\mathcal{E}_{16} = E_7 & \parbox{90pt}{\begin{picture}(100,60)
\put(-15,25){\begin{picture}(0,0)
\multiput(25,10)(15,0){6}{\circle*{2}}
\put(70,25){\circle*{2}}
\put(25,12){$\ast$}
\thicklines
\put(25,10){\line(1,0){75}}
\put(70,10){\line(0,1){15}}
\put(25,3){\makebox(0,0){$0$}}
\put(25,-2){\makebox(0,0){$8$}}
\put(25,-7){\makebox(0,0){$16$}}
\put(40,3){\makebox(0,0){$1$}}
\put(40,-2){\makebox(0,0){$7$}}
\put(40,-7){\makebox(0,0){$9$}}
\put(40,-12){\makebox(0,0){$15$}}
\put(55,3){\makebox(0,0){$2$}}
\put(55,-2){\makebox(0,0){$6$}}
\put(55,-7){\makebox(0,0){$8$}}
\put(55,-12){\makebox(0,0){$10$}}
\put(55,-17){\makebox(0,0){$14$}}
\put(70,3){\makebox(0,0){$3$}}
\put(70,-2){\makebox(0,0){$5$}}
\put(70,-7){\makebox(0,0){$7$}}
\put(70,-12){\makebox(0,0){$9$}}
\put(70,-17){\makebox(0,0){$11$}}
\put(70,-22){\makebox(0,0){$13$}}
\put(85,3){\makebox(0,0){$4$}}
\put(85,-2){\makebox(0,0){$6$}}
\put(85,-7){\makebox(0,0){$10$}}
\put(85,-12){\makebox(0,0){$12$}}
\put(100,3){\makebox(0,0){$5$}}
\put(100,-2){\makebox(0,0){$11$}}
\put(70,30){\makebox(0,0){$4,8,12$}}
\end{picture}}
\end{picture}} & \parbox{90pt}{\begin{picture}(100,60)
\put(-30,25){\begin{picture}(0,0)
\multiput(25,10)(15,0){6}{\circle*{2}}
\put(70,25){\circle*{2}}
\put(25,12){$\ast$}
\thicklines
\put(25,10){\line(1,0){75}}
\put(70,10){\line(0,1){15}}
\put(25,1){\makebox(0,0){$[1]$}}
\put(40,1){\makebox(0,0){$[2]$}}
\put(55,1){\makebox(0,0){$[3]$}}
\put(70,1){\makebox(0,0){$[4]$}}
\put(85,1){\makebox(0,0){$\frac{[6]}{[2]}$}}
\put(100,1){\makebox(0,0){$\frac{[4]}{[3]}$}}
\put(70,35){\makebox(0,0){$\frac{[4]}{[2]}$}}
\end{picture}}
\end{picture}}  & \begin{array}{rcl} 
|\mathcal{E}| &=&   2([2]^2 +[4]^2 +[4]^2/[3]^2) \\
{} &=& 38.46\\
|A_{17}| &=& \frac{| D_{10} |  \, | D_{10} |}{| J |} \\ 
| D_{10} |  &=& \frac{| A_{17} |}{2} 
\end{array} & [1]+[9]+[17] = 7.76   & \begin{array}{l}
| J |  = \frac{| D_{10}|}{2} = \frac{| A_{17}|}{4} \\
{} {} \\
| J  |   = [1]^2 + [3]^2 + [5]^2 \\
\quad \; \;  + \, \lbrack 7 \rbrack^2 +  \frac{[9]^2}{4}  + \frac{[9]^2}{4} 
\end{array}  \\
{} & {} & {} & {} & {} & {} \\
\underline{\mathcal{E}_{28}} = \underline{E_8} & \parbox{100pt}{\begin{picture}(100,85)
\put(5,40){\begin{picture}(0,0)
\multiput(0,20)(15,0){7}{\circle*{2}}
\put(60,35){\circle*{2}}
\put(0,22){$\ast$}
\thicklines
\put(0,20){\line(1,0){90}}
\put(60,20){\line(0,1){15}}
\put(0,13){\makebox(0,0){$0$}}
\put(0,8){\makebox(0,0){$10$}}
\put(0,3){\makebox(0,0){$18$}}
\put(0,-2){\makebox(0,0){$28$}}
\put(15,13){\makebox(0,0){$1$}}
\put(15,8){\makebox(0,0){$9$}}
\put(15,3){\makebox(0,0){$11$}}
\put(15,-2){\makebox(0,0){$17$}}
\put(15,-7){\makebox(0,0){$19$}}
\put(15,-12){\makebox(0,0){$27$}}
\put(30,13){\makebox(0,0){$2$}}
\put(30,8){\makebox(0,0){$8$}}
\put(30,3){\makebox(0,0){$10$}}
\put(30,-2){\makebox(0,0){$12$}}
\put(30,-7){\makebox(0,0){$16$}}
\put(30,-12){\makebox(0,0){$18$}}
\put(30,-17){\makebox(0,0){$20$}}
\put(30,-22){\makebox(0,0){$26$}}
\put(45,13){\makebox(0,0){$3$}}
\put(45,8){\makebox(0,0){$7$}}
\put(45,3){\makebox(0,0){$9$}}
\put(45,-2){\makebox(0,0){$11$}}
\put(45,-7){\makebox(0,0){$13$}}
\put(45,-12){\makebox(0,0){$15$}}
\put(45,-17){\makebox(0,0){$17$}}
\put(45,-22){\makebox(0,0){$19$}}
\put(45,-27){\makebox(0,0){$21$}}
\put(45,-32){\makebox(0,0){$25$}}
\put(60,13){\makebox(0,0){$4$}}
\put(60,8){\makebox(0,0){$6$}}
\put(60,3){\makebox(0,0){$8$}}
\put(60,-2){\makebox(0,0){$10$}}
\put(60,-7){\makebox(0,0){$12$}}
\put(60,-12){\makebox(0,0){$(14)_2$}}
\put(60,-17){\makebox(0,0){$16$}}
\put(60,-22){\makebox(0,0){$18$}}
\put(60,-27){\makebox(0,0){$20$}}
\put(60,-32){\makebox(0,0){$22$}}
\put(60,-37){\makebox(0,0){$24$}}
\put(75,13){\makebox(0,0){$5$}}
\put(75,8){\makebox(0,0){$7$}}
\put(75,3){\makebox(0,0){$11$}}
\put(75,-2){\makebox(0,0){$13$}}
\put(75,-7){\makebox(0,0){$15$}}
\put(75,-12){\makebox(0,0){$17$}}
\put(75,-17){\makebox(0,0){$21$}}
\put(75,-22){\makebox(0,0){$23$}}
\put(90,13){\makebox(0,0){$6$}}
\put(90,8){\makebox(0,0){$12$}}
\put(90,3){\makebox(0,0){$16$}}
\put(90,-2){\makebox(0,0){$22$}}
\put(60,40){\makebox(0,0){$5,9,13,15,19,23$}}
\end{picture}}
\end{picture}} & \parbox{100pt}{\begin{picture}(100,85)
\put(-15,40){\begin{picture}(0,0)
\multiput(0,20)(15,0){7}{\circle*{2}}
\put(60,35){\circle*{2}}
\put(0,22){$\ast$}
\thicklines
\put(0,20){\line(1,0){90}}
\put(60,20){\line(0,1){15}}
\put(0,11){\makebox(0,0){$[1]$}}
\put(15,11){\makebox(0,0){$[2]$}}
\put(30,11){\makebox(0,0){$[3]$}}
\put(45,11){\makebox(0,0){$[4]$}}
\put(60,11){\makebox(0,0){$[5]$}}
\put(75,11){\makebox(0,0){$\frac{[7]}{[2]}$}}
\put(90,11){\makebox(0,0){$\frac{[5]}{[3]}$}}
\put(60,45){\makebox(0,0){$\frac{[5]}{[2]}$}}
\end{picture}}
\end{picture}} & \frac{1}{2} \left(15 \left(3+\sqrt{5}\right)+\sqrt{30 \left(65+29 \sqrt{5}\right)}\right) & \frac{1}{2} \left(3 \left(5+\sqrt{5}\right)+\sqrt{150+66 \sqrt{5}}\right) & \frac{1}{2} \left(5+\sqrt{5}\right)\\
\hline
\end{array}$$}
\caption{Quantum dimensions for $sl(2)$ cases.}
\end{table}
\end{landscape}



\begin{table}[H]
{\tiny $$\begin{array}{|c|c|c|c|c|}
\hline
{} & {} & {} & {} & {}  \\
Graph & \kappa = k+2 & r_E, r_A,r_O  &  \mathcal{Z} & {\mathcal F}  \\
{} & {} & {} & {} & {}  \\
\hline
{} & {} & {} & {} & {}  \\
\underline{\mathcal{A}_k} = \underline{A_{r=k+1}} & \boldsymbol{ k+2} &k+1, k+1,k+1 & {\displaystyle \sum_{i=1}^{k+1} |i|^2} &   \lambda_0 \\
{} & {} & {} & {} & {}  \\
\underline{\mathcal{D}_k} = \underline{D_{r_{even}=\frac{k}{2}+2}} & \boldsymbol{2r-2} &\frac{k}{2}+2, k+1,k+4  & {\displaystyle \sum_{i\, \text{odd}=1}^{\frac{k}{2}-1} |i + (k+i)|^2 + 2 |\frac{k}{2}+1|^2}
& \lambda_0 \oplus \lambda_{k}  \\
{} & {} & {} & {} & {}  \\
\mathcal{D}_k = D_{r_{odd} = \frac{k}{2}+2} & 2r-2 &\frac{k}{2}+2,  k+1,k+1    &  {\displaystyle \sum_{i\, \text{odd}=1}^{k+1} |i|^2 +|\frac{k}{2}+1|^2 + 
\sum_{i \, \text{even}=2}^{\frac{k}{2}-1}  \, i(\overline{k+i}) + (k+i)\overline{i}} &  \lambda_0 \oplus \lambda_{k} \\
{} & {} & {} & {} & {}  \\
\hline
{} & {} & {} & {} & {}  \\
\underline{\mathcal{E}_{10}} = \underline{E_6} & \boldsymbol{12} &6, 11,12 &  | 1 + 7  |^2   +    | 4 + 8  |^2   +    | 5 + 11  |^2 & 
 \lambda_0 \oplus \lambda_6 \\
{} & {} & {} & {} & {}  \\
\mathcal{E}_{16} = \mathcal{D}_{16}^t = E_7 & 18 & 7,17,17 &   | 1 + 17  |^2 +  | 5 + 13  |^2 +  | 7 + 11  |^2  +  | 9  |^2  +  ((3 + 15) \; \;  \overline{9} + h.c.)
 &  \lambda_0 \oplus \lambda_{8} \oplus \lambda_{16} \\
{} & {} & {} & {} & {}  \\
\underline{\mathcal{E}_{28}} = \underline{E_8} & \boldsymbol{30} & 8,29,32 &   | 1 + 11 + 19 + 29 |^2 +  | 7 + 13 + 17 + 23 |^2   &
 \lambda_0 \oplus \lambda_{10} \oplus \lambda_{18} \oplus \lambda_{28}  \\
{} & {} & {} & {} & {}  \\
\hline
\end{array}$$}
\caption{Complements for $sl(2)$ cases.}
\end{table}
\normalsize



\begin{landscape}
\begin{table}[H]
{\tiny$$\begin{array}{|c|c|c|c|c|c|c|c|c|}
\hline
{} & {} & {} & {} & {} & {} & {} & {} & {} \\
Graph & \kappa & r_E,r_A,r_O & d_H & d_V - d_H  & d_{\mathcal B} = d_{\hat{\mathcal{B}}} & |\mathcal{E}| & |\mathcal{A} / \mathcal{E}| & |J| \\
{} & {} & {} & {} & {} & {} & {} & {} & {} \\
\hline
{} & {} & {} & {} & {} & {} & {} & {} & {} \\
\underline{\mathcal{A}_1} & 4 & 3,3,3 & 9 & 0 & 27 & 3 & 1 & |\mathcal{A}_1| \\
\underline{\mathcal{A}_2} & 5 & 6,6,6 & 45 & 0 & 351 & \frac{3}{2}(5+\sqrt{5}) & 1 & |\mathcal{A}_2| \\
\underline{\mathcal{A}_3} & 6 & 10,10,10 & 164 & 0 & 2920 & 36 & 1 & |\mathcal{A}_3| \\
\underline{\mathcal{A}_4} & 7 & 15,15,15 & 486 & 0 & 17\,766 & 106.027 & 1 & |\mathcal{A}_4| \\
\underline{\mathcal{A}_5} & 8 & 21,21,21 & 1\,242 & 0 & 85\,644 & 48(3+2\sqrt{2}) & 1 & |\mathcal{A}_5| \\
\underline{\mathcal{A}_6} & 9 & 28,28,28 & 2\,838 & 0 & 344\,826 & 671.56 & 1 & |\mathcal{A}_6| \\
\underline{\mathcal{A}_9} & 12 & 55,55,55 & 21\,307 & 0 & 10\,517\,299 & 432(7 + 4\sqrt{3}) & 1 & |\mathcal{A}_9| \\
\underline{\mathcal{A}_{21}} & 24 & 253,253,253 & 2\,729\,870 & 0 & 41\,644\,127\,980 & 288 \left(18 + 10\sqrt{3} + \sqrt{6\left( 97 + 56 \sqrt{3} \right)^2} \, \right) & 1 &  |\mathcal{A}_{21}| \\
{} & {} & {} & {} & {} & {} & {} & {} & {} \\
\underline{\mathcal{A}_{k}} & k+3 & r_E = r_A = r_O = \frac{(k+1)(k+2)}{2} & \frac{(\kappa-2)(\kappa-1)\kappa(\kappa+1)(\kappa+2)(\kappa^2+5)}{1680} & 0 & d_{\mathcal B} (\mathcal{A}_k)  & 3 \frac{\kappa^2 csc^6
(\pi/\kappa) sec^2(\pi/\kappa)}{256} & 1 & |\mathcal{A}_k| \\
{} & {} & {} & {} & {} & {} & {} & {} & {} \\
\hline
{} & {} & {} & {} & {} & {} & {} & {} & {} \\
\mathcal{A}_1^c & 4 & 1,3,3 & 3 & 0 & 3 & 1 & 3 & |\mathcal{A}_1| \\
\mathcal{A}_2^c & 5 & 2,6,6 & 15 & 0 & 39 & \frac{1}{2}(5-\sqrt{5}) & \frac{3}{2}(3+\sqrt{5}) & |\mathcal{A}_2| \\
\mathcal{A}_3^c & 6 & 2,10,10 & 36 & 0 & 144 & 2 & 18 & |\mathcal{A}_3| \\
\mathcal{A}_4^c & 7 & 3,15,15 & 102 & 0 & 798 & 2.863 & 37.034 & |\mathcal{A}_4| \\
\mathcal{A}_5^c & 8 & 3,21,21 & 204 & 0 & 2376 & 4 & 12(3+2\sqrt{2}) & |\mathcal{A}_5| \\
\mathcal{A}_6^c & 9 & 4,28,28 & 442 & 0 & 8578 & 5.446 & 123.321 & |\mathcal{A}_6| \\
\mathcal{A}_7^c & 10 & 4,36,36 & 780 & 0 & 21360 & 5+\sqrt{5} & 15(7+3\sqrt{5}) & |\mathcal{A}_7| \\
{} & {} & {} & {} & {} & {} & {} & {} & {} \\
\mathcal{A}_{k\geq 1}^c & k+3 & \lceil \frac{k+1}{2} \rceil,r_A,r_A & \left\lbrace\begin{array}{rl}
\frac{(\kappa -2)\kappa^2(\kappa +2)(\kappa^2 +4)}{1280} & \mbox{k odd} \\
\frac{(\kappa -1)(\kappa+1)((\kappa-1)^2 +4)((\kappa+1)^2+4)}{1280} & \mbox{k even}
\end{array} \right. & 0 & . & . & . & |\mathcal{A}_k| \\
{} & {} & {} & {} & {} & {} & {} & {} & {} \\
\hline
{} & {} & {} & {} & {} & {} & {} & {} & {} \\
\mathcal{D}_1 & 4 & 1,3,3 & \frac{1}{3} d_H(\mathcal{A}_1) = 3 & 0 & \frac{1}{9}d_{\mathcal{B}(\mathcal{A}_1)} = 3 & \frac{1}{3}|\mathcal{A}_1|=1 & 3 & |\mathcal{A}_1| \\
\mathcal{D}_2 & 5 & 2,6,6 & \frac{1}{3} d_H(\mathcal{A}_2) = 15 & 0 & \frac{1}{9}d_{\mathcal{B}(\mathcal{A}_2)} = 39 & \frac{1}{3}|\mathcal{A}_2|=\frac{1}{2}(5+\sqrt{5}) & 3 & |\mathcal{A}_2| \\
\mathcal{D}_4 & 7 & 5,15,15 & \frac{1}{3} d_H(\mathcal{A}_4) = 162 & 0 & \frac{1}{9}d_{\mathcal{B}(\mathcal{A}_4)} = 1974 & \frac{1}{3}|\mathcal{A}_4|=35.342 & 3 & |\mathcal{A}_4| \\
\mathcal{D}_5 & 8 & 7,21,21 & \frac{1}{3} d_H(\mathcal{A}_5) = 414 & 0 & \frac{1}{9}d_{\mathcal{B}(\mathcal{A}_5)} = 9516 & \frac{1}{3}|\mathcal{A}_5|= 16(3+2\sqrt{2}) & 3 & |\mathcal{A}_5| \\
{} & {} & {} & {} & {} & {} & {} & {} & {} \\
\mathcal{D}_{k=1,2\mbox{mod}3} & k+3 & \frac{1}{3}r_A,r_A,r_A & \frac{1}{3} d_H(\mathcal{A}_k) & 0 & \frac{1}{9}d_{\mathcal{B}(\mathcal{A}_k)} & \frac{1}{3}|\mathcal{A}_k| & 3 & |\mathcal{A}_k| \\
{} & {} & {} & {} & {} & {} & {} & {} & {} \\
\hline
{} & {} & {} & {} & {} & {} & {} & {} & {} \\
\mathcal{D}_1^c & 4 & 3,3,3 & 3\,d_H(\mathcal{A}_1^c)=9 & 0 & 9\,d_{\mathcal B}(\mathcal{A}_1^c)=27 & 3 & 1 & |\mathcal{A}_1| \\
\mathcal{D}_2^c & 5 & 6,6,6 & 3\,d_H(\mathcal{A}_2^c)=45 & 0 & 9\,d_{\mathcal B}(\mathcal{A}_2^c)=351 & \frac{3}{2}(5-\sqrt{5}) & \frac{1}{2}(3+\sqrt{5}) & |\mathcal{A}_2| \\
\mathcal{D}_4^c & 7 & 9,15,15 & 3\,d_H(\mathcal{A}_4^c)=306 & 0 & 9\,d_{\mathcal B}(\mathcal{A}_4^c)=7\,182 & 8.589 & 12.344 & |\mathcal{A}_4| \\
\mathcal{D}_5^c & 8 & 9,21,21 & 3\,d_H(\mathcal{A}_5^c)=612 & 0 & 9\,d_{\mathcal B}(\mathcal{A}_5^c)=21\,384 & 12 & 4(3+2\sqrt{2}) & |\mathcal{A}_5| \\
{} & {} & {} & {} & {} & {} & {} & {} & {} \\
\mathcal{D}_{k=1,2\mbox{mod}3}^c & k+3 & 3r_{A^c},r_A,r_A & 3\,d_H(\mathcal{A}_k^c) & 0 & 9\,d_{\mathcal B}(\mathcal{A}_k^c) & 3 |\mathcal{A}_k^c| & . & |\mathcal{A}_k| \\
{} & {} & {} & {} & {} & {} & {} & {} & {} \\
\hline

{} & {} & {} & {} & {} & {} & {} & {} & {} \\
\underline{\mathcal{D}_3} & 6 & 6,10,18 & 96 & 36 & 1\,032 & \frac{1}{3}|\mathcal{A}_3|=12 & 3 & 4 \\
\underline{\mathcal{D}_6} & 9 & 12,28,36 & 1\,218 & 174 & 64\,698 & \frac{1}{3}|\mathcal{A}_6|=223.853 & 3 & 74.618 \\
\underline{\mathcal{D}_9} & 12 & 21,55,63 & 8\,193 & 622 & 1\,573\,275 & \frac{1}{3}|\mathcal{A}_9| = 144 (7 + 4\sqrt{3}) & 3 & 48(7+4\sqrt{3}) \\
{} & {} & {} & {} & {} & {} & {} & {} & {} \\
\underline{\mathcal{D}_{k=0\mbox{mod}3}} & k+3 & \frac{r_A-1}{3}+3,r_A, 3r_E & . & . & . & \frac{1}{3}|\mathcal{A}_k| & 3 & \frac{1}{3}|\mathcal{E}|=\frac{1}{9}|\mathcal{A}_k| \\
{} & {} & {} & {} & {} & {} & {} & {} & {} \\
\hline

{} & {} & {} & {} & {} & {} & {} & {} & {} \\
\mathcal{D}_3^c & 6 & 6,10,18 & 3\,d_H(\mathcal{A}_3^c)=108 & 36 & 9\,d_{\mathcal B}(\mathcal{A}_3^c)=1\,296 & 6 & 6 & 4 \\
\mathcal{D}_6^c & 9  & 12,28,36  & 3\,d_H(\mathcal{A}_6^c)= 1326& 186 & 9\,d_{\mathcal B}(\mathcal{A}_6^c)=77\,202 & 16.338 & 41.104 & 74.618 \\
\mathcal{D}_9^c & 12 & 15,55,63 & 3\,d_H(\mathcal{A}_9^c)=6993 & . & 9\,d_{\mathcal B}(\mathcal{A}_9^c)=1\,167\,291 & 36 & 12(7+4\sqrt{3}) & 48(7+4\sqrt{3}) \\
{} & {} & {} & {} & {} & {} & {} & {} & {} \\
\mathcal{D}_{k=0\mbox{mod}3}^c & k+3 & 3r_{A^c},r_A,r_O(\mathcal{D}_k) & 3\,d_H(\mathcal{A}_k^c) & . & 9\,d_{\mathcal B}(\mathcal{A}_k^c) & 3 |\mathcal{A}_k^c| & . & \frac{1}{9}|\mathcal{A}_k| \\
{} & {} & {} & {} & {} & {} & {} & {} & {} \\
\hline

\end{array}$$}
\caption{Dimensions and quantum masses for  $sl(3)$ cases : $\mathcal{A}$, $\mathcal{D}$ series and conjugated.
 {\scriptsize $\quad \quad \quad  d_{\mathcal B} (\mathcal{A}_k) = \frac{(\kappa-2)(\kappa-1)\kappa^2(\kappa+1)(\kappa+2)(1052 + 325
\kappa^2 + 58 \kappa^4 + 5 \kappa^6)}{4435200}$}}
\end{table}
\normalsize
\end{landscape}

\begin{landscape}
\begin{table}[H]
{\scriptsize$$\begin{array}{|c|c|c|c|c|c|c|c|c|}
\hline
{} & {} & {} & {} & {} & {} & {} & {} & {} \\
Graph & \kappa & r_E,r_A,r_O & d_H & d_V - d_H  & d_{\mathcal B} = d_{\hat{\mathcal{B}}} & |\mathcal{E}| & |\mathcal{A} / \mathcal{E}| & |J| \\
{} & {} & {} & {} & {} & {} & {} & {} & {} \\
\hline
{} & {} & {} & {} & {} & {} & {} & {} & {} \\
\mathcal{D}_9^t & 12 & 17,55,63 & 7\,001 & 176 & 1\,167\,355 & 72 (2 + \sqrt{3}) & 6(2 + \sqrt{3}) & \frac{1}{9}|\mathcal{A}_9| = \frac{1}{3}|\mathcal{D}_9| = 48(7+4\sqrt{3}) \\
\mathcal{D}_9^{tc} & 12 & 11,55,63 & 4\,713 & . & 531\,435 & 36 & 12(7+4\sqrt{3}) & \frac{1}{9}|\mathcal{A}_9| \\
{} & {} & {} & {} & {} & {} & {} & {} & {} \\
\underline{\mathcal{E}_5} & 8 & 12,21,24 & 720 & 0 & 29\,376 & 12(2+\sqrt{2}) & 2(2+\sqrt{2}) & 6 \\
\mathcal{E}_5/3 & 8 & 4,21,24 & \frac{1}{3} \, d_H(\mathcal{E}{_5}) = 240 & 0 & \frac{1}{9} \, d_{\mathcal B}(\mathcal{E}_5) =  3\,264 & \frac{1}{3}|\mathcal{E}_5| = 4(2+\sqrt{2}) & 6(2+\sqrt{2}) & 6 \\
{} & {} & {} & {} & {} & {} & {} & {} & {} \\
\underline{\mathcal{E}_9} & 12 & 12,55,72 & 4\,656 & 792 & 518\,976 & 36(2+\sqrt{3}) & 12(2+\sqrt{3}) & 3 \\
\mathcal{E}_9/3 & 12 & 12,55,72 & 5\,616 & 936 & 754\,272 & \frac{1}{3}{|\mathcal{E}_9|} = 12(2+\sqrt{3}) & 36(2+\sqrt{3}) & 3 \\
{} & {} & {} & {} & {} & {} & {} & {} & {} \\
\underline{\mathcal{E}_{21}} & 24 & 24,253,288 & 288\,576 & 0 & 480\,701\,952 & 24\left(18 + 10\sqrt{3} + \sqrt{6\left( 97 + 56 \sqrt{3} \right)} \, \right) & 
12\left(18 + 10\sqrt{3} + \sqrt{6\left( 97 + 56 \sqrt{3} \right)} \, \right) & 
2 \\
{} & {} & {} & {} & {} & {} & {} & {} & {} \\
\hline
\end{array}$$}
\caption{Dimensions and quantum masses for exceptional $sl(3)$ cases.}
\end{table}
\end{landscape}
\normalsize



\begin{figure}[H] 
\begin{center} 
\unitlength0.12cm

\begin{picture}(110,60)
\put(0,0){\line(1,0){110}}
\put(0,0){\line(0,1){60}}
\put(110,60){\line(-1,0){110}}
\put(110,60){\line(0,-1){60}}

\put(0,0){\begin{picture}(0,0)
%

\put(10.5,29.5){\line(1,0){19}}
\put(10.5,30.5){\line(1,0){19}}
\put(20,29.5){\vector(1,0){0.5}}
\put(20,30.5){\vector(1,0){0.5}}
\put(29.5,31){\line(-1,2){9}}
\put(29.5,31){\vector(-1,2){5}}
\put(19.5,49){\line(-1,-2){9}}
\put(19.5,49){\vector(-1,-2){5}}
\put(10,10){\line(0,1){18.75}}
\put(30,10){\line(0,1){18.75}}
\put(10,10){\line(1,1){19.25}}
\put(30,10){\line(-1,1){19.25}}
\put(20,10){\line(1,2){9.5}}
\put(20,10){\line(-1,2){9.5}}
\put(10,23.5){\vector(0,1){0.5}}
\put(30,22.5){\vector(0,-1){0.5}}
\put(22.5,22.5){\vector(-1,-1){0.5}}
\put(17,23){\vector(-1,1){0.5}}
\put(26.5,23){\vector(-1,-2){0.5}}
\put(13.5,23){\vector(-1,2){0.5}}

\put(20,50){\circle{2}}
\put(10,10){\circle{2}}
\put(20,10){\circle{2}}
\put(30,10){\circle{2}}
\put(10,30){\circle*{1.2}}
\put(10,30){\circle{2}}
\put(30,30){\circle*{2}}

\put(20,55){\makebox(0,0){$0$}}  
\put(5,30){\makebox(0,0){$1$}}  
\put(35,30){\makebox(0,0){$2$}}  
\put(10,5){\makebox(0,0){$3$}}  
\put(20,5){\makebox(0,0){$3^{'}$}}  
\put(30,5){\makebox(0,0){$3^{''}$}}  

\end{picture}}

\put(55,22){$\small
\begin{array}{|l|rll|}
\hline
{} & {} & {} & {} \\
\mbox{q-dim} & \mathcal{D}_3& \hookleftarrow & {\cal A}_3 \\ 
{} & { } & { } & { } \\
\hline
{} & { } & { } & { } \\
\lbrack 1 \rbrack =1 & 0 & \hookleftarrow & (0,0) , (3,0) , (0,3) \\  
\lbrack 3 \rbrack =2 & 1 & \hookleftarrow & (1,0) , (2,1) , (0,2) \\  
\lbrack 3 \rbrack  & 2 & \hookleftarrow & (0,1) , (2,0) , (1,2) \\ 
\lbrack 1 \rbrack  & 3 & \hookleftarrow & (1,1)  \\ 
\lbrack 1 \rbrack  & 3' & \hookleftarrow & (1,1)  \\  
\lbrack 1 \rbrack  & 3'' & \hookleftarrow & (1,1)  \\
{} & { } & { } & { } \\
\hline
\end{array}
$}
\put(55,57){\makebox(0,0){\scriptsize $
\mathcal{Z} = \vert [1,1] + [4,1] + [1,4] \vert^2 + 3\,  \vert [1,1] \vert^2 $}}
\put(95,57){\makebox(0,0){\scriptsize with $[a,b] = (a-1,b-1).$}}
\put(70,49){\makebox(0,0){\scriptsize $\mathcal F = \lambda_{(0,0)} \oplus \lambda_{3,0} \oplus \lambda_{0,3} $}}
\end{picture}

\end{center} 
\caption{The ${\cal D}_3$ graph, quantum dimensions and ${\cal D}_3 \hookleftarrow {\cal A}_3$ induction rules.} 
\end{figure}

\begin{figure}[H] 
\begin{center} 
\unitlength0.085cm

\begin{picture}(150,110)
\put(0,0){\line(1,0){150}}
\put(0,0){\line(0,1){110}}
\put(150,110){\line(-1,0){150}}
\put(150,110){\line(0,-1){110}}

\put(15,50){\unitlength0.22mm \begin{picture}(0,0)
\put(0,0){\makebox(0,0){12}} 
\put(40,0){\makebox(0,0){26}} 
\put(80,0){\makebox(0,0){42}} 
\put(120,0){\makebox(0,0){60}} 
\put(20,30){\makebox(0,0){26}} 
\put(60,30){\makebox(0,0){60}} 
\put(100,30){\makebox(0,0){94}} 
\put(40,60){\makebox(0,0){94}}
\put(80,60){\makebox(0,0){144}}
\put(60,90){\makebox(0,0){60}}
\put(20,150){\makebox(0,0){\Large $\mathcal{A}_3$}}
\put(200,150){\makebox(0,0){\Large $Oc(\mathcal{D}_3)$}}
\end{picture}}

\put(50,0){\begin{picture}(90,110)

\put(0,0){\begin{picture}(0,0)
\put(10.5,29.5){\line(1,0){19}}
\put(10.5,30.5){\line(1,0){19}}
\put(20,29.5){\vector(1,0){0.5}}
\put(20,30.5){\vector(1,0){0.5}}
\put(29.5,31){\line(-1,2){9}}
\put(29.5,31){\vector(-1,2){5}}
\put(19.5,49){\line(-1,-2){9}}
\put(19.5,49){\vector(-1,-2){5}}
\put(10,10){\line(0,1){18.75}}
\put(30,10){\line(0,1){18.75}}
\put(10,10){\line(1,1){19.25}}
\put(30,10){\line(-1,1){19.25}}
\put(20,10){\line(1,2){9.5}}
\put(20,10){\line(-1,2){9.5}}
\put(10,23.5){\vector(0,1){0.5}}
\put(30,22.5){\vector(0,-1){0.5}}
\put(22.5,22.5){\vector(-1,-1){0.5}}
\put(17,23){\vector(-1,1){0.5}}
\put(26.5,23){\vector(-1,-2){0.5}}
\put(13.5,23){\vector(-1,2){0.5}}
\put(20,50){\circle{2}}
\put(10,10){\circle{2}}
\put(20,10){\circle{2}}
\put(30,10){\circle{2}}
\put(10,30){\circle*{1.2}}
\put(10,30){\circle{2}}
\put(30,30){\circle*{2}}
\put(20,55){\makebox(0,0){$6$}}  
\put(5,30){\makebox(0,0){$10$}}  
\put(35,30){\makebox(0,0){$10$}}  
\put(10,5){\makebox(0,0){$6$}}  
\put(20,5){\makebox(0,0){$6$}}  
\put(30,5){\makebox(0,0){$6$}}  
\end{picture}}

\put(25,50){\begin{picture}(0,0)
\put(10.5,29.5){\line(1,0){19}}
\put(10.5,30.5){\line(1,0){19}}
\put(20,29.5){\vector(1,0){0.5}}
\put(20,30.5){\vector(1,0){0.5}}
\put(29.5,31){\line(-1,2){9}}
\put(29.5,31){\vector(-1,2){5}}
\put(19.5,49){\line(-1,-2){9}}
\put(19.5,49){\vector(-1,-2){5}}
\put(10,10){\line(0,1){18.75}}
\put(30,10){\line(0,1){18.75}}
\put(10,10){\line(1,1){19.25}}
\put(30,10){\line(-1,1){19.25}}
\put(20,10){\line(1,2){9.5}}
\put(20,10){\line(-1,2){9.5}}
\put(10,23.5){\vector(0,1){0.5}}
\put(30,22.5){\vector(0,-1){0.5}}
\put(22.5,22.5){\vector(-1,-1){0.5}}
\put(17,23){\vector(-1,1){0.5}}
\put(26.5,23){\vector(-1,-2){0.5}}
\put(13.5,23){\vector(-1,2){0.5}}
\put(20,50){\circle{2}}
\put(10,10){\circle{2}}
\put(20,10){\circle{2}}
\put(30,10){\circle{2}}
\put(10,30){\circle*{1.2}}
\put(10,30){\circle{2}}
\put(30,30){\circle*{2}}
\put(20,55){\makebox(0,0){$6$}}  
\put(5,30){\makebox(0,0){$10$}}  
\put(35,30){\makebox(0,0){$10$}}  
\put(10,5){\makebox(0,0){$6$}}  
\put(20,5){\makebox(0,0){$6$}}  
\put(30,5){\makebox(0,0){$6$}}  
\end{picture}}

\put(50,0){\begin{picture}(0,0)
\put(10.5,29.5){\line(1,0){19}}
\put(10.5,30.5){\line(1,0){19}}
\put(20,29.5){\vector(1,0){0.5}}
\put(20,30.5){\vector(1,0){0.5}}
\put(29.5,31){\line(-1,2){9}}
\put(29.5,31){\vector(-1,2){5}}
\put(19.5,49){\line(-1,-2){9}}
\put(19.5,49){\vector(-1,-2){5}}
\put(10,10){\line(0,1){18.75}}
\put(30,10){\line(0,1){18.75}}
\put(10,10){\line(1,1){19.25}}
\put(30,10){\line(-1,1){19.25}}
\put(20,10){\line(1,2){9.5}}
\put(20,10){\line(-1,2){9.5}}
\put(10,23.5){\vector(0,1){0.5}}
\put(30,22.5){\vector(0,-1){0.5}}
\put(22.5,22.5){\vector(-1,-1){0.5}}
\put(17,23){\vector(-1,1){0.5}}
\put(26.5,23){\vector(-1,-2){0.5}}
\put(13.5,23){\vector(-1,2){0.5}}
\put(20,50){\circle{2}}
\put(10,10){\circle{2}}
\put(20,10){\circle{2}}
\put(30,10){\circle{2}}
\put(10,30){\circle*{1.2}}
\put(10,30){\circle{2}}
\put(30,30){\circle*{2}}
\put(20,55){\makebox(0,0){$6$}}  
\put(5,30){\makebox(0,0){$10$}}  
\put(35,30){\makebox(0,0){$10$}}  
\put(10,5){\makebox(0,0){$6$}}  
\put(20,5){\makebox(0,0){$6$}}  
\put(30,5){\makebox(0,0){$6$}}  
\end{picture}}

\end{picture}}

\end{picture}

\end{center} 
\caption{Dimensions $d_n$ and $d_x$ of the blocks for ${\cal D}_3$.} 
\end{figure}
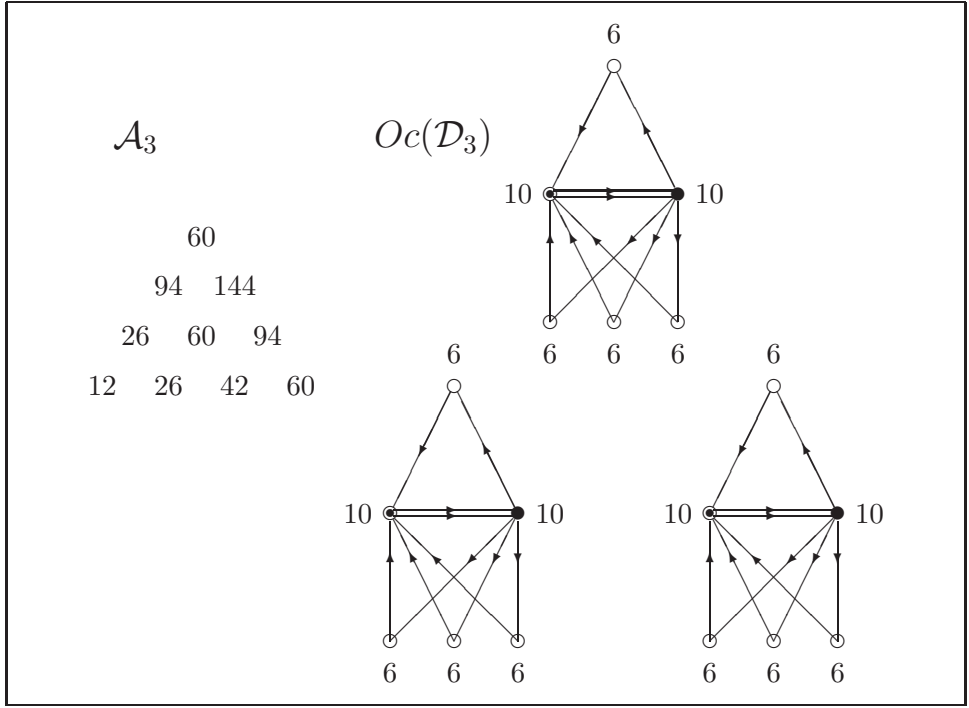



\begin{figure}[H] 
\begin{center} 
\unitlength 0.30mm 
 
\begin{picture}(560,280)(0,0)

\put(0,0){\line(1,0){560}}
\put(0,280){\line(1,0){560}}
\put(0,0){\line(0,1){280}}
\put(560,0){\line(0,1){280}}

\put(135,260){\makebox(0,0){\scriptsize $\begin{array}{rcl}
\mathcal{Z} &=& \vert [1,1] + [3,3] \vert^2 +  \vert [1,3] + [4,3] \vert^2 +
\vert [2,3] + [6,1] \vert^2 \\
{} &+&  \vert [4,1] + [1,4] \vert^2 +  \vert
[3,2] + [1,6] \vert^2 +  \vert [3,1] + [3,4] \vert^2
\end{array}$ }}

\put(105,235){\makebox(0,0){\scriptsize with $[a,b] = (a-1,b-1).$}}

\put(385,260){\makebox(0,0){\scriptsize $\mathcal F = \lambda_{(0,0)} \oplus \lambda_{2,2}$}}

\put(25,20){\begin{picture}(180,180)(0,0) 
 
\put(0,45){\begin{picture}(60,45) 
\put(0,0){\circle{7}} 
\put(60,0){\circle*{4}} 
\put(60,0){\circle{8}} 
\put(30,45){\circle*{7}} 
\put(0,0){\vector(1,0){32.5}} 
\put(30,0){\line(1,0){30}} 
\put(60,0){\vector(-2,3){16.5}} 
\put(45,22.5){\line(-2,3){15}} 
\put(30,45){\vector(-2,-3){16.5}} 
\put(15,22.5){\line(-2,-3){15}} 
\end{picture}} 
 
\put(120,45){\begin{picture}(60,45) 
\put(0,0){\circle{7}} 
\put(60,0){\circle*{4}} 
\put(60,0){\circle{8}} 
\put(30,45){\circle*{7}} 
\put(0,0){\vector(1,0){32.5}} 
\put(30,0){\line(1,0){30}} 
\put(60,0){\vector(-2,3){16.5}} 
\put(45,22.5){\line(-2,3){15}} 
\put(30,45){\vector(-2,-3){16.5}} 
\put(15,22.5){\line(-2,-3){15}} 
\end{picture}} 
 
\put(60,135){\begin{picture}(60,45) 
\put(0,0){\circle{7}} 
\put(60,0){\circle*{4}} 
\put(60,0){\circle{8}} 
\put(30,45){\circle*{7}} 
\put(0,0){\vector(1,0){32.5}} 
\put(30,0){\line(1,0){30}} 
\put(60,0){\vector(-2,3){16.5}} 
\put(45,22.5){\line(-2,3){15}} 
\put(30,45){\vector(-2,-3){16.5}} 
\put(15,22.5){\line(-2,-3){15}} 
\end{picture}}

\put(0,90){\begin{picture}(60,45) 
\put(0,45){\circle*{4}} 
\put(0,45){\circle{8}} 
\put(60,45){\vector(-1,0){32.5}} 
\put(30,45){\line(-1,0){30}} 
\put(30,0){\vector(2,3){16.5}} 
\put(45,22.5){\line(2,3){15}} 
\put(0,45){\vector(2,-3){16.5}} 
\put(15,22.5){\line(2,-3){15}} 
\end{picture}} 
 
\put(120,90){\begin{picture}(60,45) 
\put(60,45){\circle{7}} 
\put(60,45){\vector(-1,0){32.5}} 
\put(30,45){\line(-1,0){30}} 
\put(30,0){\vector(2,3){16.5}} 
\put(45,22.5){\line(2,3){15}} 
\put(0,45){\vector(2,-3){16.5}} 
\put(15,22.5){\line(2,-3){15}} 
\end{picture}} 
 
\put(60,0){\begin{picture}(60,45) 
\put(30,0){\circle*{7}} 
\put(60,45){\vector(-1,0){32.5}} 
\put(30,45){\line(-1,0){30}} 
\put(30,0){\vector(2,3){16.5}} 
\put(45,22.5){\line(2,3){15}} 
\put(0,45){\vector(2,-3){16.5}} 
\put(15,22.5){\line(2,-3){15}} 
\end{picture}} 
 
\put(60,135){\vector(0,-1){47.5}} 
\put(60,90){\line(0,-1){45}} 
\put(60,45){\vector(2,1){47.2}} 
\put(105,67.5){\line(2,1){45}} 
\put(150,90){\vector(-2,1){47.2}} 
\put(105,112.5){\line(-2,1){45}} 
 
\put(120,45){\vector(0,1){47.5}} 
\put(120,90){\line(0,1){45}} 
\put(120,135){\vector(-2,-1){47.2}} 
\put(75,112.5){\line(-2,-1){45}} 
\put(30,90){\vector(2,-1){47.2}} 
\put(75,67.5){\line(2,-1){45}}

\put(-12,45){\makebox(0,0){$1_0$}} 
\put(192,45){\makebox(0,0){$1_4$}} 
\put(-12,135){\makebox(0,0){$1_1$}} 
\put(192,135){\makebox(0,0){$1_3$}} 
\put(90,-12){\makebox(0,0){$1_5$}} 
\put(90,192){\makebox(0,0){$1_2$}} 
\put(55,35){\makebox(0,0){$2_1$}} 
 
\put(125,35){\makebox(0,0){$2_0$}} 
\put(55,145){\makebox(0,0){$2_3$}} 
\put(125,145){\makebox(0,0){$2_4$}} 
\put(20,90){\makebox(0,0){$2_2$}} 
\put(160,90){\makebox(0,0){$2_5$}} 

\end{picture}}

\put(280,110){\small{$\begin{array}{|l|rcl|}
\hline
\mbox{q-dim} & \mathcal{E}_5& \hookleftarrow & {\cal A}_5 \\ 
\hline
\lbrack 1 \rbrack =1 & 1_0 & \hookleftarrow & (0,0) , (2,2) \\  
\lbrack 1 \rbrack  & 1_1 & \hookleftarrow & (0,2) , (3,2) \\  
\lbrack 1 \rbrack  & 1_2 & \hookleftarrow & (1,2) , (5,0) \\ 
\lbrack 1 \rbrack  & 1_3 & \hookleftarrow & (3,0) , (0,3) \\ 
\lbrack 1 \rbrack  & 1_4 & \hookleftarrow & (2,1) , (0,5) \\  
\lbrack 1 \rbrack  & 1_5 & \hookleftarrow & (2,0) , (2,3) \\
{} & { } & { } & { } \\
\lbrack 3 \rbrack =1+\sqrt{2} \quad  & 2_0 & \hookleftarrow & (1,1) , (3,0) , (2,2) , (1,4) \\ 
\lbrack 3 \rbrack  & 2_1 & \hookleftarrow & (1,0) , (2,1) , (1,3) , (3,2) \\ 
\lbrack 3 \rbrack  & 2_2 & \hookleftarrow & (0,1) , (1,2) , (3,1) , (2,3) \\ 
\lbrack 3 \rbrack  & 2_3 & \hookleftarrow & (1,1) , (0,3) , (2,2) , (4,1) \\ 
\lbrack 3 \rbrack  & 2_4 & \hookleftarrow & (0,2) , (2,1) , (4,0) , (1,3) \\ 
\lbrack 3 \rbrack  & 2_5 & \hookleftarrow & (2,0) , (1,2) , (3,1) , (0,4)  \\
\hline
\end{array} $}}

\end{picture}

\label{gr_E5} 
\caption{The ${\cal E}_5$ graph, quantum dimensions and ${\cal E}_5 \hookleftarrow {\cal A}_5$ induction rules.} 
\end{center} 
\end{figure}
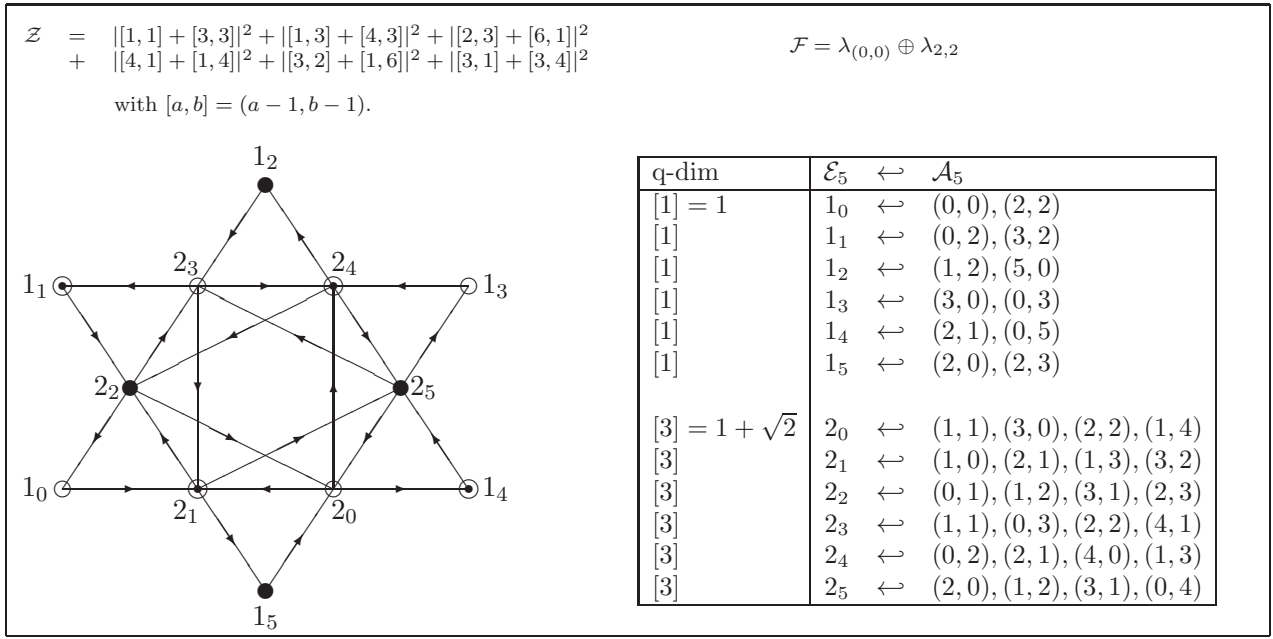

\begin{figure}[hhh] 
\begin{center} 
\unitlength 0.25mm

\begin{picture}(690,450)

\put(0,0){\line(1,0){690}}
\put(0,0){\line(0,1){450}}
\put(0,450){\line(1,0){690}}
\put(690,450){\line(0,-1){450}}

\put(200,15){\begin{picture}(400,410) 
 
\put(20,20){\begin{picture}(360,370)

\qbezier[10](0,325)(15,340)(0,350)
\qbezier[10](0,325)(-15,340)(0,350)
\qbezier[10](0,235)(15,250)(0,260)
\qbezier[10](0,235)(-15,250)(0,260)
\qbezier[10](90,190)(105,205)(90,215)
\qbezier[10](90,190)(75,205)(90,215)
\qbezier[10](90,370)(105,385)(90,395)
\qbezier[10](90,370)(75,385)(90,395)
\qbezier[10](180,325)(195,340)(180,350)
\qbezier[10](180,325)(165,340)(180,350)
\qbezier[10](180,235)(195,250)(180,260)
\qbezier[10](180,235)(165,250)(180,260)
\qbezier[10](240,135)(255,150)(240,160)
\qbezier[10](240,135)(225,150)(240,160)
\qbezier[10](300,135)(315,150)(300,160)
\qbezier[10](300,135)(285,150)(300,160)
\qbezier[10](210,90)(225,105)(210,115)
\qbezier[10](210,90)(195,105)(210,115)
\qbezier[10](330,90)(345,105)(330,115)
\qbezier[10](330,90)(315,105)(330,115)
\qbezier[10](240,45)(255,30)(240,20)
\qbezier[10](240,45)(225,30)(240,20)
\qbezier[10](300,45)(315,30)(300,20)
\qbezier[10](300,45)(285,30)(300,20)
\qbezier[80](60,235)(60,80)(180,135)
\qbezier[80](30,280)(30,90)(270,180)
\qbezier[80](120,235)(100,80)(180,45)
\qbezier[130](60,325)(250,470)(360,135)
\qbezier[140](120,325)(260,450)(360,45)
\qbezier[150](150,280)(700,80)(270,0)


\put(0,190){\begin{picture}(180,180)

\put(0,45){\begin{picture}(60,45) 
\put(0,0){\circle{7}} 
\put(0,0){\circle{13}} 
\put(60,0){\circle*{4}} 
\put(60,0){\circle{8}} 
\put(30,45){\circle*{7}} 
\put(0,0){\vector(1,0){32.5}} 
\put(30,0){\line(1,0){30}} 
\put(60,0){\vector(-2,3){16.5}} 
\put(45,22.5){\line(-2,3){15}} 
\put(30,45){\vector(-2,-3){16.5}} 
\put(15,22.5){\line(-2,-3){15}} 
\end{picture}} 
 
\put(120,45){\begin{picture}(60,45) 
\put(0,0){\circle{7}} 
\put(60,0){\circle*{4}} 
\put(60,0){\circle{8}} 
\put(60,0){\circle{13}} 
\put(30,45){\circle*{7}} 
\put(0,0){\vector(1,0){32.5}} 
\put(30,0){\line(1,0){30}} 
\put(60,0){\vector(-2,3){16.5}} 
\put(45,22.5){\line(-2,3){15}} 
\put(30,45){\vector(-2,-3){16.5}} 
\put(15,22.5){\line(-2,-3){15}} 
\end{picture}} 
 
\put(60,135){\begin{picture}(60,45) 
\put(0,0){\circle{7}} 
\put(60,0){\circle*{4}} 
\put(60,0){\circle{8}} 
\put(30,45){\circle*{7}} 
\put(30,45){\circle{13}} 
\put(0,0){\vector(1,0){32.5}} 
\put(30,0){\line(1,0){30}} 
\put(60,0){\vector(-2,3){16.5}} 
\put(45,22.5){\line(-2,3){15}} 
\put(30,45){\vector(-2,-3){16.5}} 
\put(15,22.5){\line(-2,-3){15}} 
\end{picture}}

\put(0,90){\begin{picture}(60,45) 
\put(0,45){\circle*{4}} 
\put(0,45){\circle{8}} 
\put(0,45){\circle{13}} 
\put(60,45){\vector(-1,0){32.5}} 
\put(30,45){\line(-1,0){30}} 
\put(30,0){\vector(2,3){16.5}} 
\put(45,22.5){\line(2,3){15}} 
\put(0,45){\vector(2,-3){16.5}} 
\put(15,22.5){\line(2,-3){15}} 
\end{picture}} 
 
\put(120,90){\begin{picture}(60,45) 
\put(60,45){\circle{7}} 
\put(60,45){\circle{13}} 
\put(60,45){\vector(-1,0){32.5}} 
\put(30,45){\line(-1,0){30}} 
\put(30,0){\vector(2,3){16.5}} 
\put(45,22.5){\line(2,3){15}} 
\put(0,45){\vector(2,-3){16.5}} 
\put(15,22.5){\line(2,-3){15}} 
\end{picture}} 
 
\put(60,0){\begin{picture}(60,45) 
\put(30,0){\circle*{7}} 
\put(30,0){\circle{13}} 
\put(60,45){\vector(-1,0){32.5}} 
\put(30,45){\line(-1,0){30}} 
\put(30,0){\vector(2,3){16.5}} 
\put(45,22.5){\line(2,3){15}} 
\put(0,45){\vector(2,-3){16.5}} 
\put(15,22.5){\line(2,-3){15}} 
\end{picture}} 
 
\put(60,135){\vector(0,-1){47.5}} 
\put(60,90){\line(0,-1){45}} 
\put(60,45){\vector(2,1){47.2}} 
\put(105,67.5){\line(2,1){45}} 
\put(150,90){\vector(-2,1){47.2}} 
\put(105,112.5){\line(-2,1){45}} 
 
\put(120,45){\vector(0,1){47.5}} 
\put(120,90){\line(0,1){45}} 
\put(120,135){\vector(-2,-1){47.2}} 
\put(75,112.5){\line(-2,-1){45}} 
\put(30,90){\vector(2,-1){47.2}} 
\put(75,67.5){\line(2,-1){45}}

\put(-30,47){\makebox(0,0){\scriptsize{$1_0\otimes 1_0$}}} 
\put(210,45){\makebox(0,0){\scriptsize{$1_4\otimes 1_0$}}} 
\put(-30,135){\makebox(0,0){\scriptsize{$1_1\otimes 1_0$}}} 
\put(210,135){\makebox(0,0){\scriptsize{$1_3\otimes 1_0$}}} 
\put(90,-12){\makebox(0,0){\scriptsize{$1_5\otimes 1_0$}}} 
\put(60,180){\makebox(0,0){\scriptsize{$1_2\otimes 1_0$}}} 
\put(45,35){\makebox(0,0){\scriptsize{$2_1\otimes 1_0$}}}  
\put(137,35){\makebox(0,0){\scriptsize{$2_0\otimes 1_0$}}} 
\put(43,148){\makebox(0,0){\scriptsize{$2_3\otimes 1_0$}}} 
\put(136,148){\makebox(0,0){\scriptsize{$2_4\otimes 1_0$}}} 
\put(5,92){\makebox(0,0){\scriptsize{$2_2\otimes 1_0$}}} 
\put(177,92){\makebox(0,0){\scriptsize{$2_5\otimes 1_0$}}} 

\put(90,207){\makebox(0,0){\scriptsize{(12)}}} 
\put(43,160){\makebox(0,0){\scriptsize{(24)}}} 
\put(136,160){\makebox(0,0){\scriptsize{(24)}}} 
\put(-30,147){\makebox(0,0){\scriptsize{(12)}}} 
\put(210,147){\makebox(0,0){\scriptsize{(12)}}} 
\put(5,104){\makebox(0,0){\scriptsize{(24)}}} 
\put(177,104){\makebox(0,0){\scriptsize{(24)}}} 
\put(-30,59){\makebox(0,0){\scriptsize{(12)}}} 
\put(210,57){\makebox(0,0){\scriptsize{(12)}}} 
\put(45,23){\makebox(0,0){\scriptsize{(24)}}}  
\put(137,23){\makebox(0,0){\scriptsize{(24)}}} 
\put(90,-24){\makebox(0,0){\scriptsize{(12)}}}

\end{picture}}


\put(180,0){\begin{picture}(180,180)

\put(0,45){\begin{picture}(60,45) 
\put(0,0){\circle{7}} 
\put(60,0){\circle*{4}} 
\put(60,0){\circle{8}} 
\put(30,45){\circle*{7}} 
\put(0,0){\vector(1,0){32.5}} 
\put(30,0){\line(1,0){30}} 
\put(60,0){\vector(-2,3){16.5}} 
\put(45,22.5){\line(-2,3){15}} 
\put(30,45){\vector(-2,-3){16.5}} 
\put(15,22.5){\line(-2,-3){15}} 
\end{picture}} 
 
\put(120,45){\begin{picture}(60,45) 
\put(0,0){\circle{7}} 
\put(60,0){\circle*{4}} 
\put(60,0){\circle{8}} 
\put(30,45){\circle*{7}} 
\put(0,0){\vector(1,0){32.5}} 
\put(30,0){\line(1,0){30}} 
\put(60,0){\vector(-2,3){16.5}} 
\put(45,22.5){\line(-2,3){15}} 
\put(30,45){\vector(-2,-3){16.5}} 
\put(15,22.5){\line(-2,-3){15}} 
\end{picture}} 
 
\put(60,135){\begin{picture}(60,45) 
\put(0,0){\circle{7}} 
\put(60,0){\circle*{4}} 
\put(60,0){\circle{8}} 
\put(30,45){\circle*{7}} 
\put(0,0){\vector(1,0){32.5}} 
\put(30,0){\line(1,0){30}} 
\put(60,0){\vector(-2,3){16.5}} 
\put(45,22.5){\line(-2,3){15}} 
\put(30,45){\vector(-2,-3){16.5}} 
\put(15,22.5){\line(-2,-3){15}} 
\end{picture}}

\put(0,90){\begin{picture}(60,45) 
\put(0,45){\circle*{4}} 
\put(0,45){\circle{8}} 
\put(60,45){\vector(-1,0){32.5}} 
\put(30,45){\line(-1,0){30}} 
\put(30,0){\vector(2,3){16.5}} 
\put(45,22.5){\line(2,3){15}} 
\put(0,45){\vector(2,-3){16.5}} 
\put(15,22.5){\line(2,-3){15}} 
\end{picture}} 
 
\put(120,90){\begin{picture}(60,45) 
\put(60,45){\circle{7}} 
\put(60,45){\vector(-1,0){32.5}} 
\put(30,45){\line(-1,0){30}} 
\put(30,0){\vector(2,3){16.5}} 
\put(45,22.5){\line(2,3){15}} 
\put(0,45){\vector(2,-3){16.5}} 
\put(15,22.5){\line(2,-3){15}} 
\end{picture}} 
 
\put(60,0){\begin{picture}(60,45) 
\put(30,0){\circle*{7}} 
\put(60,45){\vector(-1,0){32.5}} 
\put(30,45){\line(-1,0){30}} 
\put(30,0){\vector(2,3){16.5}} 
\put(45,22.5){\line(2,3){15}} 
\put(0,45){\vector(2,-3){16.5}} 
\put(15,22.5){\line(2,-3){15}} 
\end{picture}} 
 
\put(60,135){\vector(0,-1){47.5}} 
\put(60,90){\line(0,-1){45}} 
\put(60,45){\vector(2,1){47.2}} 
\put(105,67.5){\line(2,1){45}} 
\put(150,90){\vector(-2,1){47.2}} 
\put(105,112.5){\line(-2,1){45}} 
 
\put(120,45){\vector(0,1){47.5}} 
\put(120,90){\line(0,1){45}} 
\put(120,135){\vector(-2,-1){47.2}} 
\put(75,112.5){\line(-2,-1){45}} 
\put(30,90){\vector(2,-1){47.2}} 
\put(75,67.5){\line(2,-1){45}}

\put(-30,47){\makebox(0,0){\scriptsize{$1_0\otimes 2_0$}}} 
\put(210,45){\makebox(0,0){\scriptsize{$1_4\otimes 2_0$}}} 
\put(-30,135){\makebox(0,0){\scriptsize{$1_1\otimes 2_0$}}} 
\put(210,135){\makebox(0,0){\scriptsize{$1_3\otimes 2_0$}}} 
\put(90,-12){\makebox(0,0){\scriptsize{$1_5\otimes 2_0$}}} 
\put(90,195){\makebox(0,0){\scriptsize{$1_2\otimes 2_0$}}} 
\put(45,35){\makebox(0,0){\scriptsize{$2_1\otimes 2_0$}}}  
\put(137,35){\makebox(0,0){\scriptsize{$2_0\otimes 2_0$}}} 
\put(43,148){\makebox(0,0){\scriptsize{$2_3\otimes 2_0$}}} 
\put(136,148){\makebox(0,0){\scriptsize{$2_4\otimes 2_0$}}} 
\put(5,92){\makebox(0,0){\scriptsize{$2_2\otimes 2_0$}}} 
\put(177,92){\makebox(0,0){\scriptsize{$2_5\otimes 2_0$}}}

\put(90,207){\makebox(0,0){\scriptsize{(24)}}} 
\put(43,160){\makebox(0,0){\scriptsize{(60)}}} 
\put(136,160){\makebox(0,0){\scriptsize{(60)}}} 
\put(-30,147){\makebox(0,0){\scriptsize{(24)}}} 
\put(210,147){\makebox(0,0){\scriptsize{(24)}}} 
\put(5,104){\makebox(0,0){\scriptsize{(60)}}} 
\put(177,104){\makebox(0,0){\scriptsize{(60)}}} 
\put(-30,59){\makebox(0,0){\scriptsize{(24)}}} 
\put(210,57){\makebox(0,0){\scriptsize{(24)}}} 
\put(45,23){\makebox(0,0){\scriptsize{(60)}}}  
\put(137,23){\makebox(0,0){\scriptsize{(60)}}} 
\put(90,-24){\makebox(0,0){\scriptsize{(24)}}} 

\end{picture}} 
\end{picture}}
\end{picture}}

\put(20,150){\unitlength 0.20mm 
\begin{picture}(200,180)(0,-10)
\put(0,0){\begin{picture}(40,40)
\put(0,0){\makebox(0,0){12}} 
\put(40,0){\makebox(0,0){24}} 
\put(20,30){\makebox(0,0){12}} 
\end{picture}} 
\put(40,0){\begin{picture}(40,40) 
\put(40,0){\makebox(0,0){36}} 
\put(20,30){\makebox(0,0){46}} 
\end{picture}} 
\put(80,0){\begin{picture}(40,40) 
\put(40,0){\makebox(0,0){36}} 
\put(20,30){\makebox(0,0){60}} 
\end{picture}} 
\put(120,0){\begin{picture}(40,40) 
\put(40,0){\makebox(0,0){24}} 
\put(20,30){\makebox(0,0){48}} 
\end{picture}} 
\put(160,0){\begin{picture}(40,40) 
\put(40,0){\makebox(0,0){12}} 
\put(20,30){\makebox(0,0){24}} 
\end{picture}} 
\put(20,30){\begin{picture}(40,40) 
\put(20,30){\makebox(0,0){36}} 
\end{picture}} 
\put(60,30){\begin{picture}(40,40) 
\put(20,30){\makebox(0,0){60}} 
\end{picture}} 
\put(100,30){\begin{picture}(40,40) 
\put(20,30){\makebox(0,0){60}} 
\end{picture}} 
\put(140,30){\begin{picture}(40,40) 
\put(20,30){\makebox(0,0){36}} 
\end{picture}} 
\put(40,60){\begin{picture}(40,40) 
\put(20,30){\makebox(0,0){36}} 
\end{picture}} 
\put(80,60){\begin{picture}(40,40) 
\put(20,30){\makebox(0,0){48}} 
\end{picture}} 
\put(120,60){\begin{picture}(40,40) 
\put(20,30){\makebox(0,0){36}} 
\end{picture}} 
\put(60,90){\begin{picture}(40,40) 
\put(20,30){\makebox(0,0){24}} 
\end{picture}} 
\put(100,90){\begin{picture}(40,40) 
\put(20,30){\makebox(0,0){24}} 
\end{picture}} 
\put(80,120){\begin{picture}(40,40) 
\put(20,30){\makebox(0,0){12}} 
\end{picture}} 
\end{picture}}

\put(80,60){\makebox(0,0){\Large $\mathcal{A}_5$}}
\put(280,60){\makebox(0,0){\Large $Oc(\mathcal{E}_5)$}}

\end{picture}
\end{center} 
\caption{Dimensions $d_n$ and $d_x$ of the blocks for ${\cal E}_5$.} 
\end{figure}
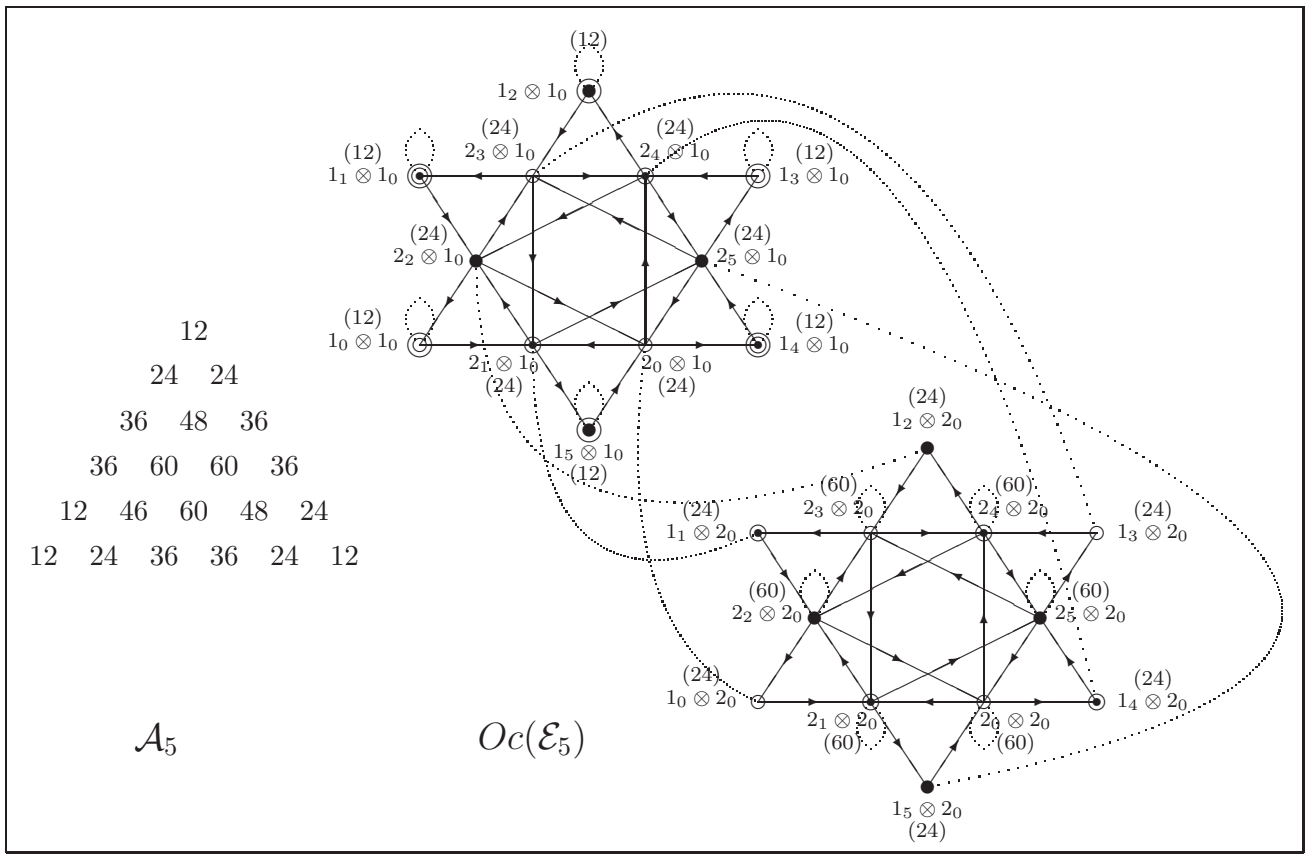



\begin{figure}[H] 
\begin{center} 
 
\unitlength0.035cm
\begin{picture}(510,230)

\put(0,0){\line(1,0){510}}
\put(0,0){\line(0,1){230}}
\put(0,230){\line(1,0){510}}
\put(510,0){\line(0,1){230}}

\put(0,190){\scriptsize \makebox{$\begin{array}{rcl} \mathcal{Z} &=& [1,1] + [1,10] + [10,1] + [2,5] + [5,2] + [5,5]||^2 \\
{} &+& 2\, |[3,3] + [3,6] + [6,3]|^2  \\
& & \\
{} & & \text{with } [a,b] = (a-1, b-1) \end{array}$}}

\put(250,200){\scriptsize \makebox{$\mathcal{F} = \lambda_{(0,0)} \oplus \lambda_{(1,4)} \oplus \lambda_{(4,1)} \oplus \lambda_{(4,4)} \oplus \lambda_{(0,9)} \oplus \lambda_{(9,0)} $}}

\put(10,50){\begin{picture}(180,130)


\put(70,60){\begin{picture}(40,30) 
\put(0,0){\circle*{3}} 
\put(0,0){\circle{7}} 
\put(40,0){\circle*{5}} 
\put(20,30){\circle{6}} 
\put(0,0){\vector(1,0){22.5}} 
\put(20,0){\line(1,0){20}} 
\put(40,0){\vector(-2,3){11.5}} 
\put(30,15){\line(-2,3){10}} 
\put(20,30){\vector(-2,-3){11.5}} 
\put(10,15){\line(-2,-3){10}} 
\end{picture}} 
 
\qbezier[920](30,0)(90,-40)(150,0) 
\put(90,-20){\vector(-1,0){0}} 
 
\put(30,0){\circle*{5}} 
\put(90,0){\circle{6}} 
\put(150,0){\circle*{3}} 
\put(150,0){\circle{7}} 
\put(10,60){\circle*{3}} 
\put(10,60){\circle{7}} 
\put(50,60){\circle*{5}} 
\put(130,60){\circle*{3}} 
\put(130,60){\circle{7}} 
\put(170,60){\circle*{5}} 
\put(0,90){\circle{6}} 
\put(180,90){\circle{6}}

\put(0,90){\vector(1,-3){6.0}} 
\put(10,60){\line(-1,3){4.5}} 
\put(10,60){\vector(1,0){22.5}} 
\put(50,60){\line(-1,0){17.5}} 
 
\put(0,90){\line(5,-3){150}} 
\put(23,76.4){\vector(-3,2){0}} 
 
\put(30,0){\line(-1,3){10.8}} 
\put(10,60){\vector(1,-3){10}} 
 
\put(110,60){\vector(-1,-3){8.6}} 
\put(90,0){\line(1,3){15}} 
\put(90,0){\vector(-1,3){13}} 
\put(70,60){\line(1,-3){10}} 
 
\put(170,60){\line(-1,-3){10.8}} 
\put(150,0){\vector(1,3){10}} 
 
\put(180,90){\line(-5,-3){150}} 
\put(154,74){\vector(-3,-2){0}} 
 
\put(130,60){\vector(1,0){22.5}} 
\put(170,60){\line(-1,0){17.5}} 
\put(170,60){\vector(1,3){6.0}} 
\put(180,90){\line(-1,-3){4.5}} 
 
\put(70,60){\line(-2,-3){40}} 
\put(70,60){\vector(-2,-3){17}} 
\put(110,60){\line(2,-3){40}} 
\put(150,0){\vector(-2,3){23}} 
 
\put(30,1.5){\line(1,0){120}} 
\put(30,-1.5){\line(1,0){120}} 
\put(30,1.5){\vector(1,0){32.5}} 
\put(30,-1.5){\vector(1,0){32.5}} 
\put(90,1.5){\vector(1,0){32.5}} 
\put(90,-1.5){\vector(1,0){32.5}} 
 
\put(90,0){\line(-4,3){80}} 
\put(90,0){\vector(-4,3){32.5}} 
\put(90,0){\line(4,3){80}} 
\put(170,60){\vector(-4,-3){52.5}} 
 
\put(50,60){\line(2,-3){40}} 
\put(50,60){\vector(2,-3){22}} 
\put(90,0){\line(2,3){40}} 
\put(90,0){\vector(2,3){22}} 
 
\put(80,41.7){\vector(-3,2){0}} 
\put(95,38.7){\vector(-3,-2){0}} 
 
\put(0,100){\makebox(0,0){$1_0$}} 
\put(90,100){\makebox(0,0){$0_0$}} 
\put(180,100){\makebox(0,0){$2_0$}} 
 
\put(1,66){\makebox(0,0){$1_1$}} 
\put(66,68){\makebox(0,0){$0_1$}} 
\put(126,68){\makebox(0,0){$2_1$}} 
 
\put(52,68){\makebox(0,0){$1_2$}} 
\put(113,68){\makebox(0,0){$0_2$}} 
\put(180,65){\makebox(0,0){$2_2$}} 
 
\put(30,-7){\makebox(0,0){$3_2$}} 
\put(90,-7){\makebox(0,0){$3_0$}} 
\put(152,-7){\makebox(0,0){$3_1$}}

\end{picture}}

\put(210,100){
${\tiny
\begin{array}{|l|rcl|}
\hline 
& & & \\
\mbox{q-dim} & \mathcal{E}_9 &\hookleftarrow& \mathcal{A}_9 \\
 & & & \\
\hline
 & & & \\
\lbrack 1\rbrack =1 & 0_0 &\hookleftarrow& (0, 0), (4, 1), (1, 4), (4, 4), (9, 0), (0, 9) \\
\lbrack 1\rbrack & 1_0 &\hookleftarrow& (2, 2), (5, 2), (2, 5) \\
\lbrack 1\rbrack & 2_0 &\hookleftarrow& (2, 2), (5, 2), (2, 5) \\
3+2\sqrt{3} & 3_0 &\hookleftarrow& (1, 1), (3, 0), (0, 3), (6, 0), (0, 6), (7, 1), (1, 7), (6, 3), (3, 6) , \\
 & & & (2, 2)_2, (4, 1)_2, (1, 4)_2, (5, 2)_2, (2, 5)_2, (4, 4)_2 , (3,3)_3 \\
 & & & \\
\lbrack 3 \rbrack =1+\sqrt{3} & 0_1 &\hookleftarrow& (1, 0), (4, 0), (1, 3), (3, 2), (0, 5), (5, 1), (2, 4), (4, 3), (
  3, 5), \\
 & & &  (0, 8), (8, 1), (5, 4) \\
\lbrack 3 \rbrack & 1_1 &\hookleftarrow& (2, 1), (1, 3), (3, 2), (5, 1), (2, 4), (4, 3), (1, 6), (6, 2), (3, 5) \\
\lbrack 3 \rbrack  & 2_1 &\hookleftarrow& (2, 1), (1, 3), (3, 2), (5, 1), (2, 4), (4, 3), (1, 6), (6, 2), (3, 5) \\
3+\sqrt{3} & 3_1 &\hookleftarrow& (0, 2), (2, 1), (4, 0), (1, 3), (0, 5), (5, 1), (7, 0), (1, 6), (
  6, 2), \\
 & & & (3, 5), (5, 4), (2, 7) , (3, 2)_2, (2, 4)_2, (4, 3)_2 \\
 & & & \\
\lbrack 3 \rbrack  & 0_2 &\hookleftarrow& (0, 1), (3, 1), (0, 4), (5, 0), (2, 3), (4, 2), (1, 5), (3, 4), (
  8, 0), \\
 & & & (5, 3), (4, 5), (1, 8) \\
\lbrack 3 \rbrack  & 1_2 &\hookleftarrow& (1, 2), (3, 1), (2, 3), (4, 2), (1, 5), (6, 1), (3, 4), (5, 3), (2, 6) \\
\lbrack 3 \rbrack  & 2_2 &\hookleftarrow& (1, 2), (3, 1), (2, 3), (4, 2), (1, 5), (6, 1), (3, 4), (5, 3), (2, 6) \\
3+\sqrt{3} & 3_2 &\hookleftarrow& (2, 0), (1, 2), (3, 1), (0, 4), (5, 0), (1, 5), (6, 1), (0, 7), (
  5, 3), \\
 & & & (2, 6), (7, 2), (4, 5) , (2, 3)_2, (4, 2)_2, (3, 4)_2 \\
 & & & \\
\hline
\end{array}}$}
\end{picture}

\end{center} 
\caption{The ${\cal E}_9$ graph, quantum dimensions and ${\cal E}_9 \hookleftarrow {\cal A}_9$ induction rules.} 
\label{grE9} 
\end{figure}
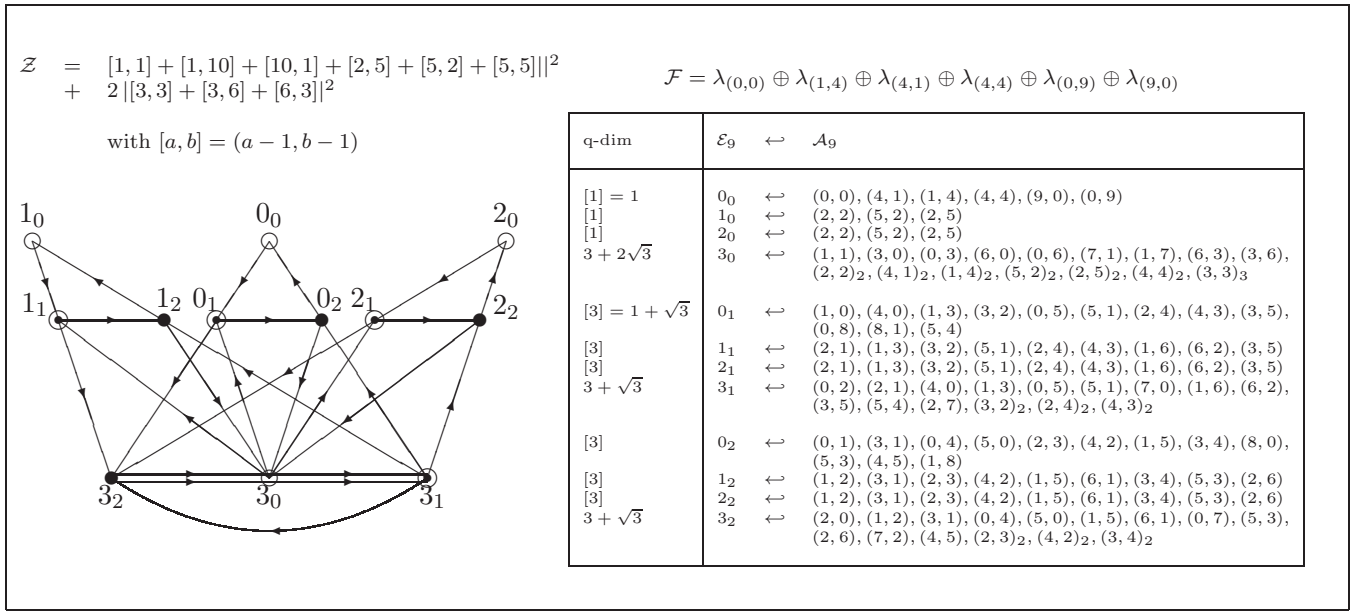

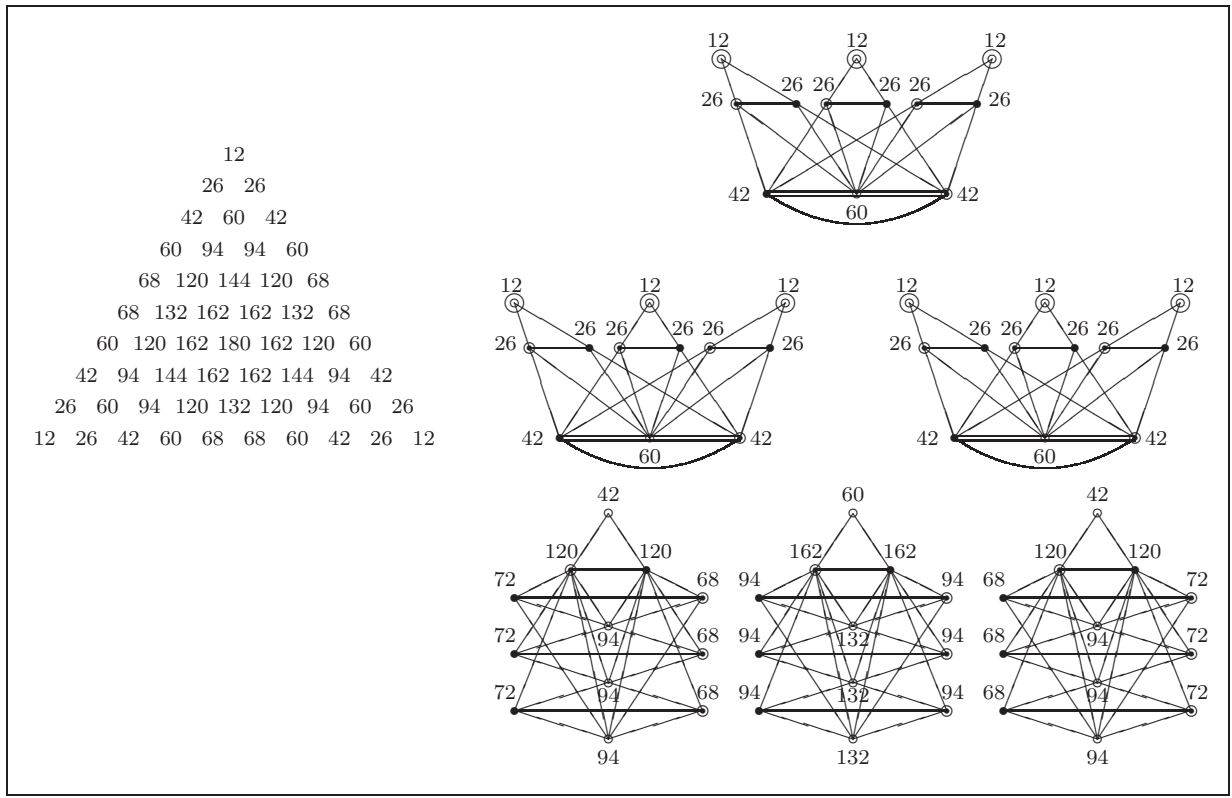
\begin{figure}[hhh] 
\unitlength=0.25mm 
\begin{center} 

\begin{picture}(650,420)

\put(0,0){\line(1,0){650}}
\put(650,0){\line(0,1){420}}
\put(0,420){\line(1,0){650}}
\put(0,0){\line(0,1){420}}

\put(20,190){\unitlength0.014cm
\begin{picture}(360,270)
\put(0,0){\makebox(0,0){\scriptsize 12}} 
\put(40,0){\makebox(0,0){\scriptsize 26}} 
\put(80,0){\makebox(0,0){\scriptsize 42}} 
\put(120,0){\makebox(0,0){\scriptsize 60}} 
\put(160,0){\makebox(0,0){\scriptsize 68}} 
\put(200,0){\makebox(0,0){\scriptsize 68}} 
\put(240,0){\makebox(0,0){\scriptsize 60}} 
\put(280,0){\makebox(0,0){\scriptsize 42}} 
\put(320,0){\makebox(0,0){\scriptsize 26}} 
\put(360,0){\makebox(0,0){\scriptsize 12}} 
\put(20,30){\makebox(0,0){\scriptsize 26}} 
\put(60,30){\makebox(0,0){\scriptsize 60}} 
\put(100,30){\makebox(0,0){\scriptsize 94}} 
\put(140,30){\makebox(0,0){\scriptsize 120}} 
\put(180,30){\makebox(0,0){\scriptsize 132}} 
\put(220,30){\makebox(0,0){\scriptsize 120}} 
\put(260,30){\makebox(0,0){\scriptsize 94}} 
\put(300,30){\makebox(0,0){\scriptsize 60}} 
\put(340,30){\makebox(0,0){\scriptsize 26}} 
\put(40,60){\makebox(0,0){\scriptsize 42}}
\put(80,60){\makebox(0,0){\scriptsize 94}}
\put(120,60){\makebox(0,0){\scriptsize 144}}
\put(160,60){\makebox(0,0){\scriptsize 162}}
\put(200,60){\makebox(0,0){\scriptsize 162}}
\put(240,60){\makebox(0,0){\scriptsize 144}}
\put(280,60){\makebox(0,0){\scriptsize 94}}
\put(320,60){\makebox(0,0){\scriptsize 42}}
\put(60,90){\makebox(0,0){\scriptsize 60}}
\put(100,90){\makebox(0,0){\scriptsize 120}}
\put(140,90){\makebox(0,0){\scriptsize 162}}
\put(180,90){\makebox(0,0){\scriptsize 180}}
\put(220,90){\makebox(0,0){\scriptsize 162}}
\put(260,90){\makebox(0,0){\scriptsize 120}}
\put(300,90){\makebox(0,0){\scriptsize 60}}
\put(80,120){\makebox(0,0){\scriptsize 68}}
\put(120,120){\makebox(0,0){\scriptsize 132}}
\put(160,120){\makebox(0,0){\scriptsize 162}}
\put(200,120){\makebox(0,0){\scriptsize 162}}
\put(240,120){\makebox(0,0){\scriptsize 132}}
\put(280,120){\makebox(0,0){\scriptsize 68}}
\put(100,150){\makebox(0,0){\scriptsize 68}}
\put(140,150){\makebox(0,0){\scriptsize 120}}
\put(180,150){\makebox(0,0){\scriptsize 144}}
\put(220,150){\makebox(0,0){\scriptsize 120}}
\put(260,150){\makebox(0,0){\scriptsize 68}}
\put(120,180){\makebox(0,0){\scriptsize 60}}
\put(160,180){\makebox(0,0){\scriptsize 94}}
\put(200,180){\makebox(0,0){\scriptsize 94}}
\put(240,180){\makebox(0,0){\scriptsize 60}}
\put(140,210){\makebox(0,0){\scriptsize 42}}
\put(180,210){\makebox(0,0){\scriptsize 60}}
\put(220,210){\makebox(0,0){\scriptsize 42}}
\put(160,240){\makebox(0,0){\scriptsize 26}}
\put(200,240){\makebox(0,0){\scriptsize 26}}
\put(180,270){\makebox(0,0){\scriptsize 12}}


\end{picture}}

\put(250,10){\begin{picture}(400,420)

\put(130,290){\unitlength=0.20mm 
\begin{picture}(180,130)
\put(0,25){\begin{picture}(0,0)
\put(70,60){\begin{picture}(40,30) 
\put(0,0){\circle*{3}} 
\put(0,0){\circle{7}} 
\put(40,0){\circle*{5}} 
\put(20,30){\circle{6}} 
\put(20,30){\circle{12}} 
\put(0,0){\line(1,0){40}}
\put(40,0){\line(-2,3){20}} 
\put(20,30){\line(-2,-3){20}} 
\end{picture}} 
\qbezier[920](30,0)(90,-40)(150,0) 
\put(30,0){\circle*{5}} 
\put(90,0){\circle{6}} 
\put(150,0){\circle*{3}} 
\put(150,0){\circle{7}} 
\put(10,60){\circle*{3}} 
\put(10,60){\circle{7}} 
\put(50,60){\circle*{5}} 
\put(130,60){\circle*{3}} 
\put(130,60){\circle{7}} 
\put(170,60){\circle*{5}} 
\put(0,90){\circle{6}} 
\put(180,90){\circle{6}} 
\put(0,90){\circle{12}} 
\put(180,90){\circle{12}} 
\put(0,90){\line(1,-3){30.0}} 
\put(10,60){\line(1,0){40}}
\put(0,90){\line(5,-3){150}} 
\put(110,60){\line(-1,-3){20}} 
\put(90,0){\line(-1,3){20}} 
\put(180,90){\line(-5,-3){150}} 
\put(130,60){\line(1,0){40}} 
\put(180,90){\line(-1,-3){30}} 
\put(70,60){\line(-2,-3){40}} 
\put(110,60){\line(2,-3){40}} 
\put(30,1.5){\line(1,0){120}} 
\put(30,-1.5){\line(1,0){120}} 
\put(90,0){\line(-4,3){80}} 
\put(90,0){\line(4,3){80}} 
\put(50,60){\line(2,-3){40}} 
\put(90,0){\line(2,3){40}} 
\put(-2,102){\makebox(0,0){\scriptsize 12}} 
\put(90,102){\makebox(0,0){\scriptsize 12}} 
\put(182,102){\makebox(0,0){\scriptsize 12}} 
\put(-6,63){\makebox(0,0){\scriptsize 26}} 
\put(68,73){\makebox(0,0){\scriptsize 26}} 
\put(132,73){\makebox(0,0){\scriptsize 26}} 
\put(47,73){\makebox(0,0){\scriptsize 26}} 
\put(112,73){\makebox(0,0){\scriptsize 26}} 
\put(185,63){\makebox(0,0){\scriptsize 26}} 
\put(12,0){\makebox(0,0){\scriptsize 42}} 
\put(90,-12){\makebox(0,0){\scriptsize 60}} 
\put(164,0){\makebox(0,0){\scriptsize 42}} 
\end{picture}}
\end{picture}}

\put(20,160){\unitlength=0.20mm 
\begin{picture}(180,130)
\put(0,25){\begin{picture}(0,0)
\put(70,60){\begin{picture}(40,30) 
\put(0,0){\circle*{3}} 
\put(0,0){\circle{7}} 
\put(40,0){\circle*{5}} 
\put(20,30){\circle{6}} 
\put(20,30){\circle{12}} 
\put(0,0){\line(1,0){40}}
\put(40,0){\line(-2,3){20}} 
\put(20,30){\line(-2,-3){20}} 
\end{picture}} 
\qbezier[920](30,0)(90,-40)(150,0) 
\put(30,0){\circle*{5}} 
\put(90,0){\circle{6}} 
\put(150,0){\circle*{3}} 
\put(150,0){\circle{7}} 
\put(10,60){\circle*{3}} 
\put(10,60){\circle{7}} 
\put(50,60){\circle*{5}} 
\put(130,60){\circle*{3}} 
\put(130,60){\circle{7}} 
\put(170,60){\circle*{5}} 
\put(0,90){\circle{6}} 
\put(180,90){\circle{6}} 
\put(0,90){\circle{12}} 
\put(180,90){\circle{12}} 
\put(0,90){\line(1,-3){30.0}} 
\put(10,60){\line(1,0){40}}
\put(0,90){\line(5,-3){150}} 
\put(110,60){\line(-1,-3){20}} 
\put(90,0){\line(-1,3){20}} 
\put(180,90){\line(-5,-3){150}} 
\put(130,60){\line(1,0){40}} 
\put(180,90){\line(-1,-3){30}} 
\put(70,60){\line(-2,-3){40}} 
\put(110,60){\line(2,-3){40}} 
\put(30,1.5){\line(1,0){120}} 
\put(30,-1.5){\line(1,0){120}} 
\put(90,0){\line(-4,3){80}} 
\put(90,0){\line(4,3){80}} 
\put(50,60){\line(2,-3){40}} 
\put(90,0){\line(2,3){40}} 
\put(-2,102){\makebox(0,0){\scriptsize 12}} 
\put(90,102){\makebox(0,0){\scriptsize 12}} 
\put(182,102){\makebox(0,0){\scriptsize 12}} 
\put(-6,63){\makebox(0,0){\scriptsize 26}} 
\put(68,73){\makebox(0,0){\scriptsize 26}} 
\put(132,73){\makebox(0,0){\scriptsize 26}} 
\put(47,73){\makebox(0,0){\scriptsize 26}} 
\put(112,73){\makebox(0,0){\scriptsize 26}} 
\put(185,63){\makebox(0,0){\scriptsize 26}} 
\put(12,0){\makebox(0,0){\scriptsize 42}} 
\put(90,-12){\makebox(0,0){\scriptsize 60}} 
\put(164,0){\makebox(0,0){\scriptsize 42}} 
\end{picture}}
\end{picture}}

\put(230,160){\unitlength=0.20mm 
 \begin{picture}(180,130)
\put(0,25){\begin{picture}(0,0)
\put(70,60){\begin{picture}(40,30) 
\put(0,0){\circle*{3}} 
\put(0,0){\circle{7}} 
\put(40,0){\circle*{5}} 
\put(20,30){\circle{6}} 
\put(20,30){\circle{12}} 
\put(0,0){\line(1,0){40}}
\put(40,0){\line(-2,3){20}} 
\put(20,30){\line(-2,-3){20}} 
\end{picture}} 
\qbezier[920](30,0)(90,-40)(150,0) 
\put(30,0){\circle*{5}} 
\put(90,0){\circle{6}} 
\put(150,0){\circle*{3}} 
\put(150,0){\circle{7}} 
\put(10,60){\circle*{3}} 
\put(10,60){\circle{7}} 
\put(50,60){\circle*{5}} 
\put(130,60){\circle*{3}} 
\put(130,60){\circle{7}} 
\put(170,60){\circle*{5}} 
\put(0,90){\circle{6}} 
\put(180,90){\circle{6}} 
\put(0,90){\circle{12}} 
\put(180,90){\circle{12}} 
\put(0,90){\line(1,-3){30.0}} 
\put(10,60){\line(1,0){40}}
\put(0,90){\line(5,-3){150}} 
\put(110,60){\line(-1,-3){20}} 
\put(90,0){\line(-1,3){20}} 
\put(180,90){\line(-5,-3){150}} 
\put(130,60){\line(1,0){40}} 
\put(180,90){\line(-1,-3){30}} 
\put(70,60){\line(-2,-3){40}} 
\put(110,60){\line(2,-3){40}} 
\put(30,1.5){\line(1,0){120}} 
\put(30,-1.5){\line(1,0){120}} 
\put(90,0){\line(-4,3){80}} 
\put(90,0){\line(4,3){80}} 
\put(50,60){\line(2,-3){40}} 
\put(90,0){\line(2,3){40}} 
\put(-2,102){\makebox(0,0){\scriptsize 12}} 
\put(90,102){\makebox(0,0){\scriptsize 12}} 
\put(182,102){\makebox(0,0){\scriptsize 12}} 
\put(-6,63){\makebox(0,0){\scriptsize 26}} 
\put(68,73){\makebox(0,0){\scriptsize 26}} 
\put(132,73){\makebox(0,0){\scriptsize 26}} 
\put(47,73){\makebox(0,0){\scriptsize 26}} 
\put(112,73){\makebox(0,0){\scriptsize 26}} 
\put(185,63){\makebox(0,0){\scriptsize 26}} 
\put(12,0){\makebox(0,0){\scriptsize 42}} 
\put(90,-12){\makebox(0,0){\scriptsize 60}} 
\put(164,0){\makebox(0,0){\scriptsize 42}} 
\end{picture}}
\end{picture}}

\unitlength=0.25mm 

\put(0,0){\begin{picture}(140,160)
\put(70,140){\circle{4}}
\put(70,80){\circle{4}}
\put(70,50){\circle{4}}
\put(70,20){\circle{4}}
\put(90,110){\circle*{4}}
\put(20,95){\circle*{4}}
\put(20,65){\circle*{4}}
\put(20,35){\circle*{4}}
\put(50,110){\circle*{3}}
\put(120,95){\circle*{3}}
\put(120,65){\circle*{3}}
\put(120,35){\circle*{3}}
\put(50,110){\circle{5.5}}
\put(120,95){\circle{5.5}}
\put(120,65){\circle{5.5}}
\put(120,35){\circle{5.5}}

\put(70,150){\makebox(0,0){\scriptsize 42}}
\put(45,120){\makebox(0,0){\scriptsize 120}}
\put(95,120){\makebox(0,0){\scriptsize 120}}
\put(15,105){\makebox(0,0){\scriptsize 72}}
\put(15,75){\makebox(0,0){\scriptsize 72}}
\put(15,45){\makebox(0,0){\scriptsize 72}}
\put(123,105){\makebox(0,0){\scriptsize 68}}
\put(123,75){\makebox(0,0){\scriptsize 68}}
\put(123,45){\makebox(0,0){\scriptsize 68}}
\put(70,73){\makebox(0,0){\scriptsize 94}}
\put(70,43){\makebox(0,0){\scriptsize 94}}
\put(70,10){\makebox(0,0){\scriptsize 94}}

\drawline(50,110)(90,110) 
\drawline(50,110)(70,140) 
\drawline(70,140)(90,110) 
\drawline(20,95)(120,95) 
\drawline(20,65)(120,65) 
\drawline(20,35)(120,35)
\drawline(50,110)(70,80) 
\drawline(50,110)(70,50) 
\drawline(50,110)(70,20) 
\drawline(90,110)(70,80) 
\drawline(90,110)(70,50) 
\drawline(90,110)(70,20)  
\drawline(50,110)(20,95) 
\drawline(50,110)(20,65) 
\drawline(50,110)(20,35) 
\drawline(90,110)(120,95) 
\drawline(90,110)(120,65) 
\drawline(90,110)(120,35) 
\drawline(70,20)(20,35) 
\drawline(70,20)(20,95) 
\drawline(70,20)(120,35) 
\drawline(70,20)(120,95)
\drawline(70,50)(20,65) 
\drawline(70,50)(20,35)
\drawline(70,50)(120,65) 
\drawline(70,50)(120,35)
\drawline(70,80)(20,65) 
\drawline(70,80)(20,95)
\drawline(70,80)(120,65) 
\drawline(70,80)(120,95)

\end{picture}}

\put(130,0){\begin{picture}(140,160)
\put(70,140){\circle{4}}
\put(70,80){\circle{4}}
\put(70,50){\circle{4}}
\put(70,20){\circle{4}}
\put(90,110){\circle*{4}}
\put(20,95){\circle*{4}}
\put(20,65){\circle*{4}}
\put(20,35){\circle*{4}}
\put(50,110){\circle*{3}}
\put(120,95){\circle*{3}}
\put(120,65){\circle*{3}}
\put(120,35){\circle*{3}}
\put(50,110){\circle{5.5}}
\put(120,95){\circle{5.5}}
\put(120,65){\circle{5.5}}
\put(120,35){\circle{5.5}}

\drawline(50,110)(90,110) 
\drawline(50,110)(70,140) 
\drawline(70,140)(90,110) 
\drawline(20,95)(120,95) 
\drawline(20,65)(120,65) 
\drawline(20,35)(120,35)
\drawline(50,110)(70,80) 
\drawline(50,110)(70,50) 
\drawline(50,110)(70,20) 
\drawline(90,110)(70,80) 
\drawline(90,110)(70,50) 
\drawline(90,110)(70,20)  
\drawline(50,110)(20,95) 
\drawline(50,110)(20,65) 
\drawline(50,110)(20,35) 
\drawline(90,110)(120,95) 
\drawline(90,110)(120,65) 
\drawline(90,110)(120,35) 
\drawline(70,20)(20,35) 
\drawline(70,20)(20,95) 
\drawline(70,20)(120,35) 
\drawline(70,20)(120,95)
\drawline(70,50)(20,65) 
\drawline(70,50)(20,35)
\drawline(70,50)(120,65) 
\drawline(70,50)(120,35)
\drawline(70,80)(20,65) 
\drawline(70,80)(20,95)
\drawline(70,80)(120,65) 
\drawline(70,80)(120,95)

\put(70,150){\makebox(0,0){\scriptsize 60}}
\put(45,120){\makebox(0,0){\scriptsize 162}}
\put(95,120){\makebox(0,0){\scriptsize 162}}
\put(15,105){\makebox(0,0){\scriptsize 94}}
\put(15,75){\makebox(0,0){\scriptsize 94}}
\put(15,45){\makebox(0,0){\scriptsize 94}}
\put(123,105){\makebox(0,0){\scriptsize 94}}
\put(123,75){\makebox(0,0){\scriptsize 94}}
\put(123,45){\makebox(0,0){\scriptsize 94}}
\put(70,73){\makebox(0,0){\scriptsize 132}}
\put(70,43){\makebox(0,0){\scriptsize 132}}
\put(70,10){\makebox(0,0){\scriptsize 132}}

\end{picture}}

\put(260,0){\begin{picture}(140,160)

\put(70,140){\circle{4}}
\put(70,80){\circle{4}}
\put(70,50){\circle{4}}
\put(70,20){\circle{4}}
\put(90,110){\circle*{4}}
\put(20,95){\circle*{4}}
\put(20,65){\circle*{4}}
\put(20,35){\circle*{4}}
\put(50,110){\circle*{3}}
\put(120,95){\circle*{3}}
\put(120,65){\circle*{3}}
\put(120,35){\circle*{3}}
\put(50,110){\circle{5.5}}
\put(120,95){\circle{5.5}}
\put(120,65){\circle{5.5}}
\put(120,35){\circle{5.5}}

\drawline(50,110)(90,110) 
\drawline(50,110)(70,140) 
\drawline(70,140)(90,110) 
\drawline(20,95)(120,95) 
\drawline(20,65)(120,65) 
\drawline(20,35)(120,35)
\drawline(50,110)(70,80) 
\drawline(50,110)(70,50) 
\drawline(50,110)(70,20) 
\drawline(90,110)(70,80) 
\drawline(90,110)(70,50) 
\drawline(90,110)(70,20)  
\drawline(50,110)(20,95) 
\drawline(50,110)(20,65) 
\drawline(50,110)(20,35) 
\drawline(90,110)(120,95) 
\drawline(90,110)(120,65) 
\drawline(90,110)(120,35) 
\drawline(70,20)(20,35) 
\drawline(70,20)(20,95) 
\drawline(70,20)(120,35) 
\drawline(70,20)(120,95)
\drawline(70,50)(20,65) 
\drawline(70,50)(20,35)
\drawline(70,50)(120,65) 
\drawline(70,50)(120,35)
\drawline(70,80)(20,65) 
\drawline(70,80)(20,95)
\drawline(70,80)(120,65) 
\drawline(70,80)(120,95)

\put(70,150){\makebox(0,0){\scriptsize 42}}
\put(45,120){\makebox(0,0){\scriptsize 120}}
\put(95,120){\makebox(0,0){\scriptsize 120}}
\put(15,105){\makebox(0,0){\scriptsize 68}}
\put(15,75){\makebox(0,0){\scriptsize 68}}
\put(15,45){\makebox(0,0){\scriptsize 68}}
\put(123,105){\makebox(0,0){\scriptsize 72}}
\put(123,75){\makebox(0,0){\scriptsize 72}}
\put(123,45){\makebox(0,0){\scriptsize 72}}
\put(70,73){\makebox(0,0){\scriptsize 94}}
\put(70,43){\makebox(0,0){\scriptsize 94}}
\put(70,10){\makebox(0,0){\scriptsize 94}}

\end{picture}}

\end{picture}}

\end{picture}

\end{center} 
\caption{Dimensions $d_n$ and $d_x$ of the blocks for ${\cal E}_9$.} 
\label{dim-grE9} 
\end{figure}



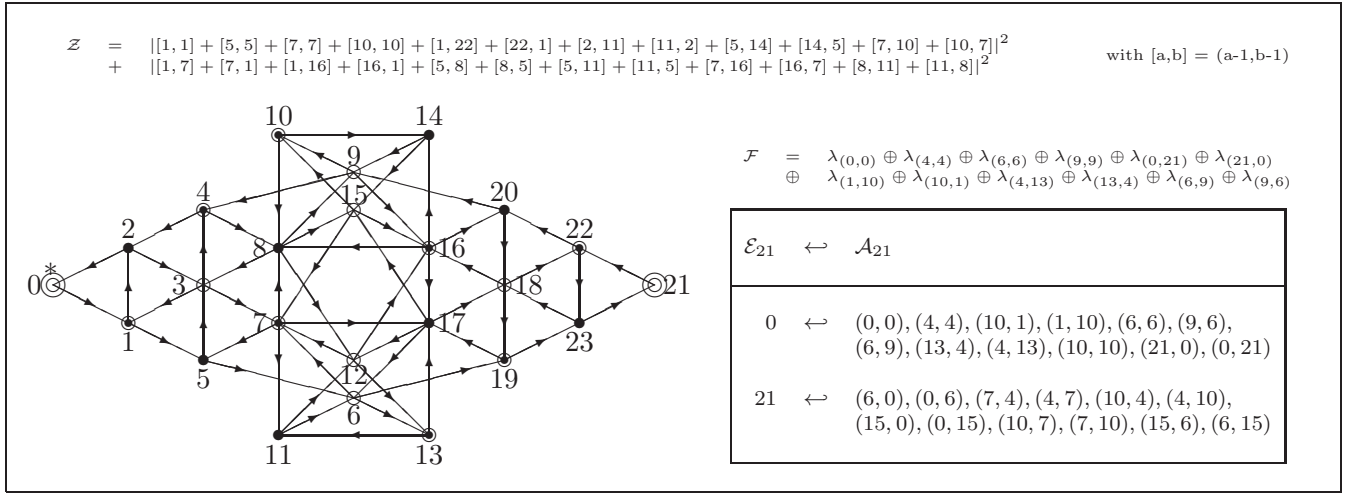
\begin{figure}[H] 
\begin{center} 
\unitlength0.25mm

\begin{picture}(710,260)

\put(0,0){\line(1,0){710}}
\put(0,260){\line(1,0){710}}
\put(0,0){\line(0,1){260}}
\put(710,0){\line(0,1){260}}

\put(25,230){\tiny $\begin{array}{rcl}
\mathcal{Z} &=& |[1,1] + [5,5] + [7,7] + [10,10] + [1,22] + [22,1] + [2,11] + [11,2] + [5,14] + [14,5] + [7,10] + [10,7] |^2 \\
{} & + & |[1,7] + [7,1] + [1,16] + [16,1] + [5,8] + [8,5] + [5,11] + [11,5] + [7,16] + [16,7] + [8,11] + [11,8] |^2
\end{array}$}

\put(585,230){\tiny with [a,b] = (a-1,b-1)}

\put(385,170){\tiny $\begin{array}{rcl}
\mathcal{F} &=& \lambda_{(0,0)} \oplus \lambda_{(4,4)} \oplus \lambda_{(6,6)} \oplus \lambda_{(9,9)} \oplus \lambda_{(0,21)} \oplus \lambda_{(21,0)} \\
{} & \oplus & \lambda_{(1,10)} \oplus \lambda_{(10,1)} \oplus \lambda_{(4,13)} \oplus \lambda_{(13,4)} \oplus \lambda_{(6,9)} \oplus \lambda_{(9,6)}
\end{array}$}

\put(385,80){\scriptsize $\begin{array}{|rcl|}
\hline
{} & {} & {} \\
\mathcal{E}_{21} &\hookleftarrow & \mathcal{A}_{21} \\
{} & {} & {} \\
\hline
{} & {} & {} \\
0 & \hookleftarrow & (0, 0), (4, 4), (10, 1), (1, 10), (6, 6), (9, 6), \\
& & (6, 9), (13, 4), (4, 13), (10, 10), (21, 0), (0, 21) \\
& & \\
21 & \hookleftarrow & (6, 0), (0, 6), (7, 4), (4, 7), (10, 4), (4, 10), \\
& & (15, 0), (0, 15), (10, 
    7), (7, 10), (15, 6), (6, 15) \\
{} & {} & {} \\
\hline
\end{array}$}

\put(15,20){\begin{picture}(340,180)

\put(10,10){\begin{picture}(320,160)

 
\put(0,60){\begin{picture}(40,40) 
\put(0,20){\circle{7}} 
\put(40,0){\circle*{4}} 
\put(40,0){\circle{7}} 
\put(40,40){\circle*{5}} 
\put(0,20){\line(2,-1){40}} 
\put(0,20){\line(2,1){40}} 
\put(40,0){\line(0,1){40}} 
\put(40,40){\vector(-2,-1){22.5}} 
\put(0,20){\vector(2,-1){22.5}} 
\put(40,0){\vector(0,1){22.5}} 
\end{picture}} 
 
\put(40,40){\begin{picture}(40,40) 
\put(40,0){\circle*{5}} 
\put(40,40){\circle{7}} 
\put(0,20){\line(2,-1){40}} 
\put(0,20){\line(2,1){40}} 
\put(40,0){\line(0,1){40}} 
\put(40,40){\vector(-2,-1){22.5}} 
\put(0,20){\vector(2,-1){22.5}} 
\put(40,0){\vector(0,1){22.5}} 
\end{picture}} 
 
\put(40,80){\begin{picture}(40,40) 
\put(40,40){\circle*{4}} 
\put(40,40){\circle{7}} 
\put(0,20){\line(2,-1){40}} 
\put(0,20){\line(2,1){40}} 
\put(40,0){\line(0,1){40}} 
\put(40,40){\vector(-2,-1){22.5}} 
\put(0,20){\vector(2,-1){22.5}} 
\put(40,0){\vector(0,1){22.5}} 
\end{picture}} 
 
\put(80,60){\begin{picture}(40,40) 
\put(40,0){\circle*{4}} 
\put(40,0){\circle{7}} 
\put(40,40){\circle*{5}} 
\put(0,20){\line(2,-1){40}} 
\put(0,20){\line(2,1){40}} 
\put(40,0){\line(0,1){40}} 
\put(40,40){\vector(-2,-1){22.5}} 
\put(0,20){\vector(2,-1){22.5}} 
\put(40,0){\vector(0,1){22.5}} 
\end{picture}}

\put(80,40){\line(4,-1){80}} 
\put(80,40){\vector(4,-1){22.5}} 
\put(80,120){\line(4,1){80}} 
\put(160,140){\vector(-4,-1){62.5}}

\put(280,60){\begin{picture}(40,40) 
\put(0,0){\circle*{5}} 
\put(40,20){\circle{7}} 
\put(0,40){\circle*{4}} 
\put(0,40){\circle{7}} 
\put(0,0){\line(0,1){40}} 
\put(0,0){\line(2,1){40}} 
\put(0,40){\line(2,-1){40}} 
\put(0,0){\vector(2,1){22.5}} 
\put(40,20){\vector(-2,1){22.5}} 
\put(0,40){\vector(0,-1){22.5}} 
\end{picture}} 
 
\put(240,40){\begin{picture}(40,40) 
\put(0,0){\circle*{4}} 
\put(0,0){\circle{7}} 
\put(0,40){\circle{7}} 
\put(0,0){\line(0,1){40}} 
\put(0,0){\line(2,1){40}} 
\put(0,40){\line(2,-1){40}} 
\put(0,0){\vector(2,1){22.5}} 
\put(40,20){\vector(-2,1){22.5}} 
\put(0,40){\vector(0,-1){22.5}} 
\end{picture}} 
 
\put(240,80){\begin{picture}(40,40) 
\put(0,40){\circle*{5}} 
\put(0,0){\line(0,1){40}} 
\put(0,0){\line(2,1){40}} 
\put(0,40){\line(2,-1){40}} 
\put(0,0){\vector(2,1){22.5}} 
\put(40,20){\vector(-2,1){22.5}} 
\put(0,40){\vector(0,-1){22.5}} 
\end{picture}} 
 
\put(200,60){\begin{picture}(40,40) 
\put(0,0){\circle*{5}} 
 
\put(0,40){\circle*{4}} 
\put(0,40){\circle{7}} 
\put(0,0){\line(0,1){40}} 
\put(0,0){\line(2,1){40}} 
\put(0,40){\line(2,-1){40}} 
\put(0,0){\vector(2,1){22.5}} 
\put(40,20){\vector(-2,1){22.5}} 
\put(0,40){\vector(0,-1){22.5}} 
\end{picture}} 
\put(160,20){\line(4,1){80}} 
\put(160,20){\vector(4,1){62.5}} 
 
\put(240,120){\line(-4,1){80}} 
\put(240,120){\vector(-4,1){22.5}}

\put(120,0){\circle*{5}} 
\put(200,0){\circle*{4}} 
\put(200,0){\circle{7}} 
\put(160,20){\circle{7}} 
\put(160,40){\circle{7}} 
\put(160,120){\circle{7}} 
\put(160,140){\circle{7}} 
\put(120,160){\circle*{4}} 
\put(120,160){\circle{7}} 
\put(200,160){\circle*{5}}

\put(200,0){\line(-1,0){80}} 
\put(200,0){\vector(-1,0){42.5}} 
\put(120,0){\line(2,1){40}} 
\put(120,0){\line(1,1){40}} 
\put(120,0){\vector(2,1){21.5}} 
\put(120,0){\vector(1,1){21.5}} 
\put(160,20){\line(2,-1){40}} 
\put(160,20){\vector(2,-1){21.5}} 
\put(160,40){\line(1,-1){40}} 
\put(160,40){\vector(1,-1){21.5}} 
\put(200,100){\line(-1,0){80}} 
\put(200,100){\vector(-1,0){42.5}} 
\put(120,100){\line(2,1){40}} 
\put(120,100){\line(1,1){40}} 
 
\put(120,100){\vector(2,1){21.5}} 
\put(120,100){\vector(1,1){21.5}} 
\put(160,120){\line(2,-1){40}} 
\put(160,120){\vector(2,-1){21.5}} 
\put(160,140){\line(1,-1){40}} 
\put(160,140){\vector(1,-1){21.5}}

\put(120,160){\line(1,0){80}} 
\put(120,160){\vector(1,0){42.5}} 
\put(200,160){\line(-2,-1){40}} 
\put(200,160){\line(-1,-1){40}} 
\put(200,160){\vector(-2,-1){21.5}} 
\put(200,160){\vector(-1,-1){21.5}} 
\put(160,140){\line(-2,1){40}} 
\put(160,140){\vector(-2,1){21.5}} 
\put(160,120){\line(-1,1){40}} 
\put(160,120){\vector(-1,1){21.5}} 
 
\put(120,60){\line(1,0){80}} 
\put(120,60){\vector(1,0){42.5}} 
\put(200,60){\line(-2,-1){40}} 
\put(200,60){\line(-1,-1){40}} 
\put(200,60){\vector(-2,-1){21.5}} 
\put(200,60){\vector(-1,-1){21.5}} 
\put(160,40){\line(-2,1){40}} 
\put(160,40){\vector(-2,1){21.5}} 
\put(160,20){\line(-1,1){40}} 
\put(160,20){\vector(-1,1){21.5}}

\put(120,0){\line(0,1){160}} 
\put(200,0){\line(0,1){160}} 
\put(120,60){\vector(0,1){22.5}} 
\put(120,60){\vector(0,-1){22.5}} 
\put(120,160){\vector(0,-1){42.5}} 
\put(200,100){\vector(0,-1){22.5}} 
\put(200,100){\vector(0,1){22.5}} 
\put(200,0){\vector(0,1){42.5}}

\put(200,60){\line(-2,3){40}} 
\put(200,60){\vector(-2,3){22.5}} 
\put(160,120){\line(-2,-3){40}} 
\put(160,120){\vector(-2,-3){22.5}}

\put(120,100){\line(2,-3){40}} 
\put(120,100){\vector(2,-3){22.5}} 
\put(160,40){\line(2,3){40}} 
\put(160,40){\vector(2,3){22.5}} 
 
\put(120,60){\line(-2,-1){40}} 
\put(120,60){\vector(-2,-1){22.5}} 
\put(80,120){\line(2,-1){40}} 
\put(80,120){\vector(2,-1){22.5}} 
 
\put(240,40){\line(-2,1){40}} 
\put(240,40){\vector(-2,1){22.5}} 
\put(200,100){\line(2,1){40}} 
\put(200,100){\vector(2,1){22.5}}

\put(-10,80){\makebox(0,0){0}} 
\put(67,80){\makebox(0,0){3}} 
\put(253,80){\makebox(0,0){18}} 
\put(332,80){\makebox(0,0){21}} 
 
\put(40,50){\makebox(0,0){1}} 
\put(40,110){\makebox(0,0){2}} 
 
\put(80,30){\makebox(0,0){5}} 
\put(80,130){\makebox(0,0){4}} 
\put(120,-10){\makebox(0,0){11}} 
\put(120,170){\makebox(0,0){10}} 
\put(200,-10){\makebox(0,0){13}} 
\put(200,170){\makebox(0,0){14}} 
\put(240,30){\makebox(0,0){19}} 
\put(240,130){\makebox(0,0){20}} 
\put(280,50){\makebox(0,0){23}} 
\put(280,110){\makebox(0,0){22}} 
\put(160,11){\makebox(0,0){6}} 
\put(160,32){\makebox(0,0){12}} 
\put(160,128){\makebox(0,0){15}} 
\put(160,149){\makebox(0,0){9}} 
\put(110,60){\makebox(0,0){7}} 
\put(110,100){\makebox(0,0){8}} 
\put(212,60){\makebox(0,0){17}} 
\put(212,100){\makebox(0,0){16}}

\put(0,80){\circle{12}} 
\put(320,80){\circle{12}} 
 
\put(-5,85){$\ast$}
\end{picture}}
\end{picture}}
 
\end{picture} 
\end{center} 
\caption{The $\mathcal{E}_{21}$ graph and $\mathcal{E}_{21} \hookleftarrow \mathcal{A}_{21}$ induction rules (for vertices $\in J$).} 
\end{figure}

\begin{figure}[H] 
\begin{center} 
\mbox{\scalebox{1.0}{\includegraphics{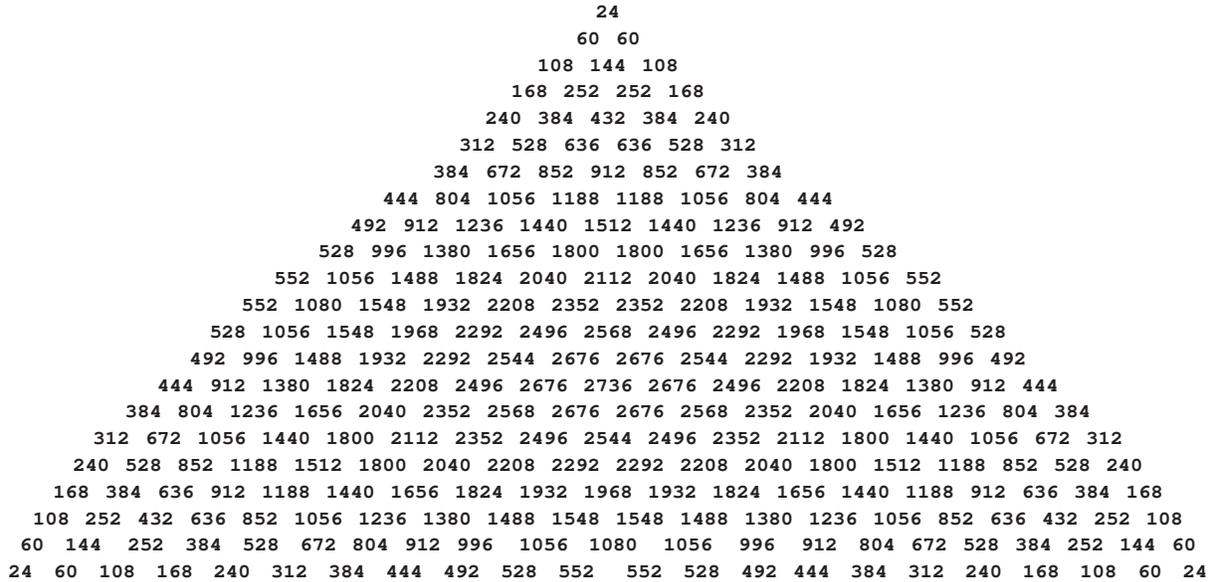}}} 
\caption{Dimension $d_n$ of the blocks labelled by vertices of the $\mathcal{A}_{21}$ graph for $\mathcal{E}_{21}$.} 
\end{center} 
\end{figure}


\begin{landscape}

\begin{figure}[H] 
\begin{center} 
\unitlength0.14mm

\begin{picture}(1590,710)

\put(0,0){\line(1,0){1590}}
\put(0,710){\line(1,0){1590}}
\put(0,0){\line(0,1){710}}
\put(1590,0){\line(0,1){710}}

\put(25,25){\begin{picture}(340,180)
\put(10,10){\begin{picture}(320,160)
\put(-20,160){$\mathbf{a \otimes 4}$}
\put(0,80){\makebox(0,0){\scriptsize 108}} 
\put(80,80){\makebox(0,0){\scriptsize 804}}
\put(160,20){\makebox(0,0){\scriptsize 948}} 
\put(160,140){\makebox(0,0){\scriptsize 948}} 
\put(160,40){\makebox(0,0){\scriptsize 1236}} 
\put(160,120){\makebox(0,0){\scriptsize 1236}} 
\put(240,80){\makebox(0,0){\scriptsize 804}} 
\put(320,80){\makebox(0,0){\scriptsize 108}} 
\put(40,60){\makebox(0,0){\scriptsize 312}} 
\put(80,120){\makebox(0,0){\scriptsize 600}} 
\put(120,60){\makebox(0,0){\scriptsize 1440}}
\put(120,160){\makebox(0,0){\scriptsize 744}} 
\put(200,0){\makebox(0,0){\scriptsize 744}} 
\put(200,100){\makebox(0,0){\scriptsize 1440}} 
\put(240,40){\makebox(0,0){\scriptsize 600}} 
\put(280,100){\makebox(0,0){\scriptsize 312}} 
\put(40,100){\makebox(0,0){\scriptsize 312}}
\put(80,40){\makebox(0,0){\scriptsize 600}} 
\put(120,100){\makebox(0,0){\scriptsize 1440}}
\put(120,0){\makebox(0,0){\scriptsize 744}} 
\put(200,160){\makebox(0,0){\scriptsize 744}} 
\put(200,60){\makebox(0,0){\scriptsize 1440}}  
\put(240,120){\makebox(0,0){\scriptsize 600}} 
\put(280,60){\makebox(0,0){\scriptsize 312}} 
\end{picture}}
\end{picture}}

\put(1225,25){\begin{picture}(340,180)
\put(10,10){\begin{picture}(320,160)
\put(-20,160){$\mathbf{a \otimes 5}$}
\put(0,80){\makebox(0,0){\scriptsize 108}} 
\put(80,80){\makebox(0,0){\scriptsize 804}}
\put(160,20){\makebox(0,0){\scriptsize 948}} 
\put(160,140){\makebox(0,0){\scriptsize 948}} 
\put(160,40){\makebox(0,0){\scriptsize 1236}} 
\put(160,120){\makebox(0,0){\scriptsize 1236}} 
\put(240,80){\makebox(0,0){\scriptsize 804}} 
\put(320,80){\makebox(0,0){\scriptsize 108}} 
\put(40,60){\makebox(0,0){\scriptsize 312}} 
\put(80,120){\makebox(0,0){\scriptsize 600}} 
\put(120,60){\makebox(0,0){\scriptsize 1440}}
\put(120,160){\makebox(0,0){\scriptsize 744}} 
\put(200,0){\makebox(0,0){\scriptsize 744}} 
\put(200,100){\makebox(0,0){\scriptsize 1440}} 
\put(240,40){\makebox(0,0){\scriptsize 600}} 
\put(280,100){\makebox(0,0){\scriptsize 312}} 
\put(40,100){\makebox(0,0){\scriptsize 312}}
\put(80,40){\makebox(0,0){\scriptsize 600}} 
\put(120,100){\makebox(0,0){\scriptsize 1440}}
\put(120,0){\makebox(0,0){\scriptsize 744}} 
\put(200,160){\makebox(0,0){\scriptsize 744}} 
\put(200,60){\makebox(0,0){\scriptsize 1440}}  
\put(240,120){\makebox(0,0){\scriptsize 600}} 
\put(280,60){\makebox(0,0){\scriptsize 312}}  
\end{picture}}
\end{picture}}

\put(25,265){\begin{picture}(340,180)
\put(10,10){\begin{picture}(320,160)
\put(-20,160){$\mathbf{a \otimes 1}$}
\put(0,80){\makebox(0,0){\scriptsize 60}} 
\put(80,80){\makebox(0,0){\scriptsize 420}}
\put(160,20){\makebox(0,0){\scriptsize 492}} 
\put(160,140){\makebox(0,0){\scriptsize 492}} 
\put(160,40){\makebox(0,0){\scriptsize 636}} 
\put(160,120){\makebox(0,0){\scriptsize 636}} 
\put(240,80){\makebox(0,0){\scriptsize 420}} 
\put(320,80){\makebox(0,0){\scriptsize 60}} 
\put(40,60){\makebox(0,0){\scriptsize 168}} 
\put(80,120){\makebox(0,0){\scriptsize 312}} 
\put(120,60){\makebox(0,0){\scriptsize 744}}
\put(120,160){\makebox(0,0){\scriptsize 384}} 
\put(200,0){\makebox(0,0){\scriptsize 384}} 
\put(200,100){\makebox(0,0){\scriptsize 744}} 
\put(240,40){\makebox(0,0){\scriptsize 312}} 
\put(280,100){\makebox(0,0){\scriptsize 168}} 
\put(40,100){\makebox(0,0){\scriptsize 168}}
\put(80,40){\makebox(0,0){\scriptsize 312}} 
\put(120,100){\makebox(0,0){\scriptsize 744}}
\put(120,0){\makebox(0,0){\scriptsize 384}} 
\put(200,160){\makebox(0,0){\scriptsize 384}} 
\put(200,60){\makebox(0,0){\scriptsize 744}}  
\put(240,120){\makebox(0,0){\scriptsize 312}} 
\put(280,60){\makebox(0,0){\scriptsize 168}} 
\end{picture}}
\end{picture}}

\put(1225,265){\begin{picture}(340,180)
\put(10,10){\begin{picture}(320,160)
\put(-20,160){$\mathbf{a \otimes 2}$}
\put(0,80){\makebox(0,0){\scriptsize 60}} 
\put(80,80){\makebox(0,0){\scriptsize 420}}
\put(160,20){\makebox(0,0){\scriptsize 492}} 
\put(160,140){\makebox(0,0){\scriptsize 492}} 
\put(160,40){\makebox(0,0){\scriptsize 636}} 
\put(160,120){\makebox(0,0){\scriptsize 636}} 
\put(240,80){\makebox(0,0){\scriptsize 420}} 
\put(320,80){\makebox(0,0){\scriptsize 60}} 
\put(40,60){\makebox(0,0){\scriptsize 168}} 
\put(80,120){\makebox(0,0){\scriptsize 312}} 
\put(120,60){\makebox(0,0){\scriptsize 744}}
\put(120,160){\makebox(0,0){\scriptsize 384}} 
\put(200,0){\makebox(0,0){\scriptsize 384}} 
\put(200,100){\makebox(0,0){\scriptsize 744}} 
\put(240,40){\makebox(0,0){\scriptsize 312}} 
\put(280,100){\makebox(0,0){\scriptsize 168}} 
\put(40,100){\makebox(0,0){\scriptsize 168}}
\put(80,40){\makebox(0,0){\scriptsize 312}} 
\put(120,100){\makebox(0,0){\scriptsize 744}}
\put(120,0){\makebox(0,0){\scriptsize 384}} 
\put(200,160){\makebox(0,0){\scriptsize 384}} 
\put(200,60){\makebox(0,0){\scriptsize 744}}  
\put(240,120){\makebox(0,0){\scriptsize 312}} 
\put(280,60){\makebox(0,0){\scriptsize 168}} 
\end{picture}}
\end{picture}}

\put(825,505){\begin{picture}(340,180)
\put(10,10){\begin{picture}(320,160)
\put(-20,160){$\mathbf{a \otimes 6}$}
\put(0,80){\makebox(0,0){\scriptsize 168}} 
\put(80,80){\makebox(0,0){\scriptsize 1272}}
\put(160,20){\makebox(0,0){\scriptsize 1512}} 
\put(160,140){\makebox(0,0){\scriptsize 1512}} 
\put(160,40){\makebox(0,0){\scriptsize 1968}} 
\put(160,120){\makebox(0,0){\scriptsize 1968}} 
\put(240,80){\makebox(0,0){\scriptsize 1272}} 
\put(320,80){\makebox(0,0){\scriptsize 168}} 
\put(40,60){\makebox(0,0){\scriptsize 492}} 
\put(80,120){\makebox(0,0){\scriptsize 948}} 
\put(120,60){\makebox(0,0){\scriptsize 2292}}
\put(120,160){\makebox(0,0){\scriptsize 1188}} 
\put(200,0){\makebox(0,0){\scriptsize 1188}} 
\put(200,100){\makebox(0,0){\scriptsize 2292}} 
\put(240,40){\makebox(0,0){\scriptsize 948}} 
\put(280,100){\makebox(0,0){\scriptsize 492}} 
\put(40,100){\makebox(0,0){\scriptsize 492}}
\put(80,40){\makebox(0,0){\scriptsize 948}} 
\put(120,100){\makebox(0,0){\scriptsize 2292}}
\put(120,0){\makebox(0,0){\scriptsize 1188}} 
\put(200,160){\makebox(0,0){\scriptsize 1188}} 
\put(200,60){\makebox(0,0){\scriptsize 2292}}  
\put(240,120){\makebox(0,0){\scriptsize 948}} 
\put(280,60){\makebox(0,0){\scriptsize 492}} 
\end{picture}}
\end{picture}}

\put(1225,505){\begin{picture}(340,180)
\put(10,10){\begin{picture}(320,160)
\put(-20,160){$\mathbf{a \otimes 12}$}
\put(0,80){\makebox(0,0){\scriptsize 216}} 
\put(80,80){\makebox(0,0){\scriptsize 1656}}
\put(160,20){\makebox(0,0){\scriptsize 1968}} 
\put(160,140){\makebox(0,0){\scriptsize 1968}} 
\put(160,40){\makebox(0,0){\scriptsize 2568}} 
\put(160,120){\makebox(0,0){\scriptsize 2568}} 
\put(240,80){\makebox(0,0){\scriptsize 1656}} 
\put(320,80){\makebox(0,0){\scriptsize 216}} 
\put(40,60){\makebox(0,0){\scriptsize 636}} 
\put(80,120){\makebox(0,0){\scriptsize 1236}} 
\put(120,60){\makebox(0,0){\scriptsize 2988}}
\put(120,160){\makebox(0,0){\scriptsize 1548}} 
\put(200,0){\makebox(0,0){\scriptsize 1548}} 
\put(200,100){\makebox(0,0){\scriptsize 2988}} 
\put(240,40){\makebox(0,0){\scriptsize 1236}} 
\put(280,100){\makebox(0,0){\scriptsize 636}} 
\put(40,100){\makebox(0,0){\scriptsize 636}}
\put(80,40){\makebox(0,0){\scriptsize 1236}} 
\put(120,100){\makebox(0,0){\scriptsize 2988}}
\put(120,0){\makebox(0,0){\scriptsize 1548}} 
\put(200,160){\makebox(0,0){\scriptsize 1548}} 
\put(200,60){\makebox(0,0){\scriptsize 2988}}  
\put(240,120){\makebox(0,0){\scriptsize 1236}} 
\put(280,60){\makebox(0,0){\scriptsize 636}} 
\end{picture}}
\end{picture}}

\put(25,505){\begin{picture}(340,180)
\put(10,10){\begin{picture}(320,160)
\put(-20,160){$\mathbf{a \otimes 0}$}
\put(0,80){\makebox(0,0){\scriptsize 24}} 
\put(80,80){\makebox(0,0){\scriptsize 144}}
\put(160,20){\makebox(0,0){\scriptsize 168}} 
\put(160,140){\makebox(0,0){\scriptsize 168}} 
\put(160,40){\makebox(0,0){\scriptsize 216}} 
\put(160,120){\makebox(0,0){\scriptsize 216}} 
\put(240,80){\makebox(0,0){\scriptsize 144}} 
\put(320,80){\makebox(0,0){\scriptsize 24}} 
\put(40,60){\makebox(0,0){\scriptsize 60}} 
\put(80,120){\makebox(0,0){\scriptsize 108}} 
\put(120,60){\makebox(0,0){\scriptsize 252}}
\put(120,160){\makebox(0,0){\scriptsize 132}} 
\put(200,0){\makebox(0,0){\scriptsize 132}} 
\put(200,100){\makebox(0,0){\scriptsize 252}} 
\put(240,40){\makebox(0,0){\scriptsize 108}} 
\put(280,100){\makebox(0,0){\scriptsize 60}} 
\put(40,100){\makebox(0,0){\scriptsize 60}}
\put(80,40){\makebox(0,0){\scriptsize 108}} 
\put(120,100){\makebox(0,0){\scriptsize 252}}
\put(120,0){\makebox(0,0){\scriptsize 132}} 
\put(200,160){\makebox(0,0){\scriptsize 132}} 
\put(200,60){\makebox(0,0){\scriptsize 252}}  
\put(240,120){\makebox(0,0){\scriptsize 108}} 
\put(280,60){\makebox(0,0){\scriptsize 60}} 
\end{picture}}
\end{picture}}

\put(425,505){\begin{picture}(340,180)
\put(10,10){\begin{picture}(320,160)
\put(-20,160){$\mathbf{a \otimes 3}$}
\put(0,80){\makebox(0,0){\scriptsize 144}} 
\put(80,80){\makebox(0,0){\scriptsize 1080}}
\put(160,20){\makebox(0,0){\scriptsize 1272}} 
\put(160,140){\makebox(0,0){\scriptsize 1272}} 
\put(160,40){\makebox(0,0){\scriptsize 1656}} 
\put(160,120){\makebox(0,0){\scriptsize 1656}} 
\put(240,80){\makebox(0,0){\scriptsize 1080}} 
\put(320,80){\makebox(0,0){\scriptsize 144}} 
\put(40,60){\makebox(0,0){\scriptsize 420}} 
\put(80,120){\makebox(0,0){\scriptsize 804}} 
\put(120,60){\makebox(0,0){\scriptsize 1932}}
\put(120,160){\makebox(0,0){\scriptsize 996}} 
\put(200,0){\makebox(0,0){\scriptsize 996}} 
\put(200,100){\makebox(0,0){\scriptsize 1932}} 
\put(240,40){\makebox(0,0){\scriptsize 804}} 
\put(280,100){\makebox(0,0){\scriptsize 420}} 
\put(40,100){\makebox(0,0){\scriptsize 420}}
\put(80,40){\makebox(0,0){\scriptsize 804}} 
\put(120,100){\makebox(0,0){\scriptsize 1932}}
\put(120,0){\makebox(0,0){\scriptsize 996}} 
\put(200,160){\makebox(0,0){\scriptsize 996}} 
\put(200,60){\makebox(0,0){\scriptsize 1932}}  
\put(240,120){\makebox(0,0){\scriptsize 804}} 
\put(280,60){\makebox(0,0){\scriptsize 420}} 
\end{picture}}
\end{picture}}

\put(425,265){\begin{picture}(340,180)
\put(10,10){\begin{picture}(320,160)
\put(-20,160){$\mathbf{a \otimes 7}$}
\put(0,80){\makebox(0,0){\scriptsize 252}} 
\put(80,80){\makebox(0,0){\scriptsize 1932}}
\put(160,20){\makebox(0,0){\scriptsize 2292}} 
\put(160,140){\makebox(0,0){\scriptsize 2292}} 
\put(160,40){\makebox(0,0){\scriptsize 2988}} 
\put(160,120){\makebox(0,0){\scriptsize 2988}} 
\put(240,80){\makebox(0,0){\scriptsize 1932}} 
\put(320,80){\makebox(0,0){\scriptsize 252}} 
\put(40,60){\makebox(0,0){\scriptsize 744}} 
\put(80,120){\makebox(0,0){\scriptsize 1440}} 
\put(120,60){\makebox(0,0){\scriptsize 3480}}
\put(120,160){\makebox(0,0){\scriptsize 1800}} 
\put(200,0){\makebox(0,0){\scriptsize 1800}} 
\put(200,100){\makebox(0,0){\scriptsize 3480}} 
\put(240,40){\makebox(0,0){\scriptsize 1440}} 
\put(280,100){\makebox(0,0){\scriptsize 744}} 
\put(40,100){\makebox(0,0){\scriptsize 744}}
\put(80,40){\makebox(0,0){\scriptsize 1440}} 
\put(120,100){\makebox(0,0){\scriptsize 3480}}
\put(120,0){\makebox(0,0){\scriptsize 1800}} 
\put(200,160){\makebox(0,0){\scriptsize 1800}} 
\put(200,60){\makebox(0,0){\scriptsize 3480}}  
\put(240,120){\makebox(0,0){\scriptsize 1440}} 
\put(280,60){\makebox(0,0){\scriptsize 744}} 
\end{picture}}
\end{picture}}

\put(825,265){\begin{picture}(340,180)
\put(10,10){\begin{picture}(320,160)
\put(-20,160){$\mathbf{a \otimes 8}$}
\put(0,80){\makebox(0,0){\scriptsize 252}} 
\put(80,80){\makebox(0,0){\scriptsize 1932}}
\put(160,20){\makebox(0,0){\scriptsize 2292}} 
\put(160,140){\makebox(0,0){\scriptsize 2292}} 
\put(160,40){\makebox(0,0){\scriptsize 2988}} 
\put(160,120){\makebox(0,0){\scriptsize 2988}} 
\put(240,80){\makebox(0,0){\scriptsize 1932}} 
\put(320,80){\makebox(0,0){\scriptsize 252}} 
\put(40,60){\makebox(0,0){\scriptsize 744}} 
\put(80,120){\makebox(0,0){\scriptsize 1440}} 
\put(120,60){\makebox(0,0){\scriptsize 3480}}
\put(120,160){\makebox(0,0){\scriptsize 1800}} 
\put(200,0){\makebox(0,0){\scriptsize 1800}} 
\put(200,100){\makebox(0,0){\scriptsize 3480}} 
\put(240,40){\makebox(0,0){\scriptsize 1440}} 
\put(280,100){\makebox(0,0){\scriptsize 744}} 
\put(40,100){\makebox(0,0){\scriptsize 744}}
\put(80,40){\makebox(0,0){\scriptsize 1440}} 
\put(120,100){\makebox(0,0){\scriptsize 3480}}
\put(120,0){\makebox(0,0){\scriptsize 1800}} 
\put(200,160){\makebox(0,0){\scriptsize 1800}} 
\put(200,60){\makebox(0,0){\scriptsize 3480}}  
\put(240,120){\makebox(0,0){\scriptsize 1440}} 
\put(280,60){\makebox(0,0){\scriptsize 744}} 
\end{picture}}
\end{picture}}

\put(425,25){\begin{picture}(340,180)
\put(10,10){\begin{picture}(320,160)
\put(-20,160){$\mathbf{a \otimes 10}$}
\put(0,80){\makebox(0,0){\scriptsize 132}} 
\put(80,80){\makebox(0,0){\scriptsize 996}}
\put(160,20){\makebox(0,0){\scriptsize 1188}} 
\put(160,140){\makebox(0,0){\scriptsize 1188}} 
\put(160,40){\makebox(0,0){\scriptsize 1548}} 
\put(160,120){\makebox(0,0){\scriptsize 1548}} 
\put(240,80){\makebox(0,0){\scriptsize 996}} 
\put(320,80){\makebox(0,0){\scriptsize 132}} 
\put(40,60){\makebox(0,0){\scriptsize 384}} 
\put(80,120){\makebox(0,0){\scriptsize 744}} 
\put(120,60){\makebox(0,0){\scriptsize 1800}}
\put(120,160){\makebox(0,0){\scriptsize 936}} 
\put(200,0){\makebox(0,0){\scriptsize 936}} 
\put(200,100){\makebox(0,0){\scriptsize 1800}} 
\put(240,40){\makebox(0,0){\scriptsize 744}} 
\put(280,100){\makebox(0,0){\scriptsize 384}} 
\put(40,100){\makebox(0,0){\scriptsize 384}}
\put(80,40){\makebox(0,0){\scriptsize 744}} 
\put(120,100){\makebox(0,0){\scriptsize 1800}}
\put(120,0){\makebox(0,0){\scriptsize 936}} 
\put(200,160){\makebox(0,0){\scriptsize 936}} 
\put(200,60){\makebox(0,0){\scriptsize 1800}}  
\put(240,120){\makebox(0,0){\scriptsize 744}} 
\put(280,60){\makebox(0,0){\scriptsize 384}} 
\end{picture}}
\end{picture}}

\put(825,25){\begin{picture}(340,180)
\put(10,10){\begin{picture}(320,160)
\put(-20,160){$\mathbf{a \otimes 11}$}
\put(0,80){\makebox(0,0){\scriptsize 132}} 
\put(80,80){\makebox(0,0){\scriptsize 996}}
\put(160,20){\makebox(0,0){\scriptsize 1188}} 
\put(160,140){\makebox(0,0){\scriptsize 1188}} 
\put(160,40){\makebox(0,0){\scriptsize 1548}} 
\put(160,120){\makebox(0,0){\scriptsize 1548}} 
\put(240,80){\makebox(0,0){\scriptsize 996}} 
\put(320,80){\makebox(0,0){\scriptsize 132}} 
\put(40,60){\makebox(0,0){\scriptsize 384}} 
\put(80,120){\makebox(0,0){\scriptsize 744}} 
\put(120,60){\makebox(0,0){\scriptsize 1800}}
\put(120,160){\makebox(0,0){\scriptsize 936}} 
\put(200,0){\makebox(0,0){\scriptsize 936}} 
\put(200,100){\makebox(0,0){\scriptsize 1800}} 
\put(240,40){\makebox(0,0){\scriptsize 744}} 
\put(280,100){\makebox(0,0){\scriptsize 384}} 
\put(40,100){\makebox(0,0){\scriptsize 384}}
\put(80,40){\makebox(0,0){\scriptsize 744}} 
\put(120,100){\makebox(0,0){\scriptsize 1800}}
\put(120,0){\makebox(0,0){\scriptsize 936}} 
\put(200,160){\makebox(0,0){\scriptsize 936}} 
\put(200,60){\makebox(0,0){\scriptsize 1800}}  
\put(240,120){\makebox(0,0){\scriptsize 744}} 
\put(280,60){\makebox(0,0){\scriptsize 384}} 
\end{picture}}
\end{picture}}

\end{picture}

\end{center}
\caption{Dimensions $d_x$ of the blocks labelled by $Oc(\mathcal{E}_{21})$ for $\mathcal{E}_{21}$.}
\end{figure}

\end{landscape}



 \end{document}